\documentclass[a4paper,onecolumn,11pt,accepted=2022-09-18]{quantumarticle}
\pdfoutput=1

\usepackage[numbers]{natbib}
\usepackage{amsmath,amssymb}
\usepackage[utf8]{inputenc}
\usepackage[english]{babel}
\usepackage[T1]{fontenc}
\usepackage{hyperref}

\usepackage{tikz}

\usepackage{amsthm}
\usepackage{latexsym}
\usepackage{amsfonts}
\usepackage{bbm,dsfont}
\usepackage{graphicx,xcolor}
\usepackage{braket}
\usepackage{hyperref}
\usepackage[normalem]{ulem}
\usepackage{braket}
\definecolor{bottle_green}{RGB}{0,106,78}
\definecolor{celadon_green}{RGB}{47,132,124}
\definecolor{emerald}{RGB}{80,220,100}
\definecolor{jade}{RGB}{0,168,107}
\hypersetup{colorlinks,
linkcolor=bottle_green,
citecolor=celadon_green,
urlcolor=emerald,
anchorcolor=jade}
\newcommand{\R}{\mathbb{R}}
\newcommand{\C}{\mathbb{C}}
\newcommand{\N}{\mathbb{N}}

\newcommand{\BH}{\mathcal{B}(\hil)}
\newcommand{\hil}{\mathcal{H}}
\newcommand{\Id}{\mathbb{I}}
\newcommand{\tr}[1]{\mathrm{Tr}\left[ {#1} \right]}

\newcommand{\rmd}{\mathrm{d}}

\newcommand{\cqstate}{\varrho}
\newcommand{\pb}[2]{\left\{ {#1} , {#2} \right\}}
\newcommand{\acom}[2]{\left\{ {#1} , {#2} \right\}_{+}}

\def\ab{^{\alpha\beta}}

\newcommand{\hq}{H_Q}
\newcommand{\htot}{H}
\newcommand{\proj}[1]{|#1\rangle\!\langle#1|}

\begin{document}
\title{Objective trajectories in hybrid classical-quantum dynamics}
\author{Jonathan Oppenheim}
\affiliation{Department of Physics and Astronomy, University College London, Gower Street, London WC1E 6BT, United Kingdom}
\author{Carlo Sparaciari}
\affiliation{Department of Physics and Astronomy, University College London, Gower Street, London WC1E 6BT, United Kingdom}
\author{Barbara \v{S}oda}
\affiliation{Department of Physics and Astronomy, University College London, Gower Street, London WC1E 6BT, United Kingdom}
\affiliation{Dept. of Physics,
University of Waterloo, Waterloo, Ontario, Canada}
\affiliation{Perimeter Institute for Theoretical Physics, Waterloo, Ontario, Canada}
\author{Zachary Weller-Davies}
\affiliation{Department of Physics and Astronomy, University College London, Gower Street, London WC1E 6BT, United Kingdom}
\date{October 12th 2022}
\begin{abstract}
Consistent dynamics which couples classical and quantum degrees of freedom exists, provided it is stochastic. This dynamics is linear in the hybrid state, completely positive and trace preserving. One application of this is to study the back-reaction of quantum fields on space-time which does not suffer from the pathologies of the semi-classical equations. Here we introduce several toy models in which to study hybrid classical-quantum evolution, including a qubit coupled to a particle in a potential, and a quantum harmonic oscillator coupled to a classical one. We present an unravelling approach to calculate the dynamics, and provide code to numerically simulate it. Unlike the purely quantum case, the trajectories (or histories) of this unravelling can be unique, conditioned on the classical degrees of freedom for discrete realisations of the dynamics, when different jumps in the classical degrees of freedom are accompanied by the action of unique operators on the quantum system.
As a result, the ``measurement postulate'' of quantum theory is not needed; quantum systems become classical because they interact with a fundamentally classical field.

\end{abstract}
\maketitle
\section{Introduction}
We are often interested in the dynamics of composite systems where one system can be considered classical while the other must be treated quantum mechanically. In quantum thermodynamics and quantum chemistry, we often have small molecules interacting with a large thermal reservoir which can be treated classically. In measurement theory the quantum system interacts with a macroscopic device which can be considered classical. In gravity, macroscopic objects such as evaporating black holes radiate thermally, and we imagine there is a regime where we can treat the black hole space-time classically even though the radiation must still be described by quantum field theory.
 
There is a lot of debate in the literature on whether one can consistently couple quantum systems and classical ones, and many ways of doing so which although useful in some regimes, are pathological~\cite{bohr1933question,cecile2011role,dewitt1962definition,eppley1977necessity,caro1999impediments,terno2006inconsistency,salcedo1996absence,sahoo2004mixing,salcedo2012statistical,barcelo2012hybrid,marletto2017we}. Many proposals for such dynamics~\cite{boucher1988semiclassical,diosi2000quantum,anderson1995quantum,kapral1999mixed,kapral2006progress,prezhdo1997mixing,PhysRevD.18.4580,PhysRevD.20.857,bondar2019koopman,dowker2008dynamical} are not completely positive or haven't been shown to be, and since both the density matrix of a quantum system and the phase-space density of a classical system are positive functions or matrices, such maps lead to negative probabilities, and are at best an approximation to the true dynamics. Other proposals, such as semi-classical gravity, are non-linear and thus don't respect the statistical nature of the density matrix~\cite{page1981indirect,gisin1989stochastic}. Another approach, inspired by quantum measurement and control~\cite{wiseman2009quantum}, sources classical degrees of freedom such as the Newtonian potential via feedback and measurement of quantum matter~\cite{Kafri_2014,Kafri_2015,Tilloy_2016,Tilloy_2017}.

If on the other hand, one doesn't wish to introduce auxiliary processes such as measurement or collapse, then one can still construct consistent evolution laws which couple quantum and classical degrees of freedom. Two such evolution laws were introduced in~\cite{blanchard1993interaction,diosi1995quantum}. These dynamics are completely positive, trace preserving, and preserve the split between classical and quantum degrees of freedom. These dynamics are special cases of the master equation~\eqref{eq:CQ_dynamics_general}, derived in ~\cite{oppenheim_post-quantum_2018,UCLPawula}. These evolution laws have been used to study the collisionless Boltzmann equation~\cite{alicki2003completely} and even proposed as a fundamental theory coupling quantum matter to classical Newtonian gravity~\cite{Di_si_2011,poulinKITP} and quantum fields to classical space-time in the context of General Relativity\cite{oppenheim_post-quantum_2018}. Because we don't have a quantum theory of gravity, the prospect of understanding how classical space-time and quantum fields interact is exciting, whether the dynamics are fundamental or merely effective. In addition to the potential to better understand black hole evaporation, an effective theory opens up the possibility of gaining new insights into cosmology, since the quantum nature of vacuum fluctuations are important in structure formation in an expanding universe.

We will not discuss gravity in the present work, nor address the question of what degrees of freedom can or should be considered classical. Rather the aim here is to introduce basic techniques, and study some simple systems in order to build up an intuition for how classical systems couple to quantum ones. In particular, we shall see that such coupling not only leads to decoherence of the quantum system~\cite{hall2005interacting,poulinPC,oppenheim_post-quantum_2018, Tilloy_2016}, but also a ''collapse'' of the wavefunction, in the sense that the quantum system jumps to a pure state which can be uniquely determined from the classical degrees of freedom. 
In this sense, we will see that the state of the system can be considered to have some objectivity, unlike the purely quantum case where a decomposition of the system's density matrix into pure states (or evolution into a history) is not unique~\cite{Kent_1998,dowker1995properties}. We also study another feature of hybrid dynamics, namely a trade-off between decoherence and diffusion. Quantum systems which have long coherence times necessarily have high diffusion in the classical degrees of freedom. Here, we shall exhibit systems which have this feature. That there is necessarily a trade-off between decoherence of the quantum system and diffusion of the classical one, will be shown in \cite{dec_Vs_diff}.

We shall here consider a qubit and a quantum harmonic oscillator that are coupled to classical degrees of freedom that are described by a phase-space manifold $\mathcal{M}$, and for simplicity in the following we take $\mathcal{M} \equiv \R^{2n}$, the space of position and momentum  $z=q,p$ for $n$ particles. The quantum degrees of freedom are described by a Hilbert space $\hil$, which might be finite or infinite dimensional. 
Given the Hilbert space, we define the set of positive semi-definite operators as $S(\hil)$. The object describing the state of this composite system at a given time is the map $\cqstate : \mathcal{M} \rightarrow S(\hil)$ such that, when the quantum degrees of freedom are traced out, returns a valid probability distribution over the phase-space. We will call this the hybrid-density, and in this paper we are using natural units, setting $\hbar = 1$. The hybrid density is a probability density over a phase space, taking values in the collective variable $z=(q,p)$. As such, it need not be subnormalized for each $z$ but it must be a normalized distribution once integrated over all classical configurations. Mathematically, we have that the distribution $\text{prob}(z) = \tr{\cqstate(z)}$ is such that $\text{prob}(z) \geq 0$ for all $z \in \mathcal{M}$, and it is normalized $\int_{\mathcal{M}} \rmd z \, \text{prob}(z) = 1$. From the above property it is easy to show that, if the classical degrees of freedom are traced out, the resulting state $\rho = \int_{\mathcal{M}} \rmd z \, \cqstate(z)$ is a valid quantum state (i.e., a positive semi-definite operator with unit trace).
The simplest such system is the hybrid qubit, whose state can be written as
\begin{equation}
\cqstate(q,p) =
\begin{pmatrix}
u_0(q,p) & c(q,p) \\
c^{\star}(q,p) & u_1(q,p)
\end{pmatrix}
\nonumber
\end{equation}
If we integrate over phase space, we obtain the density matrix of the qubit, while if we take the trace of the density matrix, we are left with the phase-space density of a classical particle.
 
 A master equation which governs the dynamics of this hybrid state must posses several necessary properties; it is linear in the state of the composite classical-quantum system, and it is completely positive and trace preserving, so that it's action preserves the statistical interpretation of $\cqstate(q,p)$. A valid quantum-classical state is transformed to another valid quantum-classical state, with the master equation preserving the separation between classical and quantum degrees of freedom.
Two generators of this kind of dynamics have previously been proposed. In~\cite{blanchard1995event}, a master equation with diagonal coupling to Lindblad operators was considered, which reduces to the GKSL or Lindblad equation under suitable assumptions. This dynamics was shown to be the most general master equation in the case of discrete classical variables and bounded Lindblad operators by embedding the classical system in Hilbert space~\cite{poulinKITP}. However, for continuous classical degrees of freedom, one generally requires an uncountable number of Lindblad operators. D{\'i}osi, instead, considered a master equation exhibiting a diffusion term, which generates a continuous dynamics for the hybrid system~\cite{diosi1995quantum}. These two CQ master equations are special cases of Equation ~\eqref{eq:CQ_dynamics_general}~\cite{oppenheim_post-quantum_2018}.\footnote{We have since shown in~\cite{UCLPawula}, that there are two classes of CQ master equation 
which can be generated from the master equation~\eqref{eq:CQ_dynamics_general}. One class is continuous in the classical degrees of freedom, while another exhibits finite jumps in the classical phase space, the later of which we study here. Both classes of dynamics must generate diffusion in phase space and decoherence of the quantum system \cite{dec_Vs_diff}.}

It and other hybrid dynamics have also been proposed as a mechanism which leads to fundamental decoherence of quantum states \cite{hall2005interacting,Tilloy_2016,Tilloy_2017,poulinKITP,oppenheim_post-quantum_2018}  
 In this paper, we are interested in understanding the main features of the discontinuous dynamics generated by the discrete master equation, given in Eq.~\eqref{eq:CQ_master_ham}.  In order to study these properties, we generalise a well-known tools from open quantum systems, known as the \emph{unravelling} of the master equation~\cite{belavkin, dalibard_cas_mol, gardiner_par_zol, gisin1984, gisin1997}.
In open quantum systems~\cite{alicki_book,breuer-pet}, one studies the evolution of a quantum system coupled to an external environment.  The most general Markovian evolution of the quantum system given some reasonable assumptions is the GKSL or Lindblad equation~\cite{lindbald,gorini_koss_sud}, whose action over the set of quantum states forms a semi-group~\cite{davies}. The dynamics of a quantum state $\rho$ are given by
\begin{align}
    \frac{\partial \, \rho}{\partial t}
=
-i \left[ H , \rho \right]
+\sum_{\alpha,\beta} \lambda^{\alpha\beta}
L_{\alpha} \rho L^{\dagger}_{\beta}
-
\frac{1}{2}
\sum_{\alpha,\beta} \lambda\ab \left( z \right)
\acom{ L^{\dagger}_{\beta} L_{\alpha} }{ \rho }
\label{eq:gksl}
\end{align}
    with $H$ the Hamiltonian of the system, $\acom{ \cdot
}{ \cdot }$ is the anti-commutator and $L_{\alpha}, L_{\beta} \in \BH$ are {\it Lindblad operators}. Such dynamics arise, for example, when a quantum system is weakly coupled to a very large environment.

The unravelling is a technique that allows us to numerically simulate the evolution of a quantum system generated by the GKSL equation~\cite{belavkin, dalibard_cas_mol, gardiner_par_zol, carmichael, paper_on_unravelling2, paper_on_unravelling3}. Instead of considering the evolution of the full density matrix describing the system, in the unravelling one consider a single pure state, which is evolved according to a stochastic dynamics. Multiple paths in the Hilbert space are thus generated by the stochastic dynamics, and by averaging over the paths we recover the evolution of the quantum system. This procedure is computationally advantageous compared to evolving the full density matrix, in particular when the dimension of the system is large. Moreover, the unravelling technique provides a different perspective on the evolution of open quantum systems; indeed, the evolution of the system can be understood as generated by a continuous, deterministic dynamics (when the system does not interact with the environment), scattered by discrete jumps of the wave-function (when system and environment interact), which occur stochastically in time~\cite{gardiner_par_zol}. 

In the quantum case, there is generally not a unique unravelling for the master equation, so one cannot regard a particular trajectory as actually {\it happening}. Indeed, the decomposition of the dynamics in terms of Lindblad operators $L_\alpha$ and a Hamiltonian $H$ in Eq.~\eqref{eq:gksl}, is not in general unique. One can rewrite the $L_\alpha$ in terms of some other basis of operators. This presents a problem~\cite{Kent_1998,dowker1995properties} for interpretations of quantum theory which take the trajectories as giving a theory of microscopic state reduction~\cite{wiseman1996quantum}, and likewise for the decohering histories approach\cite{omnes1992consistent,griffiths1993consistent,gell1993classical,diosi1995decoherent}. In contrast, by extending the tools of unravelling to the master equation which generates the dynamics of classical-quantum systems, we shall find that, under the assumption that a unique classical shift is associated with a different Lindblad operator $L_{\alpha}$, trajectories are unique when one conditions on the classical degrees of freedom. This imbues the unravelled trajectories with an ontological significance and allows one to interpret the state of the system as ''collapsing'' to a particular state. This collapse of the wave-function is here generated by the interaction between
classical and quantum degrees of freedom as in \cite{hall2005interacting,Tilloy_2016,Tilloy_2017,poulinKITP,oppenheim_post-quantum_2018}, rather than the introduction of an ad-hoc field as in spontaneous collapse models~\cite{ghirardi_unified_1986,pearle_combining_1989,ghirardi_markov_1990,bassi2013models}. Objective trajectories in unitary quantum theory have previously been studied in \cite{zurek1982environment,paz1993environment,zurek2006decoherence,Brun_2002}, where the environment plays this role of the classical system in the limit where we take it's size to infinity.
Classical systems can be in definite states, and we find that a classical-quantum interaction  causes the quantum system to inherit this property. This provides a new resolution of the measurement problem of quantum theory -- quantum systems become classical because they interact with another system which is itself intrinsically classical (e.g. space-time as in \cite{oppenheim_post-quantum_2018}).

A simple example illustrates this point. Consider the purely quantum case of a qubit in an equal superposition of basis states $\ket{0}$ and $\ket{1}$. If the quantum state $\rho$ undergoes pure decoherence via the Linblad equation
\begin{align}
\frac{\partial \rho}{\partial t}=\frac{\lambda}{2}\left[\proj{0},\left[\rho,\proj{0}\right]\right]+\frac{\lambda}{2}\left[\proj{1},\left[\rho,\proj{1}\right]\right]
\label{eq:decoherence}
\end{align}
then it's state at any time is given by 
\begin{align}
    \rho(t)=e^{-\lambda t}\proj{+}+(1-e^{-\lambda t})\frac{1}{2}\Id
    \label{eq:continuous_decoherence}
\end{align}
While we might be inclined to describe the state of the system as evolving continuously from $\ket{+}:=\left(\ket{0}+\ket{1}\right)/\sqrt{2}$ to the equal mixture $\Id/2$,
others may
describe the system's evolution in terms of its state starting in $\ket{+}$ and then at some random time, suddenly {\it collapsing} to the $\ket{0}$ or $\ket{1}$ state at a rate given by $\lambda$. Still others may describe it as suddenly collapsing to any other two orthogonal states. There is however no physical meaning to these statements, since there is no way to perform an experiment which would distinguish between these different descriptions. In quantum theory, the density matrix completely determines the system, and any decomposition into an ensemble of states is arbitrary and has no physical meaning in this case. In contrast to this, the evolution law we will consider allows for the sudden change of the quantum state to be accompanied by a change in a classical degree of freedom. Since a classical degree of freedom can be monitored without disturbing the system, one can perform measurements on the quantum system conditioned on when the classical system changes, to verify that the quantum system has undergone a sudden jump and to what state, unambiguously. This example is discussed in Subsection \ref{ssec:SG_diag_lin}. 

However, one needn't make such an interpretational commitment. The unravelling can merely be considered to be a calculations tool which allows us to better simulate the dynamics of some simple hybrid systems and we present an algorithm and code for this purpose. We study both the hybrid qubit and hybrid oscillator numerically and analytically, and explore a relation between the diffusion of the classical degrees of freedom in phase space and the rate of decoherence of the quantum degrees of freedom. Interestingly, this relation provides a way to experimentally test whether the master equation can be used to construct a "post-" quantum theory of gravity~\cite{oppenheim_post-quantum_2018}.  We also study the energy of the composite system, defined in terms of the sum of the classical and quantum Hamiltonian operators. Because this composite energy is not the generator of time-translations, there is no reason to expect it to be conserved~\cite{bps} -- Noether's theorem doesn't apply if the dynamics is not unitary. Nonetheless, insofar as one can have a classical-quantum Hamiltonian, one might want to keep violations of its conservation small. There are various mechanisms one can employ for this purpose which we will briefly mention in the discussion. In the gravitational setting, the more natural question to consider is whether the constraints are preserved in time, a discussion which we reserve for the future~\cite{UCLconstraints}.
\par
The paper is structured as follow. In Sec.~\ref{sec:unravelling} we briefly review the hybrid master equation describing the evolution of classical-quantum systems, and we extend the unravelling technique to this setting. In Sec.~\ref{sec:SG_evo} we study two different toy models where the classical degrees of freedom interact with a finite-dimensional quantum system (a qubit). We use the unravelling technique to simulate the hybrid dynamics, and to identify its main features. In Sec.~\ref{sec:HO_evo} we consider the case in which the quantum system is infinite-dimensional (a harmonic oscillator), and we repeat our analysis in this setting, with a special focus on the decoherence of this system. We conclude in Sec.~\ref{sec:con}, and suggest various ways these models could naturally be extended, for example, by having the quantum dynamics always dependent on phase space degrees of freedom or by adding in friction terms which would control the diffusion of the classical degrees of freedom in phase space.
\section{The classical-quantum master equation and its unravelling}
\label{sec:unravelling}
In this section, we show that the most general master equation governing the dynamics of a classical-quantum (CQ) system can be unravelled, allowing us to efficiently simulate the evolution of different finite-dimensional hybrid systems. Recasting the dynamics under the unravelling formalism provides a useful perspective for understanding the evolution of CQ systems. Indeed, from this point of view the hybrid system evolves continuously in its classical and quantum degrees of freedom, and its evolution is interrupted at random times by jumps in both the classical phase-space and the Hilbert space. It is worth noting that the jumps are here due to the interactions between the classical and the quantum system, and they are not, as in the case of open quantum systems, due to the presence of an external environment. Since one can always consider the state of a hybrid classical-quantum system to be the restriction of a pure quantum state on part of an enlarged Hilbert-space, the evolution of Eq.~\eqref{eq:CQ_dynamics_general} might in fact be the result of unitary dynamics on the entire system. However, strong constrains would be placed on this global dynamics due to the form of the reduced one, and it is doubtful that such dynamics can be made completely positive and trace preserving. Dynamics which is completely positive on the reduced state of a restricted class of system and environment states, needn't be completely positive on the full state of the system when we include the environment \cite{oppenheim_post-quantum_2018}.

Another difference between the unravelling for hybrid dynamics and the one used in open system dynamics concerns the objectiveness of the jumps. In the case of open quantum systems, a given dynamics can be obtained from an entire family of jumps~\cite{breuer-pet}. There are usually many ways in which one can decompose the dynamics in terms of jumps (Lindblad operators) in Hilbert space, and so one cannot think of the unravelling as representing actual trajectories that the system makes. In contrast, for some classes of hybrid dynamics, each classical jump in phase-space uniquely determines which Lindblad operator was applied to the system. One can then imagine continuously monitoring the classical system through time without disturbing the system, enabling one to uniquely determine the sequence of Lindblad operators, and thus the evolution of the quantum state, which remains pure, conditioned on the classical degrees of freedom. We can thus think of the classical trajectories as being ones which objectively occur.  
\par
Before we can extend the unravelling technique to the case of classical-quantum dynamics, we need to formally introduce the class of hybrid systems we consider, and the form of the master equation describing their evolution. The most general Markovian CQ dynamics in the case of bounded operators $L_\alpha$ is given by
\begin{align}
\frac{\partial \, \cqstate(z,t)}{\partial t}
=&
-i \left[ \htot(z) , \cqstate(z,t) \right]
+
\int \rmd \Delta \, \sum_{\alpha,\beta} W\ab \left( z | z - \Delta \right) \,
L_{\alpha} \, \cqstate(z - \Delta,t) \, L^{\dagger}_{\beta}\nonumber
\\
&-\frac{1}{2}
\sum_{\alpha,\beta} W\ab \left( z \right)
\acom{ L^{\dagger}_{\beta} L_{\alpha} }{ \cqstate(z,t) },
\label{eq:CQ_dynamics_general}
\end{align}
where $z = \left( q_1, p_1, q_2, p_2 , \ldots \right) \in \R^{2 n}$ is the vector of phase space
coordinates for $n$ systems, and $\cqstate(z,t)$ is a classical-quantum state at time $t$. The
Hamiltonian $\htot(z)$, which appears in the commutator with the CQ state, controls the unitary evolution
of the quantum system, and in principle depends on the classical degrees of freedom.
For each $\alpha$ and $\beta$, the rate $W\ab
\left( z | z - \Delta \right)$ is non-negative, and it governs the transition of the classical degrees of freedom from $z
- \Delta$ to $z$, as well as the jump of the quantum state due to the map $L_{\alpha} \cdot
L^{\dagger}_{\beta}$. For the master
equation to preserve the norm of the quantum state (once the classical degrees of freedom have been
traced out), we need to require that
\begin{equation}
\label{eq:relation_between_rates}
W\ab \left( z \right) = \int \rmd \Delta \, W\ab \left( z + \Delta | z \right)
\quad \forall \, \alpha, \beta.
\end{equation}
\par
The hybrid master equation~\eqref{eq:CQ_dynamics_general} is completely positive (CP) over the classical and quantum degrees of freedom, and trace-preserving (TP) in the sense that the normalisation condition
$\int dz\tr{\varrho(z)}=1$
is preserved. Furthermore, its structure is such that the separation between classical and quantum degrees of freedom is preserved i.e. $\varrho(z)$ is always a un-normalized density matrix. Finally, the master equation is clearly linear in the CQ state. This ensures that the theory is operationally non-signalling: if one considers two space-like separated systems then if the dynamics is local, local operations on one of the systems does not alter the reduced density matrix of the other systems, just like in standard quantum theory.    All these properties ensure that a CQ state is mapped into another CQ state by the dynamics generated by Eq.~\eqref{eq:CQ_dynamics_general}.

In the previous paragraph we mentioned how Eq.~\eqref{eq:CQ_dynamics_general} is the most general Markovian hybrid which is completely positive. It is important to notice that the dynamics is Markovian on the whole hybrid system, that is, on both the classical and quantum degrees of freedom. However, when we restrict our focus to either the classical or quantum degrees of freedom, this is not the case anymore. Indeed, the classical (quantum) system can act as a memory for the state of the quantum (classical) one, so that the reduced dynamics is not Markovian.

In Ref.~\cite{UCLPawula}, we have since shown that the master equation~\eqref{eq:CQ_dynamics_general} gives rise to two different classes of hybrid dynamics. On the one hand, one can obtain a Fokker-Plank type of equation, where the classical-quantum degrees of freedom evolve continuously. This dynamics has been studied in~\cite{diosi1995quantum} in the case where the dynamics is Markovian even when restricted to either the classical or quantum. One can also obtain  dynamics which, as we shall see here, has finite jumps in the classical-quantum degrees of freedom. As we stressed in the introduction, in this paper we focus on the latter kind of dynamics, and explore the properties associated with a discontinuous hybrid dynamics.

Although Eq.~\eqref{eq:CQ_dynamics_general}
gives the general form of the dynamics, we are interested in evolution which in some limit reproduces classical Hamiltonin dynamics on the entire hybrid system. This was done in~\cite{oppenheim_post-quantum_2018} by expanding $W\ab$ in terms of moments of $\Delta$ and demanding that the 1st moment reduce to the Poisson bracket of the state with some Hamiltonian. The particular form we will explore here, is given by taking
matrix $W^{\alpha\beta} \left( z | z-\Delta \right)$ defined through the following equation:
\begin{equation} 
\label{eq:CQ-hamiltonian}
\int \rmd \Delta \, \sum_{\alpha, \beta} W^{\alpha\beta} \left( z | z - \Delta \right) \,
L_{\alpha} \, \cqstate(z - \Delta,t) \, L^{\dagger}_{\beta}
=
\sum_{\alpha,\beta} \frac{1}{\tau_{\alpha\beta}}
e^{\tau_{\alpha\beta} \, \pb{ h^{\alpha\beta} (z)}{\, \cdot \,}}
\, L_{\alpha} \, \cqstate(z,t) \, L^{\dagger}_{\beta},
\end{equation}
where $\tau_{\alpha\beta} > 0$ can be understood as a rate, $\pb{ \cdot }{ \cdot }$ is the Poisson bracket, and the functions $h^{\alpha \beta} (z)$ are associated with the CQ interaction Hamiltonian 
\begin{align}
    H(z)=\sum_{\alpha, \beta} h\ab(z) L_\beta^\dagger L_\alpha.
    \label{eq:int_ham}
\end{align}

The Poisson bracket in the exponential acts on the CQ state as a linear operator:

\begin{equation}
    \sum_{\alpha,\beta} \frac{1}{\tau_{\alpha\beta}}
e^{\tau_{\alpha\beta} \, \pb{ h^{\alpha\beta} (z)}{\, \cdot \,}}
\, L_{\alpha} \, \cqstate(z,t) \, L^{\dagger}_{\beta}=\sum_{\alpha,\beta} \frac{1}{\tau_{\alpha\beta}}
\left(1+\tau_{\alpha\beta} \, \pb{ h^{\alpha\beta} (z)}{ L_{\alpha}\, \cqstate(z,t)\, L^{\dagger}_{\beta}}
\,+...\right)
\end{equation}

The resulting master equation is
\begin{equation}
\label{eq:CQ_master_ham}
\frac{\partial \, \cqstate(z,t)}{\partial t}
=
-i \left[ H(z) , \cqstate(z,t) \right]
+
\sum_{\alpha,\beta} \frac{1}{\tau_{\alpha\beta}} 
\left(
e^{\tau_{\alpha\beta} \, \pb{h^{\alpha\beta} (z)}{\, \cdot \,}} \, L_{\alpha} \, \cqstate(z,t) \, L^{\dagger}_{\beta}
-
\frac{1}{2} \acom{ L^{\dagger}_{\beta} L_{\alpha} }{ \cqstate(z,t) }
\right).
\end{equation}

The dynamics in this equation are completely determined by the choice of the free parameters $\tau_{\alpha\beta}$, and the choice of the Lindblad operators $L_{\alpha}$ and the functions $h^{\alpha\beta}(z)$.

We see that to 0'th order in the $\tau_{\alpha\beta}$, this master equation reproduces Hamiltonian dynamics in the appropriate limit. Namely, if we expand the exponential in the right hand side of the equation for small values of the parameters $\tau_{\alpha\beta}$'s, we obtain
\begin{align}
\frac{\partial \, \cqstate(z,t)}{\partial t}
=&
-i \left[ \htot(z) , \cqstate(z,t) \right]
+
\sum_{\alpha,\beta} \frac{1}{\tau_{\alpha\beta}} 
\left(
L_{\alpha} \, \cqstate(z,t) \, L^{\dagger}_{\beta}
-
\frac{1}{2} \acom{ L^{\dagger}_{\beta} L_{\alpha} }{ \cqstate(z,t) }
\right)
\nonumber\\
&
+\sum_{\alpha,\beta} \{h\ab(z),L_\alpha\cqstate(z)L^\dagger_\beta\}
+\cdots
\label{eq:cq-expanded}
\end{align}
where the parameters $\tau_{\alpha\beta}$'s play the role of the relaxation rates.
The first term of Eq.~\eqref{eq:cq-expanded} is the dynamics of the quantum system, which can in principle depend on the classical degrees of freedom. In this way, it represents the evolution of the quantum system as controlled by the classical one. 
The terms on the first line of Eq.~\eqref{eq:cq-expanded} which include a summation over $\alpha, \beta$ weighted by $\frac{1}{\tau_{\alpha \beta}}$ define purely Lindbladian dynamics and in the limit that $\tau_{\alpha\beta}\rightarrow 0$, it can lead to rapid decoherence of the quantum system. The final term is the interaction Hamiltonian dynamics we desire, namely under the trace, it is $\tr{\{\htot,\cqstate\}}$. We can think of it as the counter-part to the commutator in the sense that it gives the back-action of the quantum system on the classical one, while the commutator reflects the influence of the classical system on the quantum one.  Indeed, in the classical limit, we expect $-i[\htot(z),\cqstate]$ to act as the Poisson bracket with respect to the degrees of freedom of the quantum system, while in this limit one can verify that ${\{h\ab,L_\alpha\cqstate L^\dagger_\beta\}}$ are the additional terms of the Poisson bracket with respect to the degrees of freedom of the classical system. We can take purely classical dynamics to correspond to identity Lindblad operators which we take to be $L_{\alpha = 0} = \Id$. If we wish this classical dynamics to be purely deterministic, then we can take $\tau_{00}\rightarrow 0$, because in this limit, the only terms which remain in the expansion of the exponential are the pure Lindbladian term which acts only on the quantum system and the deterministic term in the dynamics given by $\{H_C,\cqstate\}$ with the classical Hamiltonian  defined as
\begin{equation}
H_C(z): = h^{00} (z) \, \Id.
\end{equation}

Higher order terms can lead to diffusion in phase space. For example, at first order in $\tau$, Eq.~\eqref{eq:cq-expanded} will acquire terms of the form
\begin{align}
    {\cal D}(\cqstate)=\tau D\ab_{p,ij}(z)L_\alpha\frac{\partial^2\cqstate}{\partial p_i \partial p_j}L_\beta^\dagger
    +
    \tau D\ab_{qp,ij}(z)L_\alpha\frac{\partial^2\cqstate}{\partial q_i \partial p_j}L_\beta^\dagger
    +
    \tau D\ab_{q,ij}(z)L_\alpha\frac{\partial^2\cqstate}{\partial q_i \partial q_j}L_\beta^\dagger
    \label{eq:diffusion}
\end{align}
where for ease of presentation we have taken all $\tau_{\alpha\beta}=\tau$ and used the Einstein summation convention to suppress the summation over $\alpha, \beta$ on the right hand side of Equation \eqref{eq:diffusion}. Such terms lead to diffusion in phase-space, with $D\ab_{p,ij}(z)$ being the diffusion term usually found in the Fokker-Planck equation, or arising from Langevin dynamics. When $W\ab(z|z-\Delta)$ is deterministic in $q$, but diffusive in $p$, we shall refer to it as dynamics of the {\it Langevin type}.  It is possible to also have terms appearing in the expansion which give rise to friction. We discuss such master equations in greater detail in Sec.~\ref{sec:con}. If the dynamics is of the Langevin type, then one can verify that the equation of motion for $\dot{q}$ is unchanged by the diffusive terms appearing in Eq.~\eqref{eq:cq-expanded}. Inverting this expression gives $p(q,\dot{q})$. On the other hand, master equations which contain terms $D\ab_{qp,ij}$ and $D\ab_{q,ij}(z)$ in their expansion, we shall call {\it q-diffusive} and we will discuss such master equations in more detail elsewhere. Care should be taken in any expansion of $W\ab(z|z-\Delta)$ -- although the full dynamics is completely positive, if we truncate our expansion to any finite order, it is not.  Note that because the jumps are in classical momenta, and not position, the dynamics is local and there is no signalling. However, if we considered evolution operators which jump in position, we would encounter some violation of faster-than-light signalling. To address this problem, we
have to go to a generalization of the classical-quantum theory, which involves Lorentz symmetry and quantum field theory, which we study in \cite{pathIntegral}.
\subsection{Unravelling of the CQ dynamics}
\label{sec:unravelling_CQ}
In the quantum open system formalism, the unravelling of a master equation takes the master equation governing the evolution of the density matrix, and divides it up into {\it trajectories} -- i.e. probabilities assigned to an ensemble of histories, where each history is the evolution of a pure quantum states which is described as undergoing a continuous evolution punctuated by stochastic jumps, where the system is mapped from one state to another as a result of a Lindblad operator being applied to it. Due to the stochastic nature of the unravelling, the initial state is mapped into different final states by the same evolution. If we record each evolution in time (or trajectory) and we average over them, we re-obtain the evolution law for the density matrix, as given via the master equation (in the limit of infinite trajectories and infinitesimal time steps). As noted in the Introduction, this formalism is merely one of many possible descriptions, and doesn't have physical meaning. If we consider only the quantum system, there is no measurement which can distinguish between continuous evolution of the density matrix in the Hilbert space, and one which is described in terms of jumps, nor is there physical meaning to what basis is used to describe the jumps.

On the other hand, while the unravelling of the CQ master equation proceeds in analogy with the quantum one, it can have physical content. In this section we provide an algorithm to unravel the dynamics given by Eq.~\eqref{eq:CQ_dynamics_general}, where in addition we consider a fully classical term, given by the Poisson bracket between a classical Hamiltonian $H_C(z)$ and the hybrid state,
\begin{align*}
\frac{\partial \, \cqstate(z,t)}{\partial t}
=
&-i \left[ \htot(z) , \cqstate(z,t) \right]
+
\pb{H_C(z)}{\cqstate(z,t)} \\
&+
\int \rmd \Delta \, \sum_{\alpha,\beta} W\ab \left( z | z - \Delta \right) \,
L_{\alpha} \, \cqstate(z - \Delta,t) \, L^{\dagger}_{\beta}
-
\frac{1}{2}
\sum_{\alpha,\beta} W\ab \left( z \right)
\acom{ L^{\dagger}_{\beta} L_{\alpha} }{ \cqstate(z,t) }.
\end{align*}
Specifically, we will provide the tools to generate different classical-quantum trajectories, and we will show that the average of these trajectories coincides with the evolution of the CQ state under the above master equation.
\par
In the following, we consider the case in which the matrix of rates $W\ab \left( z | z - \Delta \right)$
are diagonal in $\alpha, \beta$. If this matrix is not diagonal in the same basis at each point, we can introduce
unitary operators $u(z)$ that diagonalize it,
\begin{equation}
W'_{\gamma} \left( z | z - \Delta \right)
=
u_{\gamma,\alpha}(z) \,
W\ab \left( z | z - \Delta \right) \,
u^{\dagger}_{\beta, \gamma}(z),
\end{equation}
and use the new Lindblad operators in the unravelling procedure, $L'_{\gamma}(z) = u_{\gamma, \alpha}(z) \, L_{\beta}$.
In this way, the same Lindblad operator acts on the left and the right of the CQ state in Eq.~\eqref{eq:CQ_dynamics_general},
and we can use the same procedure explained in the following, with the difference that now the Lindblad operators additionally
depends on the phase space coordinate $z$. The central object of this study is a pure CQ state, that is,
\begin{equation}
\label{eq:CQ_pure_state}
\ket{\cqstate(z,t)} = \delta(z - \bar{z}(t)) \, \ket{\phi(t)},
\end{equation}
where $\ket{\phi(t)}$ is a quantum state independent of the classical degrees of freedom. This object represent a
classical-quantum system, centred in the point $\bar{z}(t)$ of the phase space, and described by the quantum
state $\ket{\phi(t)}$. 
\par
The first step in the unravelling procedure is the discretisation of the dynamics of the CQ state; we fix a time interval $\delta t$, and the evolution of the CQ state is given for multiples of this interval. It is worth noting that the unravelling evolution is exact at first order in this parameter, and coincides with the one given by the master equation in the limit of $\delta t \rightarrow 0$. We now outline the procedure for generating each trajectory, and we will then show that when we average over trajectories, we obtain the CQ master equation. For a given point $z$ in phase space, at each time step we either update the
state with a continuous evolution, or with a jump in the classical and quantum degrees of freedom. When the evolution
is continuous, we apply to the quantum part of the CQ state a unitary evolution given by the (non-Hermitian) \emph{effective Hamiltonian}
\begin{equation}
\label{eq:effective_ham}
H_{\text{eff}}(z) = H(z) - \frac{i}{2} \sum_{\alpha} W^{\alpha} \left( z \right) L_{\alpha}^{\dagger} L_{\alpha},
\end{equation}
which is composed by the Hamiltonian $H(z)$, which appears in the commutator of Eq.~\eqref{eq:CQ_dynamics_general}, and by the operator appearing in the anti-commutator of the same equation. The classical part of the CQ state is instead subjected to a translation in phase space generated by the classical Hamiltonian $H_C(z)$, which appears in the Poisson bracket in the master equation. In order to evolve the CQ state with a jump, we first need to choose at random a value of the tuple $\left( \alpha, \Delta \right)$ from the (un-normalised) distribution $W^{\alpha} \left( z | z - \Delta \right)$. The $\alpha$ parameter specifies which Lindblad operator $L_{\alpha}$ is performed over the quantum state, while the $\Delta$ parameter quantifies the shift performed over the classical degrees of freedom.
\par
The unravelling algorithm is then defined by the following set of instructions, that specify how to update a pure CQ state at time $t$ under continuous or jumping evolution,
\begin{equation}
\label{eq:unrav_evo}
\ket{\cqstate(z, t + \delta t)}
=
\begin{cases}
\frac{1}{\sqrt{N_0(z,t)}} \, \left( \Id - i \, \delta t \, H_{\text{eff}}(z) \right)
\ket{\cqstate(z - \delta t \, \Omega \, \nabla H_C(z), t)}
&
\text{with} \ p_0(z,t),\\
\frac{1}{\sqrt{N_{\alpha,\Delta}(z,t)}} \, L_{\alpha}
\ket{\cqstate(z - \Delta, t)}
&
\text{with} \ p_{\alpha,\Delta}(z,t).
\end{cases}
\end{equation}
At each time step, the hybrid pure state can be updated either continuously, first line in the right hand side of the above equation, or with a jump, second line. The type of evolution is chosen at random; the probability of updating the state with a continuous evolution is given by $p_0(z,t)$, while the probability of evolving it with a jump is given by $p_{\alpha,\Delta}(z,t)$, where $\alpha$ specifies the quantum jump and $\Delta$ specifies the classical one. In the continuous evolution, the quantum degrees of freedom are evolved by the unitary operator $U(z) = e^{-i \delta t \, H_{\text{eff}}(z)}$, that we have expanded at first order in $\delta t$, while the continuous shift in the classical degrees of freedom is generated by the operator $e^{\delta t \, \pb{H_C(z)}{\cdot}}$. The gradient vector $\nabla H_C(z) = \left( \partial_{q_1} H_C(z) , \partial_{p_1} H_C(z), \ldots \right)^T \in \R^{2n}$, while
\begin{equation}
\Omega = \bigoplus_{k=1}^n \,
\begin{pmatrix}
0 & 1\\
-1 & 0
\end{pmatrix} \in M_{2n,2n}(\R),
\end{equation}
is the symplectic matrix. By iterating the procedure in Eq.~\eqref{eq:unrav_evo},we can produce a single trajectory. Different iterations provide different trajectories, since at every time step the CQ state is evolved either continuously or with a jump.
\par
The pure CQ states produced by the unravelling technique are normalized. In the following, we provide the normalization coefficient for both the continuous and jumping evolution, and the probability that either evolution occurs. For simplicity, in the rest of the section we suppress the time $t$ dependence. The normalization coefficient are given as
\begin{subequations}
\label{eq:normalisation_evo}
\begin{align}
N_0(z) &= \delta(z - \bar{z})
\left(
1 - \delta t \sum_{\alpha} W^{\alpha} \left( z \right) \bra{\phi} L_{\alpha}^{\dagger} L_{\alpha} \ket{\phi}
\right) , \\
N_{\alpha, \Delta}(z) &= \delta(z - (\bar{z} + \Delta)) \bra{\phi} L_{\alpha}^{\dagger} L_{\alpha} \ket{\phi},
\quad \forall \, \alpha, \Delta,
\end{align}
\end{subequations}
where we have used the form of the CQ pure state, given in Eq.~\eqref{eq:CQ_pure_state}, and we are only considering terms in the first order of the time step $\delta t$. In the next section we explicitly derive these coefficients. At a given time step, the probability of evolving the pure CQ state continuously or with a jump is given by,
\begin{subequations}
\label{eq:probability_evo}
\begin{align}
p_0(z) &= N_0(z), \\
\label{eq:prob_jump}
p_{\alpha, \Delta}(z) &= \delta t \, W^{\alpha} \left( z | z - \Delta \right) \, N_{\alpha, \Delta}(z), \quad \forall \, \alpha, \Delta.
\end{align}
\end{subequations}
These probabilities are chosen such that the unravelling procedure, given the appropriate averaging, reproduces the dynamics of the master equation, which is discussed in the next subsection. It is easy to see that the above distribution is a valid probability distribution over the phase space. Clearly, each
probability distribution is non-negative, and the distribution is normalised,
\begin{align}
&\int \rmd z \, \left( p_0(z) + \sum_{\alpha} \int \rmd \Delta \, p_{\alpha, \Delta}(z) \right) =\nonumber\\
&=1 - \delta t  \sum_{\alpha}
\left(
\int \rmd z \, W^{\alpha} \left( z \right) \delta(z - \bar{z})
\right)
\bra{\phi}  L_{\alpha}^{\dagger} L_{\alpha} \ket{\phi} \nonumber \\
&+ \delta t
\sum_{\alpha} \int \rmd \Delta \, 
\left(
\int \rmd z \, W^{\alpha} \left( z | z - \Delta \right) \delta(z - (\bar{z} + \Delta))
\right)
\bra{\phi}  L_{\alpha}^{\dagger} L_{\alpha} \ket{\phi} \nonumber \\
&= 1 - \delta t  \sum_{\alpha}
\left(
W^{\alpha} \left( \bar{z} \right) - \int \rmd \Delta \, W^{\alpha} \left( \bar{z} + \Delta | \bar{z} \right)
\right)
\bra{\phi}  L_{\alpha}^{\dagger} L_{\alpha} \ket{\phi} = 1,
\end{align}
where the last equality follows from Eq.~\eqref{eq:relation_between_rates}.
\par
\par
Before concluding the section, let us comment on an interesting feature of the unravelling of the hybrid master equation, which is not shared by its quantum counterpart. If a unique shift $\Delta$ in the classical degrees of freedom is associated with each Lindblad operator $L_{\alpha}$, then by monitoring the classical degrees of freedom we have complete knowledge on the jumps that occur in the quantum part of the hybrid state. Furthermore, the quantum evolution conditioned on the classical degrees of freedom is unique, since only a specific sequence of jumps could produce the sequence of shifts in the classical part. The hybrid state can be said to actually make the transitions given in Eq.~\eqref{eq:unrav_evo}. This is in contrast with the unravelling for open quantum systems, where the lack of a classical system which can be monitored makes it impossible to single out different trajectories in the Hilbert space.
\par
Clearly, the above holds as long as the assumption that a unique shift $\Delta$ in the classical degrees of freedom is associated with each Lindblad operator $L_{\alpha}$ holds. In this paper, we consider toy models where we demand the assumption to hold. However, it is interesting to understand when the assumption holds in a physical setting, when there is an interaction Hamiltonian $H(z)$ as in Eq.~\eqref{eq:int_ham}. From Eq.~\eqref{eq:CQ_master_ham} we find that the classical jumps are given by the Hamiltonian components $h^{\alpha\beta}(z)$. If these components are different for each value of $\alpha$ and $\beta$, then each shift is uniquely assigned to a different Lindblad operator. Notice that additional freedom is given by the choice of the $\tau_{\alpha\beta}$'s, which contribute to the shifts. Then, in any physical situation where classical-quantum fields interact with different coupling strengths, one should expect the above assumption to hold.

We have since shown in \cite{layton2022semiclassical} that the continuous master equation has a unique unravelling if it saturates an inequality we call the decoherence-diffusion trade-off.
\subsection{From unravelling to the CQ master equation}
We can now show that the update rule presented in Eq.~\eqref{eq:unrav_evo} reproduces, at first order in $\delta t$, the same dynamics of the hybrid master equation. First, let us derive the normalization coefficients shown in Eqs.~\eqref{eq:normalisation_evo}. To compute the normalization of the continuously evolved state, we need to Taylor expand it and truncate the expansion at first order in $\delta t$. It is straightforward to show that
\begin{equation}
\label{eq:first_order_cont}
\ket{\cqstate(z - \delta t \, \Omega \, \nabla H_C(z))}
=
\left( 1 + \delta t \, \pb{H_C(z)}{ \, \cdot \, } \right)
\ket{\cqstate(z)}
+ O(\delta t^2).
\end{equation}
The normalization coefficient is then given by
\begin{align*}
&N_0(z,t) = \\
&\bra{\cqstate(z - \delta t \, \Omega \, \nabla H_C(z), t)}
\left( \Id + i \, \delta t \, H_{\text{eff}}(z)^{\dagger} \right)
\left( \Id - i \, \delta t \, H_{\text{eff}}(z) \right)
\ket{\cqstate(z - \delta t \, \Omega \, \nabla H_C(z), t)} \\
&=
1 +
\delta t 
\pb{H_C(z)}{\braket{\cqstate(z,t)|\cqstate(z,t)}}
-
\delta t \sum_{\alpha}
W^{\alpha}(z)
\bra{\cqstate(z,t)}
L_{\alpha}^{\dagger} L_{\alpha}
\ket{\cqstate(z,t)}
+ O(\delta t^2) \\
&=
1 - \delta t \sum_{\alpha} W^{\alpha} \left( z \right) \bra{\cqstate(z,t)} L_{\alpha}^{\dagger} L_{\alpha} \ket{\cqstate(z,t)} + O(\delta t^2),
\end{align*}
where the second line follows from the Taylor expansion in Eq.~\eqref{eq:first_order_cont}, from the form of the effective Hamiltonian $H_{\text{eff}}(z)$ and from Leibniz's rule of the Poisson bracket, while the last line follows from the fact that $\ket{\cqstate(z,t)}$ is normalized. The normalization of the CQ pure state after a jump is simply
\begin{equation}
N_{\alpha, \Delta}(z) =
\bra{\cqstate(z - \Delta, t)}
L_{\alpha}^{\dagger} L_{\alpha}
\ket{\cqstate(z - \Delta, t)}.
\end{equation}
\par
We can now derive the master equation~\eqref{eq:CQ_dynamics_general} (equipped with the additional fully-classical term) from the update rule given in Eq.~\eqref{eq:unrav_evo}. To so so,
we express the state $\ket{\cqstate(z, t + \delta t)}$ as a density operator, and we re-write it as a mixture
over the possible evolutions the system undergoes. Let us first consider the continuous update for the CQ
density operator, which we refer to as $\rho_0(z,t+\delta t)$, where we truncate at first order in $\delta t$,
\begin{align*}
\rho_0 (z, t+\delta t) =&
\frac{1}{N_0(z,t)} \Bigg(
\ket{\cqstate(z, t)}\bra{\cqstate(z, t)}
+
\delta t \, \pb{H_C(z)}{\ket{\cqstate(z, t)}\bra{\cqstate(z, t)}} \\
&-i \, \delta t \left[ H(z), \ket{\cqstate(z, t)}\bra{\cqstate(z, t)} \right]
- \frac{\delta t}{2} \sum_{\alpha} W^{\alpha} \left( z \right) 
\acom{ L_{\alpha}^{\dagger} L_{\alpha} }{ \ket{\cqstate(z, t)}\bra{\cqstate(z, t)} }
\Bigg),
\end{align*}
where we have used the definition of $H_{\text{eff}}(z)$, see Eq.~\eqref{eq:effective_ham},
and $[\cdot,\cdot]$ is the commutator while $\acom{\cdot}{\cdot}$ is the anti-commutator. When
the evolution is given by an $\alpha$-jump in the quantum degrees of freedom, and by a
$\Delta$-jump in the classical ones, we find the following density operator
\begin{equation}
\rho_{\alpha,\Delta} (z, t + \delta t) =
\frac{1}{N_{\alpha,\Delta}(z,t)} \, L_{\alpha}
\ket{\cqstate(z - \Delta, t)} \bra{\cqstate(z - \Delta, t)} L^{\dagger}_{\alpha}.
\end{equation}
\par
We can now express the overall evolution of the state by weighting the different updates
with the correct probabilities, given in Eq.~\eqref{eq:probability_evo}. Thus, we find that
\begin{align}
\ket{\cqstate(z, t + \delta t)}&\bra{\cqstate(z, t + \delta t)}
=\nonumber\\
&=p_0(z, t) \, \rho_0(z, t+\delta t) + \sum_{\alpha} \int \rmd \Delta \, p_{\alpha,\Delta}(z, t) \, \rho_{\alpha, \Delta} (z, t+\delta t) \nonumber \\
&=
\ket{\cqstate(z, t)}\bra{\cqstate(z, t)}
+ \delta t \, \pb{H_C(z)}{\ket{\cqstate(z, t)}\bra{\cqstate(z, t)}} \nonumber \\
&-i \, \delta t \left[ H(z), \ket{\cqstate(z, t)}\bra{\cqstate(z, t)} \right]
- \frac{\delta t}{2} \sum_{\alpha} W^{\alpha} \left( z \right) 
\acom{ L_{\alpha}^{\dagger} L_{\alpha} }{ \ket{\cqstate(z, t)}\bra{\cqstate(z, t)} } \nonumber \\
&+ \delta t \sum_{\alpha} \int \rmd \Delta \, W^{\alpha} \left( z | z - \Delta \right)
L_{\alpha} \ket{\cqstate(z - \Delta, t)} \bra{\cqstate(z - \Delta, t)} L^{\dagger}_{\alpha},
\end{align}
and if we rearrange the above equation by moving $\ket{\cqstate(z, t)}\bra{\cqstate(z, t)}$ from the right
to the left hand side, we divide by $\delta t$, and we send $\delta t \rightarrow 0$, we obtain
\begin{align}
\label{eq:master_eq_pure}
\frac{\partial \ }{\partial t} \ket{\cqstate(z, t)}\bra{\cqstate(z, t)}
=
&- i \, \left[ H(z), \ket{\cqstate(z, t)}\bra{\cqstate(z, t)} \right]
+ \pb{H_C(z)}{\ket{\cqstate(z, t)}\bra{\cqstate(z, t)}} \nonumber \\ 
&+ \int \rmd \Delta \, \sum_{\alpha} W^{\alpha} \left( z | z - \Delta \right)
L_{\alpha} \ket{\cqstate(z - \Delta, t)} \bra{\cqstate(z - \Delta, t)} L^{\dagger}_{\alpha}  \nonumber \\
&- \frac{1}{2} \sum_{\alpha} W^{\alpha} \left( z \right)
\acom{ L_{\alpha}^{\dagger} L_{\alpha} }{ \ket{\cqstate(z, t)}\bra{\cqstate(z, t)} }.
\end{align}
As a last step, we need to move from a CQ pure state $\ket{\cqstate(z, t)}$ to a (possibly mixed) CQ state
$\cqstate(z, t)$. To do so, we perform an average over the classical-quantum configurations of the initial state,
that is, we average over a phase space distribution $P_{\text{ps}}$ of initial points $\bar{z}(t=0) = \bar{z}_0$,
as well as a distribution $P_{\text{q}}$ over the initial quantum states $\ket{\phi(t=0)} = \ket{\phi_0}$. Thus,
by applying this average over the CQ pure state $\ket{\cqstate(z, t)}$, we obtain
\begin{equation}
\cqstate(z, t) = \mathbb{E}[ \ket{\cqstate(z, t)}\bra{\cqstate(z, t)} ]
=
\int \rmd \bar{z}_0 \, \rmd \phi_0 \, P_{\text{ps}}(\bar{z}_0) P_{\text{q}}(\phi_0) \, \ket{\cqstate(z, t)}\bra{\cqstate(z, t)}.
\end{equation}
If we now perform the average in Eq.~\eqref{eq:master_eq_pure}, we obtain the CQ master equation.

\section{Main features of hybrid dynamics in qubit toy models}
\label{sec:SG_evo}
In this section we present a few toy models to illustrate the main features of the hybrid dynamics generated by the master equation~\eqref{eq:CQ_dynamics_general}. The simplest hybrid dynamics we can consider is that of a spin half particle in a classical potential, which in our examples is taken to be linear. For each of the toy models we derive, both analytically and numerically, the dynamics of the corresponding hybrid system. To numerically evolve the CQ state we use the unravelling procedure presented in the previous section. We tailor the unravelling to the specific master equation used for the toy models, see Eq.~\eqref{eq:general_SG_dynamics}. This unravelling procedure is shown in Appendix~\ref{app:unravellign_code}.
\par
\begin{figure}[!ht]
\center
\includegraphics[width=0.32\textwidth]{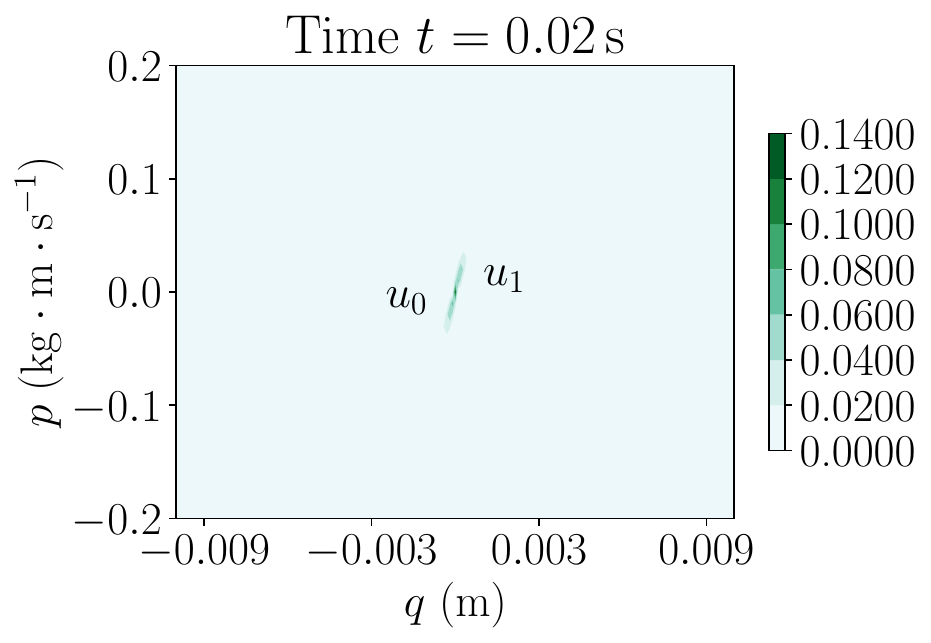}
\includegraphics[width=0.32\textwidth]{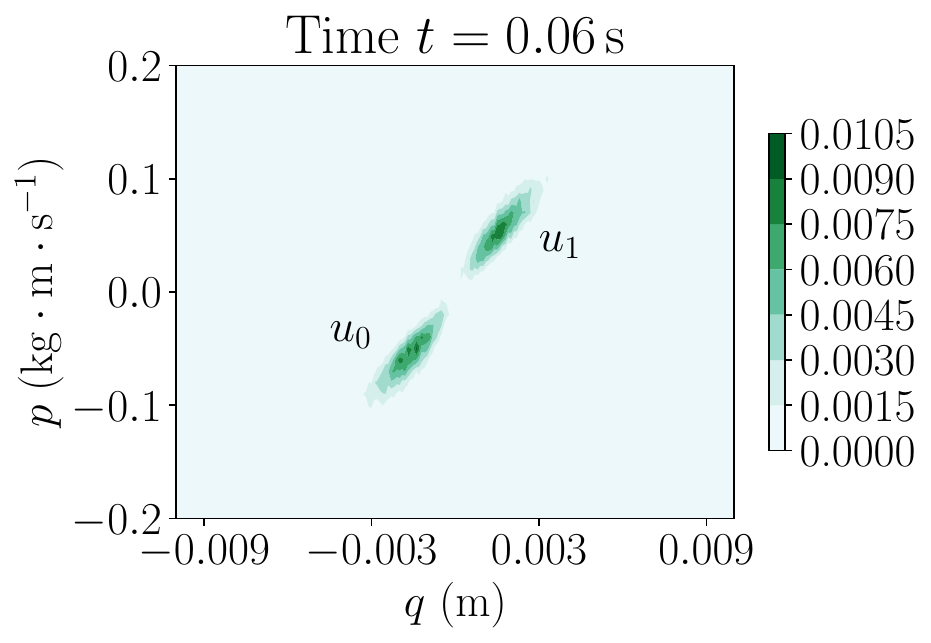}
\includegraphics[width=0.32\textwidth]{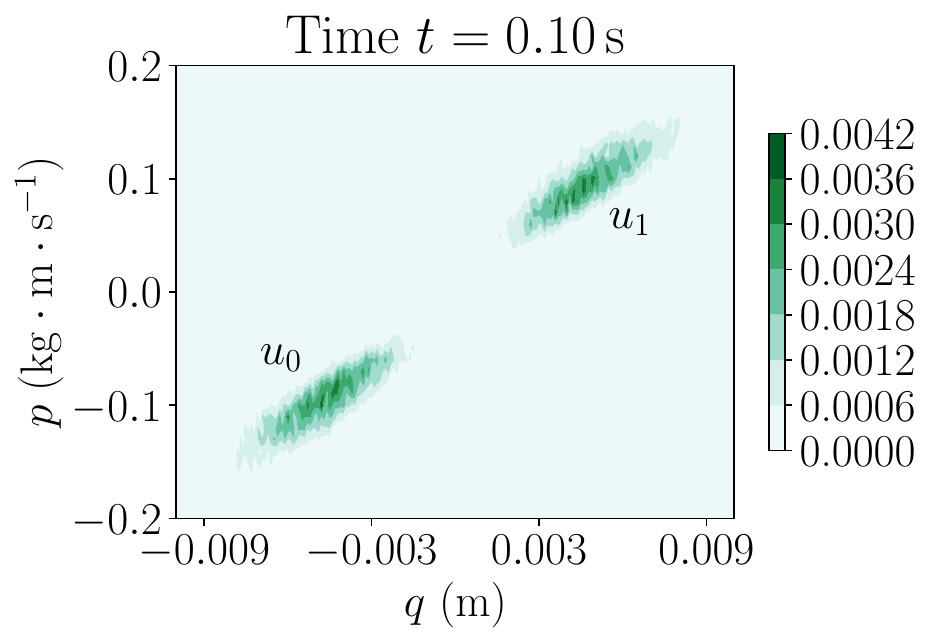}
\caption{The evolution of the population $u_0(q,p,t)$ and $u_1(q,p,t)$ in phase space, under the CQ qubit dynamics with diagonal Lindblad operators. 
The initial CQ state is $\delta(q) \delta(p) \ket{+}$, where $\ket{+} = \frac{1}{\sqrt{2}} \left( \ket{0} + \ket{1} \right)$.
The interaction with the field makes the quantum state stochastically collapse into either $\ket{0}$ or $\ket{1}$ at a rate
given by $\frac{\omega}{\tau}$. The same collapse shifts the classical momentum of the particle by $\pm B \tau$, depending
on the quantum state it projects onto. By monitoring the classical degrees of freedom, we can establish the state of the quantum
system. Indeed, a particle in the state $\ket{0}$ has negative momentum, whereas a particle in $\ket{1}$ has positive momentum. The evolution of the populations is shown for $t = 0.02 \, \mathrm{s}$, $0.06  \, \mathrm{s}$
and $0.1  \, \mathrm{s}$. The state is evolved for $t = 0.1  \, \mathrm{s}$, with time steps of $\delta t = 10^{-4}  \, \mathrm{s}$,
and the jump rate is $\tau = 10^{-2}  \, \mathrm{s}$. The constant $B = 1  \, \mathrm{J} \cdot \mathrm{s} \cdot \mathrm{m}^{-1}$,
the mass $m =  1  \, \mathrm{kg}$, and the frequencies are $\omega_0 = -\omega_1 = \omega = 1  \, \mathrm{s}^{-1}$.}
\label{fig:SG_diag_evo}
\end{figure}

The main features of the hybrid dynamics, that are highlighted in our toy models, are {\it i)} the presence of stochastic collapses of the quantum degrees of freedom, {\it ii)} a trade-off between the classical diffusion in phase space and the quantum decoherence, and {\it iii)} the fact that energy is not conserved by the dynamics. More specifically, we find that the interaction between classical and quantum degrees of freedom generates a sudden collapse in the latter. This collapse occurs when the CQ state is subjected to a jump, that is, when the classical degrees of freedom are shifted in phase space and a Lindblad operator acts over the quantum state. If the Lindblad operators are associated with unique shifts in the classical degrees of freedom, we find that the quantum dynamics can be unambiguously recorded by monitoring the classical one, see for example Fig.~\ref{fig:SG_diag_evo}. Furthermore, we find that the rate of decoherence of the quantum degrees of freedom is linked to the rate of diffusion of the classical ones; the faster the quantum state decoheres, the slower the classical degrees of freedom spread in phase space. This is not a specific feature of the toy models we study, but rather a general property of the hybrid dynamics, as shown in Ref.~\cite{dec_Vs_diff}. Finally, we find that the average energy of the toy models studied is not conserved by the hybrid dynamics. This should not come as a surprise, since the Hamiltonian operator appearing in Eq.~\eqref{eq:CQ_dynamics_general} is not the generator of the dynamics. A detailed study of the conservation laws and symmetries present in the hybrid dynamics is performed in Ref.~\cite{UCLconstraints}.
\par
Let us introduce the toy models we study in this section, namely, a single spin half particle in a classical potential. The position and momentum of the particle are taken to be the classical degrees of freedom $z = \left( q , p \right)$, while the spin of the particle ($s=\frac{1}{2}$) is the quantum one. To simplify the notation, we take the transition matrix in the hybrid master equation~\eqref{eq:CQ_master_ham} to be diagonal in $\alpha, \beta$ (if the transition matrix is not diagonal, we have shown in the previous section that we can always diagonalise it). 

For a spin-$\frac{1}{2}$ particle, the most general quantum Hamiltonian can be
written as $ \boldsymbol{\lambda} \cdot \boldsymbol{\sigma}$ where $\boldsymbol{\lambda}$ is
a vector in $\R^4$ and $\boldsymbol{\sigma} = \left\{ \Id, \sigma_x, \sigma_y, \sigma_z \right\}$ is the
vector of Pauli matrices. By having this depend on the classical degrees of freedom, we cause the classical and quantum system to interact. Let us take the interaction term of the CQ-Hamiltonian to be
\begin{equation}
\label{eq:quant_int_ham}
H_I(q,p) = B(q) \left( \omega_0 \ket{0} \bra{0} + \omega_1 \ket{1} \bra{1} \right),
\end{equation}
where the coupling constant $B(q)$ depends on the position of the particle $q$, and the operator
is diagonal in the basis $\left\{ \ket{0} , \ket{1} \right\}$. By comparing the above interaction
Hamiltonian with the general form of Eq.~\eqref{eq:int_ham}, we can easily see that we are left
some freedom in the choice of the Lindblad operators $\left\{ L_{\alpha} \right\}_{\alpha \neq 0}$
and of the functions $h^{\alpha}(z)$. By exploiting this freedom, we will study two
different CQ models describing a particle in a linear potential. In both cases, we choose a classical
Hamiltonian $H_C(q,p) = \frac{p^2}{2 m}$, that is, the one for a free particle of mass $m$. One could imagine a purely quantum evolution determined by a Hamiltonian which is not dependent on classical degrees of freedom, but in analogy with the gravitational case we will not consider it here. Finally, we demand the rates $\tau_{\alpha \neq 0}$’s to
be all equal to $\tau > 0$, so that no non-trivial Lindblad operator acts more often than the others.
\par
The CQ master equation \eqref{eq:CQ_master_ham} for the case we are considering is then given by
\begin{align}
\label{eq:general_SG_dynamics}
\frac{\partial \, \cqstate(z,t)}{\partial t}
&=
-i \left[ H_I(z) , \cqstate(z,t) \right]
+
\frac{1}{\tau_0} \left( e^{\tau_0 \, \pb{ H_C (z) }{ \, \cdot \, }} - 1 \right) \cqstate(z,t) \nonumber \\
&+
\frac{1}{\tau} \sum_{\alpha \neq 0} 
\left(
e^{\tau \, \pb{ h^{\alpha} (z) }{ \, \cdot \, }} \, L_{\alpha} \, \cqstate(z,t) \, L^{\dagger}_{\alpha}
-
\frac{1}{2} \acom{ L^{\dagger}_{\alpha} L_{\alpha} }{ \cqstate(z,t) }
\right).
\end{align}
The first term on the right is the evolution of the quantum degrees of freedom and it depends on the position of the classical particle. The next term is the purely classical evolution. If we 
take the rate $\tau_0$ to be a finite positive value, this classical evolution is stochastic, while if it tends to $0$ it is determinstic.
In the former case, we will see that the equation can be solved analytically. When $\tau_0 \rightarrow
0$, the purely classical evolution is given by the Poisson brackets between the classical Hamiltonian and the
CQ state, $\pb{ H_C (z) }{ \, \cqstate(z,t) \, }$. In this second case, we analyse numerically the dynamics generated by the master
equation, using the unravelling method shown in App.~\ref{sec:unravelling_CQ}. The final set of terms, gives the backreaction of the qubit on the classical degrees of freedom. 
\subsection{Qubit evolution with diagonal Lindblad operators in a linear potential}
\label{ssec:SG_diag_lin}
The first toy model we consider is a spin half particle interacting with a linear potential through Lindblad operators that are diagonal in the interaction Hamiltonian eigenbasis $\left\{\ket{0}, \ket{1}\right\}$. We make the following choice for the operators,
\begin{subequations}
\label{eq:lindblad_diag}
\begin{align}
L_{\alpha=1} &= \ket{0}\bra{0}, \\
L_{\alpha=2} &= \ket{1}\bra{1}.
\end{align}
\end{subequations}
Once the Lindblad operators are defined, they fix uniquely the functions $h^{\alpha}(q,p)$. Indeed, in order to identify the interaction Hamiltonian $H_I(q,p)$ in Eq.~\eqref{eq:quant_int_ham} with the expression in Eq.~\eqref{eq:int_ham}, we need to define these functions as
\begin{subequations}
\begin{align}
h^{\alpha=1}(q,p) &= \omega_0 \, B(q), \\
h^{\alpha=2}(q,p) &= \omega_1 \, B(q).
\end{align}
\end{subequations}
It is worth noting that, for a given interaction Hamiltonian, the above choice of Lindblad operators and $h^{\alpha}(q,p)$ functions is not unique. In the next section, for instance, we consider non-diagonal operators that nonetheless return the same interaction Hamiltonian, but generate a completely different dynamics. Furthermore, one could chose the same Lindblad operators as we did above, but with a different normalization. To re-obtain the correct Hamiltonian $H_I(q,p)$ we then need to scale the functions $h^{\alpha}(q,p)$ consistently. Notice that the normalization of the Lindblad operators influences the rate at which the operator is applied, while the function $h^{\alpha}(q,p)$ regulates the amplitude of the shifts in the classical degrees of freedom. This freedom in the Lindblad operators and the functions $h^{\alpha}(q,p)$ is linked to the fact that the interaction Hamiltonian is not the generator of the hybrid dynamics of Eq.~\eqref{eq:CQ_dynamics_general}, and this is highlighted by the fact that this operator is in fact not conserved by the evolution.
\par
In this example, we consider a potential linearly dependent in position, so that $B(q) = q B$, where $B$ has a constant value. We can express the CQ state $\cqstate(q,p,t)$ as
\begin{equation}
\label{eq:CQ_state_qubit}
\cqstate(q,p,t) =
\begin{pmatrix}
u_0(q,p,t) & c(q,p,t) \\
c^{\star}(q,p,t) & u_1(q,p,t)
\end{pmatrix},
\end{equation}
where $u_i(q,p,t)$ is the population of the quantum state $\ket{i}$ (for $i = 0,1$), and $c(q,p,t)$ is the coherence.
\par
We now re-write the CQ master equation given in Eq.~\eqref{eq:general_SG_dynamics}, in terms of the populations
and the coherence of the state $\cqstate(q,p,t)$. In the equations below, we take $\tau_0 \rightarrow 0$, so as to
obtain a system of non-local differential equations. When we solve the equations analytically, however, we
will require $\tau_0$ to be a positive (albeit small) constant. The dynamics of the populations and coherence is thus
given by
\begin{subequations}
\begin{align}
\label{eq:evo_pop_SG_diag}
\frac{\partial \, u_i(q,p,t)}{\partial t}
&= - \frac{p}{m} \frac{\partial \, u_i(q,p,t)}{\partial q} 
+\frac{1}{\tau} \big( u_i(q,p + \omega_i B \tau,t) - u_i(q,p,t) \big) ,  \quad i \in \left\{ 0, 1 \right\}, \\
\label{eq:evo_cohe_SG_diag}
\frac{\partial \, c(q,p,t)}{\partial t}
&= - i q B \left( \omega_0 - \omega_1 \right) c(q,p,t) - \frac{p}{m} \frac{\partial \, c(q,p,t)}{\partial q}
- \frac{1}{\tau} \, c(q,p,t).
\end{align}
\end{subequations}
It is worth noting that the equation for the coherence can be solve analytically, and the solution has the form
\begin{equation}
\label{eq:anal_sol_coher}
c(q,p,t) = \tilde{c}(q-\frac{p}{m}t) \, e^{-i B ( \omega_0 - \omega_1) \left( q - \frac{p}{2 m} t \right) t - \frac{t}{\tau}},
\end{equation}
for any function $\tilde{c}(q-\frac{p}{m}t)$ in $C^1$, the set of all continuously differentiable functions. We thus see that the coherence term decays exponentially fast. We will see that this is a result of the quantum system collapsing to the $0$ or $1$ state while making a momentum jump.
\subsubsection{Analytical and numerical evolution of populations}
In order to study the evolution of the populations analytically, we can re-express Eq.~\eqref{eq:evo_pop_SG_diag}
as a stochastic equation, see App.~\ref{app:master_stoch}. To do so, we do not send to $0$ the rate $\tau_0$ in
Eq.~\eqref{eq:general_SG_dynamics}, and we obtain the following evolution for the populations,
\begin{equation}
\label{eq:SG_stoch_pop}
\frac{\partial \, u_i(q,p,t)}{\partial t}
= 
\frac{1}{\tau_0} \big( u_i(q - \frac{p}{m} \tau_0, p, t) - u_i(q,p,t) \big)
+\frac{1}{\tau} \big( u_i(q,p + \omega_i B \tau,t) - u_i(q,p,t) \big) ,  \quad i \in \left\{ 0, 1 \right\},
\end{equation}
The solution of this equation is given in the appendix for general Hamiltonians, see Eq.~\eqref{eq:sol_stoch_two_ham}.
For the specific case we have are considering, the solution is
\begin{equation}
\label{eq:solution_SG_diag_lin}
u_i(q,p,t) = \sum_{k,n = 0}^{\infty} P_0(k) \, P_1(n) \, u_i(n,k)
\end{equation}
where $k$ is the number of position shifts, $n$ is the number of momentum shifts, and both $P_0(k)$
and $P_1(n)$ are Poisson distributions with mean value $\lambda_0 = \frac{t}{\tau_0}$ and $\lambda_1
= \frac{t}{\tau}$, respectively. The function $u_i(n,k)$ is given by
\begin{equation}
\label{eq:evo_ens_SG_diag}
u_i(k,n) = \frac{1}{\binom{n+k}{k}} \sum_{\pi \in \mathcal{S}_{n,k}} \pi
\left( 
\underbrace{e^{- \frac{p}{m} \tau_0 \partial_q} \ldots e^{- \frac{p}{m} \tau_0 \partial_q}}_k
 \,
\underbrace{e^{B \omega_i \tau \partial_p} \ldots e^{B \omega_i \tau \partial_p}}_n
\right)
u_i(q, p, 0),
\end{equation}

where $\mathcal{S}_{n,k}$ is a proper subset of the set of all permutations of the shift operators, with $k$ momentum shifts and $n$ position ones. Each element $\pi$ creates a different combination of the $n+k$ shifts.
\par
The model under consideration can additionally be solved numerically, using the unravelling method presented
in the previous section. In particular, we can use Eq.~\eqref{eq:general_SG_dynamics} to better understand the evolution
of the CQ state. In this equation, both classical and quantum degrees of freedom evolve either continuously, or
with a sudden jump. The latter evolution is the most interesting, since during a jump one of the Lindblad operators is applied to the quantum state, and this state is projected either in $\ket{0}$ or in $\ket{1}$, so that the coherence of the state is destroyed, and the classical momentum receives a kick proportional to $B \tau$, whose direction depends on the quantum state. From this point of view, this hybrid model mimics some of the features of a standard Stern-Gerlach experiment, namely the fact that when the magnetic field measures the quantum state (and collapses it into $\ket{0}$ or $\ket{1}$), it kicks the particle (that is, increase the classical momentum) either upward or downward, depending on the state the spin is collapsed into. In Fig.~\ref{fig:SG_diag_evo} we provide the numerical solution for the evolution of the populations, obtained using the unravelling technique of Sec.~\eqref{sec:unravelling_CQ}.
\begin{figure}[!ht]
\center
\includegraphics[width=.32\textwidth]{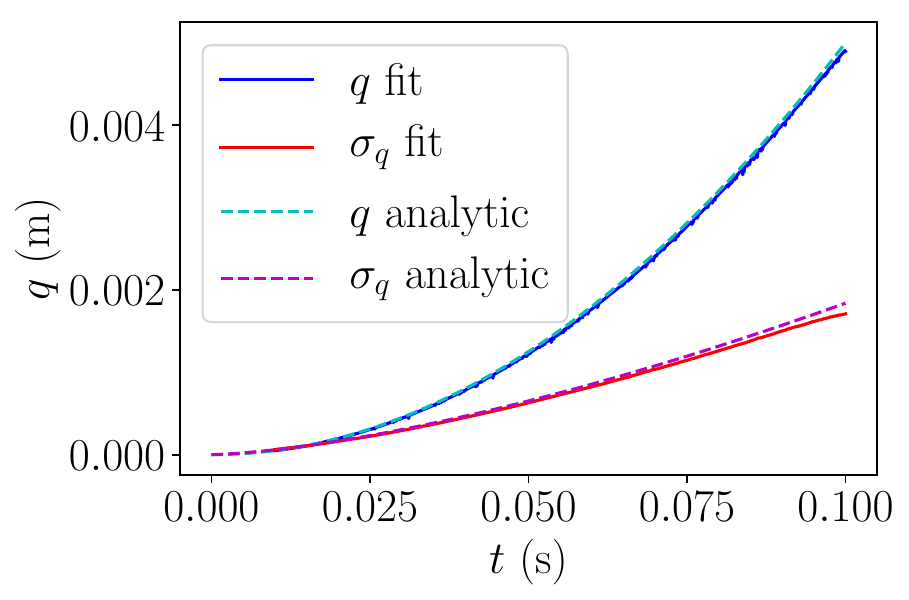}
\includegraphics[width=.32\textwidth]{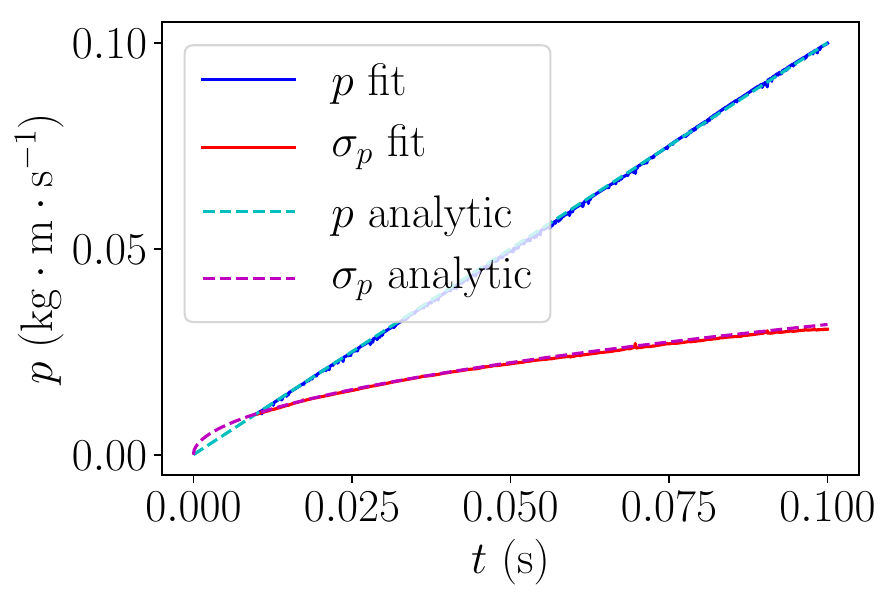}
\includegraphics[width=.32\textwidth]{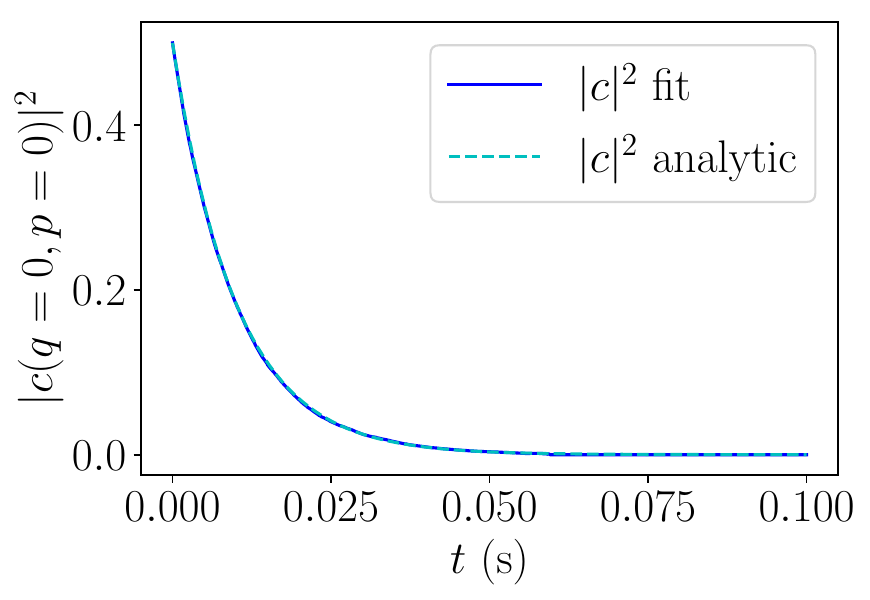}
\caption{The mean value and standard deviation of the population $u_0(q,p,t)$ in phase space.
{\bf Left.} The mean value $\langle q \rangle$ and standard deviation $\sigma_q$ of the marginal
distribution $u_0(q,t)$ (where we traced out over the momentum degree of freedom). The solid lines
are the values obtained from the numerical simulation, whereas the dashed line are obtained from
the analytical solution.  {\bf Centre.} The mean value $\langle p \rangle$ and standard deviation
$\sigma_p$ of the marginal distribution $u_0(p,t)$. {\bf Right.} The coherence $\left| c \right|^2$ of
the quantum state located in $q=0 \, \mathrm{m}$ and $p=0 \, \mathrm{kg} \cdot \mathrm{m} \cdot
\mathrm{s}^{-1}$ , as a function of time. It is worth noting that the numeric solution is
obtained with the unravelling method, in which the position is updated during the continuous evolution
step, see Eq.~\eqref{eq:cont_evo_code}. Instead, the analytic solution is obtained for the case in which the
position evolves through jumps, see Eq.~\eqref{eq:SG_stoch_pop}. In order to match the two evolution,
we have to ask the stochastic rate $\tau_0$ to be of the order of the infinitesimal time steps $\delta t$
used in the simulation. The mean value and variance of position and momentum are therefore computed
using Eqs.~\eqref{eq:SG_means} and \eqref{eq:SG_variances} for a value of $\tau_0 = \delta t$.}
\label{fig:SG_diag_variance}
\end{figure}

This is the example mentioned in the Introduction and contrasted with the purely quantum decoherence of Equation (\ref{eq:decoherence}). For the Lindblad equation, and the state of the system initially in the $\ket{+}$ state, there is no meaning to the statement that the quantum state starts suddenly collapses to $\ket{0}$ or $\ket{1}$, while here, one can continuously monitor the classical momentum without disturbing the quantum degree of freedom. Conditioned on finding a jump in momentum, one can then measure the qubit to verify that it is now in the $\ket{0}$ or $\ket{1}$ state, depending on the direction of the jump, while any previous measurement of the qubit in the $\pm$ basis, will yield the initial $\ket{+}$ state.
\subsubsection{Diffusion in phase-space and trade-off with decoherence}
We can also study the diffusion in phase-space of the populations of the CQ state, both numerically and analytically. Let us take the initial state of the CQ system to be at the origin of phase-space (position $q(t=0)=0$ and momentum $p(t=0)=0$), and in
the quantum state $\ket{+} = \frac{1}{\sqrt{2}}\left( \ket{0} + \ket{1} \right)$. In the unravelling picture when a quantum jump
occurs, the state jumps to either  $\ket{0}$ or $\ket{1}$ due to the application of the Lindblad operators of  Eq.~\eqref{eq:lindblad_diag} to the state. After
the jump, the quantum state cannot change anymore, but additional jumps do increase the
momentum of the particle. Since the initial condition is symmetric in $\ket{0}$ and $\ket{1}$, as the
evolution of the populations is, we can focus on the population $u_0(q,p,t)$. The kind of decoherence
affecting the quantum degrees of freedom in this model can be understood in terms of a leakage of
information about the quantum degrees of freedom into the classical degrees of freedom. Indeed, each
Lindblad operator $L_{\alpha}$ is here associated to a distinct classical jump in phase-space, which allows
an observer monitoring the classical degrees of freedom to know exactly in which quantum state the system
is in and when the transition occurred. Thus, the possibility of monitoring the quantum system using its position in phase space removes any
coherence in the basis $\left\{ \ket{0}, \ket{1} \right\}$. Furthermore, by conditioning over the classical
degrees of freedom, the quantum state remains pure and does not become mixed. 
\par
This is different to the standard decoherence found in the usual Lindblad equation. There, the decomposition of the dynamics into Lindblad operators is not unique, and there is no physical meaning to the jumps -- rather, the density matrix of the system evolves continuously in time with the off-diagonal matrix elements slowly decaying with time. In contrast, here, the quantum jump is accompanied by a discontinuous jump in $p$. The quantum jump thus has physical meaning, since an observer who is monitoring the classical degree of freedom can verify the quantum jump. If they observe a sudden increase (decrease) in momentum, they will expect that the quantum state is now in the $\ket{1}$ ($\ket{0}$) state, and can verify this by measuring the quantum state. The crucial ingredient here is that each classical jump corresponds to a single Lindblad operator being applied to the quantum state. 
\par
We will now see that the quantum system's coherence time is related to the diffusion in phase space. In order to obtain the diffusion constant from the numeric results, we compute the marginal distribution for position
and momentum as a function of time, and we fit them with a Gaussian distribution to extrapolate mean value
and variance. These quantities can also be computed using the solution in Eq.~\eqref{eq:solution_SG_diag_lin},
as we show in App.~\ref{app:sol_SG_diag_lin}. From the analytic solution we find that the mean value of
position and momentum is given by
\begin{subequations}
\label{eq:SG_means}
\begin{align}
\langle q \rangle(t) &= \frac{1}{2} \frac{B \omega_0}{m} t^2 + \langle q \rangle(0), \\
\langle p \rangle(t) &= B \omega_0 t + \langle p \rangle(0),
\end{align}
\end{subequations}
which is the expected solution for the position and momentum of a particle in a linear potential
given by $V(q) = q B \omega_0$. We notice that the expectation values of position and momentum
are independent of the parameter $\tau_0$, which can therefore be sent to $0$. The variance of
position and momentum can be obtained too, as we show in the Appendix~\ref{app:master_stoch}, and are given by
\begin{subequations}
\label{eq:SG_variances}
\begin{align}
\sigma_{q}^2(t) &=
\frac{1}{3} \left( \frac{B \omega_0 t}{m} \right)^2 \left( (\tau_0 + \tau) t + 5 \tau_0 \tau \right)
\xrightarrow{\tau_0 \rightarrow 0}
\frac{1}{3} \left( \frac{B \omega_0 t}{m} \right)^2 \tau t , \\
\sigma_{p}^2(t) &= (B \omega_0)^2 \tau t .
\end{align}
\end{subequations}
The variation in momentum is due to the fact that the number of momentum jumps will be normally distributed and as with a random walk, increase like $\sqrt{t}$, while the variation in position is a consequence of the momentum having a variation. We can additionally link the variance in momentum to the diffusion term appearing in the expansion of the hybrid master equation, see Eq.~\eqref{eq:diffusion}. To do so, we need to expand the exponential operator in Eq.~\eqref{eq:general_SG_dynamics} to second order, obtaining the following diffusion term,
\begin{equation}
\mathcal{D}\left( u_0(q,p,t) \right)
=
\frac{1}{\tau} \pb{\tau \, h^{\alpha=1}(q,p)}{\pb{\tau \, h^{\alpha=1}(q,p)}{u_0(q,p,t)}}
=
(B \omega_0)^2 \tau \, \frac{\partial^2}{\partial \, p^2} u_0(q,p,t).
\label{eq:SG-diffusion}
\end{equation}
It is then easy to see that the variance in momentum arises from the diffusion coefficient $D^{\alpha=1}_{p}(q,p) = (B \omega_0)^2 \tau$. This is consistent with the fact that the expansion of the master equation at first order in $\tau$ has the form of a Fokker-Planck equation with the above diffusion term.

In Fig.~\ref{fig:SG_diag_variance} we compare the numerical results with the ones obtained analytically,
and we show that indeed the numeric simulation accurately describes the evolution of the CQ state.
It is worth noting that, while the decoherence of the quantum system is inversely proportional to  $\tau$ as seen from Eq.~\eqref{eq:anal_sol_coher}, 
the diffusion in phase space is directly proportional to it. Thus, we can have two
opposite situations; for $\tau \ll 1$, the state of the quantum system quickly collapses and does not diffuses much
in phase space, following an almost Liouvillian deterministic evolution. When $\tau \gg 1$, instead, the quantum system
slowly decoheres, but the classical degrees of freedom quickly diffuse in phase space. We thus have a  trade-off between the decoherence rate and the diffusion rate. This turns out to be a standard feature of hybrid dynamics and can be
understood in terms of the moments of the Kramers-Moyal expansion of the CQ master equation, as proven in ~\cite{dec_Vs_diff}. 
\subsubsection{Energy conservation}

We now turn to the question of energy conservation in this model of hybrid dynamics. We will see that there is a violation of energy conservation at
a rate proportional to $\tau$, although it can be made arbitrarily small.
The average energy of the system can be computed using the total Hamiltonian $H(q,p) = H_C(q,p) + H_I(q,p)$, and
the solution obtained in Eq.~\eqref{eq:solution_SG_diag_lin} for the populations. Indeed, the average energy of the
CQ system is given by
\begin{equation}
\label{eq:average_energy}
\langle H \rangle (t) = 
\int \rmd q \rmd p \, \tr{H(q,p) \cqstate(q,p,t)}
= \sum_{i=0,1} \int \rmd q \rmd p \left( \frac{p^2}{2m} + q \omega_i B \right) u_i(q,p,t).
\end{equation}
Let us consider, for simplicity, that the initial state of the CQ system is $\cqstate(q,p,0) = \delta(q) \delta(p) \ket{0}\bra{0}$,
so that we only need to consider the energy contribution of the population $u_0$. Furthermore, given the specific
initial state of the system, we can express the ensemble $u_0(n,k)$, shown in Eq.~\eqref{eq:evo_ens_SG_diag}, as
$u_0(n,k) = \int \rmd \bar{q} \rmd \bar{p} \ P(\bar{q}, \bar{p} \, | \, n, k) \ u_0(\bar{q}, \bar{p})$, where
$u_0(\bar{q}, \bar{p}) = \delta(q - \bar{q}) \delta(p - \bar{p})$. The evolution of the population $u_0$ can
therefore be expressed as
\begin{equation}
u_0(q,p,t) = \int \rmd \bar{q} \rmd \bar{p}
\left(
\sum_{k,n = 0}^{\infty} P_0(k) \, P_1(n) \, P(\bar{q}, \bar{p} \, | \, n, k)
\right)
u_0(\bar{q}, \bar{p})
=
\int \rmd \bar{q} \rmd \bar{p} \, P(\bar{q}, \bar{p}) \, u_0(\bar{q}, \bar{p}),
\end{equation}
and by replacing it in Eq.~\eqref{eq:average_energy} we find that the average energy is
\begin{equation}
\langle H \rangle (t)
=
\int \rmd \bar{q} \rmd \bar{p} \, P(\bar{q}, \bar{p}) \left( \frac{\bar{p}^2}{2m} + \bar{q} \omega_0 B \right)
=
\int \rmd \bar{p} \, P(\bar{p}) \, \frac{\bar{p}^2}{2m}
+
\int \rmd \bar{q} \, P(\bar{q}) \, \bar{q} \omega_0 B,
\end{equation}
where the mean value and variance of the marginal probabilities $P(\bar{q})$ and $P(\bar{p})$ have been computed in the previous section, in Eqs.~\eqref{eq:SG_means} and \eqref{eq:SG_variances} respectively.
\par
\begin{figure}[!ht]
\center
\includegraphics[width=.5\textwidth]{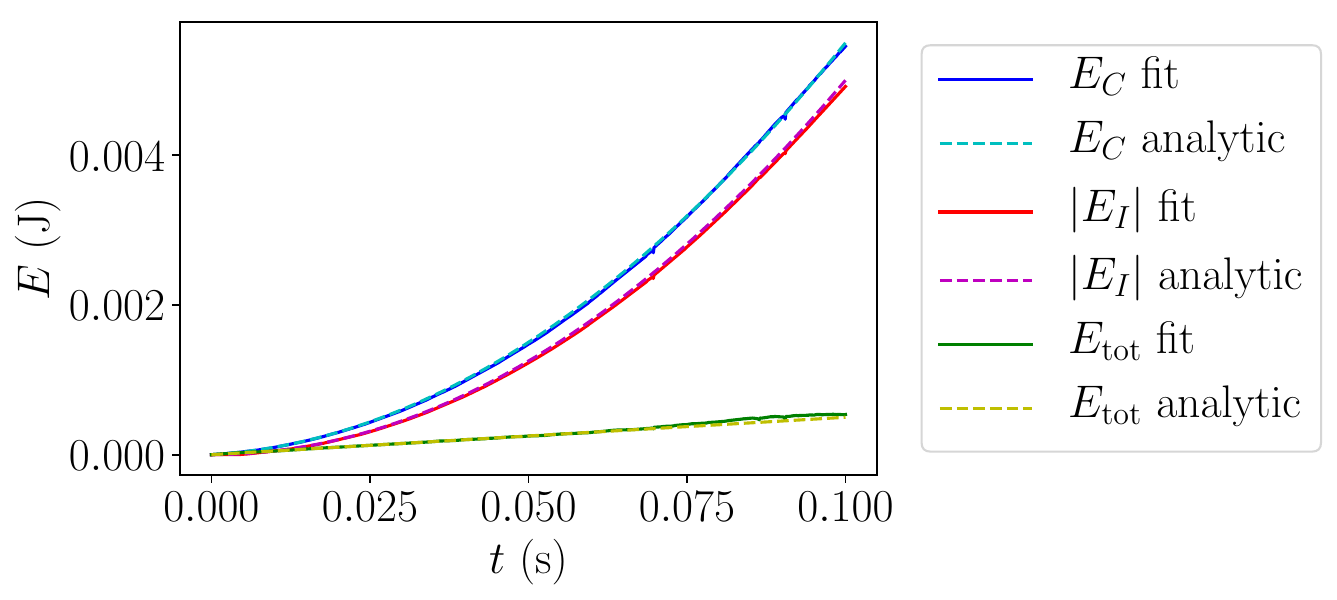}
\caption{The different contributions to the total energy of the CQ dynamics, as a function of time. The solid
lines are the numerical results, while the dashed ones are the analytical ones. The energy contribution given
by the kinetic energy, $E_C = \langle H_C \rangle$, grows faster than the one given by the potential energy,
$E_I = \langle H_I \rangle$. As a result, the total energy of the system, $E_{\mathrm{tot}} = \langle H
\rangle$, increases linearly as a function of time.}
\label{fig:energy_SG_diag}
\end{figure}

Using these results, we can show that the energy associated with the classical Hamiltonian is given by
$\langle H_C \rangle(t) = \frac{(B \omega_0 \tau)^2}{2 m} \left( \frac{t}{\tau} + \left( \frac{t}{\tau}
\right)^2 \right)$, while the quantum coupling contributes to the energy as $\langle H_I \rangle(t) =
- \frac{(B \omega_0)^2}{2 m} t^2$ (notice that in the previous section we have considered the absolute
value of the position for convenience). As a result, we find that the average energy of the system is not conserved, and instead grows linearly in time from the initial value of $\langle H \rangle (0) = 0$,
\begin{equation}
\langle H \rangle (t) = \frac{(B \omega_0)^2 \tau}{2 m} t,
\label{eq:qubitenergyviol}
\end{equation}
see Fig.~\ref{fig:energy_SG_diag}. In the limit when $\tau$ tends to $0$, one recovers
the classical evolution given by the Liouville equation, and energy is conserved, but otherwise the energy
increases at a rate proportional to the jump distance $\tau$. 

As we stressed at the beginning of the section, the fact that the average energy is not conserved does not come as a surprise. Indeed, we are taking the energy to be given by the operator $H(z)$ even though it is not the generator of the symmetry of time-translation. In this model however, the failure of the system to conserve energy is not due to changes in the quantum system -- the interaction of the qubit with the classical system merely causes it to decohere in a basis which commutes with the total CQ Hamiltonian. Instead, the term appearing in Eq.~\eqref{eq:qubitenergyviol} is due to dispersion in the momentum of the classical system. This can be prevented by adding in a friction term, as discussed in Sec.~\ref{sec:con}.

\subsection{Qubit evolution with non-diagonal Lindblad operators in linear potential}
\label{ssec:SG_nondiag}
We now consider the same hybrid system, with the difference that now the Lindblad operators are non-diagonal in the interaction Hamiltonian eigenbasis. We will see that the dynamics of this CQ system is qualitatively and quantitatively different from the one in the previous section. The form of the Lindblad operators is
\begin{subequations}
\label{eq:lindblad_non_diag}
\begin{align}
L_{\alpha=1} &= \ket{1}\bra{0}, \\
L_{\alpha=2} &= \ket{0}\bra{1}.
\end{align}
\end{subequations}
In order to obtain the interaction Hamiltonian $H_I(q,p)$ of Eq.~\eqref{eq:quant_int_ham} from these operators,
the functions $h^{\alpha}(q,p)$ are
\begin{subequations}
\begin{align}
h^{\alpha=1}(q,p) &= \omega_0 \, q B , \\
h^{\alpha=2}(q,p) &= \omega_1 \, q B,
\end{align}
\end{subequations}
where again the particle is in a linear potential, i.e., $B(q) = q B$, with $B = \text{const}$. If we express the CQ state
$\cqstate(q,p,t)$ as in Eq.~\eqref{eq:CQ_state_qubit}, and we take the rate $\tau_0 \rightarrow 0$, then the master equation
given in Eq.~\eqref{eq:general_SG_dynamics} can be expressed as a system of non-local differential equations. In
particular, the evolution of the populations is given by of the following differential equations,
\begin{equation}
\label{eq:evo_pop_SG_non_diag}
\frac{\partial \, u_i(q,p,t)}{\partial t}
= - \frac{p}{m} \frac{\partial \, u_i(q,p,t)}{\partial q} 
+\frac{1}{\tau} \big( u_{i \oplus 1}(q,p + \omega_i B \tau,t) - u_i(q,p,t) \big) ,  \quad i \in \left\{ 0, 1 \right\},
\end{equation}
where $\oplus$ is addition modulo $2$. Notice that the above equations have the same form as the ones obtained in the
previous section, see Eq.~\eqref{eq:evo_pop_SG_diag}, but now each population depends non-locally on the other. The
evolution for the coherence is unchanged from the one in Eq.~\eqref{eq:evo_cohe_SG_diag}.
\begin{figure}[!ht]
\center
\includegraphics[width=0.32\textwidth]{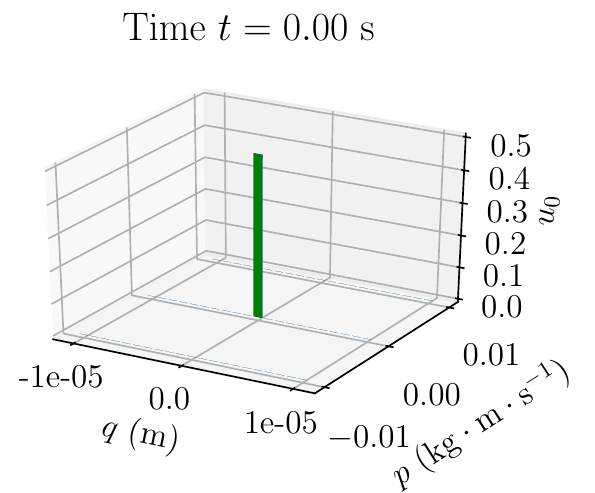}
\includegraphics[width=0.32\textwidth]{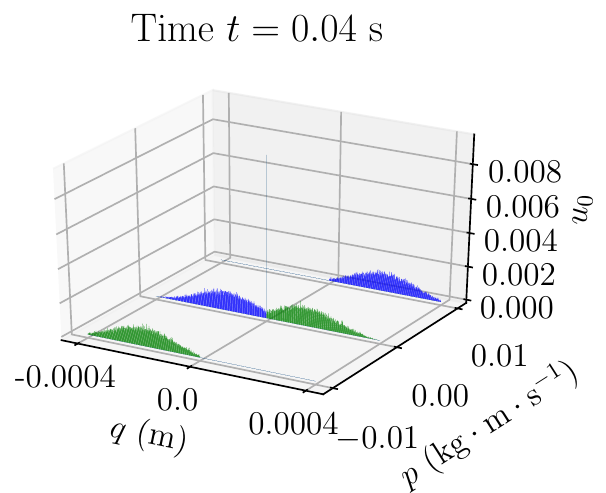}
\includegraphics[width=0.32\textwidth]{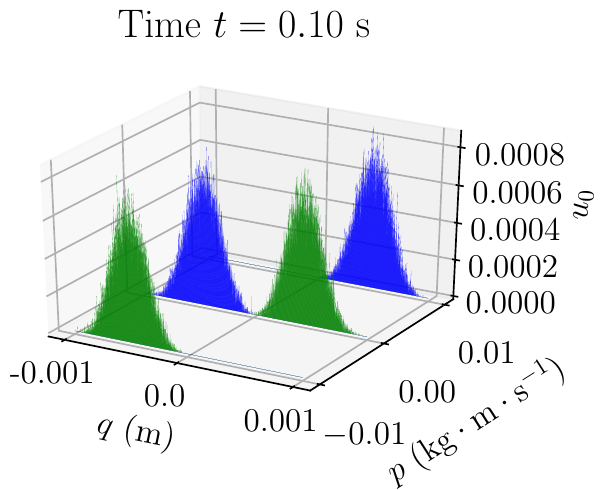}
\caption{The evolution of the population $u_0$ (blue) and $u_1$ (green) under the CQ dynamics with non-diagonal Lindblad operators. The evolution of the populations is shown for $t = 0 \, \mathrm{s}$,
$0.04 \, \mathrm{s}$ and $0.1 \, \mathrm{s}$. In this figure, we set the initial CQ state to be $\delta(q) \delta(p) \ket{+}$, where $\ket{+}
= \frac{1}{\sqrt{2}} \left( \ket{0} + \ket{1} \right)$ is a coherent superposition of the eigenstates of the interaction
Hamiltonian $H_I$. Notice that, in the rest of the section, we instead focus on the case where the initial quantum state is $\ket{0}$, which simplifies the analytical study of the dynamics; the qualitative behaviour of the phase-space evolution, however, is the same as the one shown in this figure. This interaction makes the quantum degrees of freedom stochastically collapse from $\ket{0}$ to
$\ket{1}$ and vice versa, at a rate given by $\tau$. When the quantum state is mapped as $\ket{0} \rightarrow \ket{1}$,
the classical momentum is increased by $B \tau$, while the transformation $\ket{1} \rightarrow \ket{0}$ is accompanied
by a change in momentum equal to $- B \tau$. The simulation depicted here uses time steps of $\delta t = 10^{-4} \,
\mathrm{s}$, and a jump rate $\tau = 10^{-2} \, \mathrm{s}$. The interaction constant is $B = 1 \, \mathrm{J} \cdot
\mathrm{s} \cdot \mathrm{m}^{-1}$, the mass is $m =  1 \, \mathrm{kg}$, and the frequencies are $\omega_0 =
-\omega_1 = \omega = 1 \, \mathrm{s}^{-1}$.}
\label{fig:SG_non_diag_evo}
\end{figure}
\subsubsection{Analytical and numerical evolution of populations}
To solve the model analytically we can express Eq.~\eqref{eq:evo_pop_SG_non_diag}, which describes the evolution
of the populations, as a stochastic equation, see App.~\ref{app:master_stoch}. To do so, we fix the rate $\tau_0$ in
Eq.~\eqref{eq:general_SG_dynamics} to be a (small) positive constant, so that the equation for the populations can
be expressed as
\begin{equation}
\label{eq:evo_stoc_SG_non_diag}
\frac{\partial \, u_i(q,p,t)}{\partial t}
=
\frac{1}{\tau_0} \big( u_i(q - \frac{p}{m} \tau_0, p, t) - u_i(q,p,t) \big)
+
\frac{1}{\tau} \big( u_{i \oplus 1}(q,p + \omega_i B \tau,t) - u_i(q,p,t) \big) , \ i \in \left\{ 0, 1 \right\},
\end{equation}
If $\tau_0 \rightarrow 0$, we come back to the
original equation for the populations shown in Eq.~\eqref{eq:evo_pop_SG_non_diag}. In order to solve this equation
we can make use of the tools described in appendix, see Sec.~\ref{app:stoc_sol_SG_nondiag}, and the solution we
obtain is,
\begin{equation}
u_i(q,p,t)
=
\sum_{\ell, j = 0}^{\infty} P_0(\ell) \, P_1(2j) \, u_i(\ell,j)
+
\sum_{\ell^{\prime} , j^{\prime} = 0}^{\infty} P_0(\ell^{\prime}) \, P_1(2j^{\prime}+1) \, u_{i \oplus 1}(\ell^{\prime},j^{\prime}),  \quad i \in \left\{ 0, 1 \right\},
\end{equation}
where $P_0$ and $P_1$ are Poisson distributions with mean value $\frac{t}{\tau_0}$ and $\frac{t}{\tau}$,
respectively. For the specific case under consideration, we can express the state in terms of shift operators
\begin{subequations}
\begin{align}
u_i(\ell,j) &=
\frac{1}{\binom{\ell + 2j}{\ell}}
\sum_{\pi \in \mathcal{S}^{(1)}_{\ell,j}} \pi
\left(
\left(
e^{- \frac{p}{m} \tau_0 \partial_q}
\right)^{\ell}
\left(
e^{B \omega_i \tau \partial_p}
\,
e^{B \omega_{i \oplus 1} \tau \partial_p}
\right)^j
\right)
u_i(q,p,t=0), \\
u_{i \oplus 1}(\ell,j) &=
\frac{1}{\binom{\ell + 2j + 1}{\ell}}
\sum_{\pi \in \mathcal{S}^{(2)}_{\ell,j}} \pi
\left(
\left(
e^{- \frac{p}{m} \tau_0 \partial_q}
\right)^{\ell}
\left(
e^{B \omega_i \tau \partial_p}
\,
e^{B \omega_{i \oplus 1} \tau \partial_p}
\right)^j
e^{B \omega_i \tau \partial_p}
\right)
u_{i \oplus 1}(q,p,t=0),
\end{align}
\end{subequations}
where in the first equation above, $\mathcal{S}^{(1)}_{\ell,j}$ is a proper subset of the set of all permutations of the shift operators. Each element $\pi \in \mathcal{S}^{(1)}_{\ell,j}$ creates a different combinations of the $\ell+2j$ shift operators for position and momentum, while preserving the relative order of the shift operators for the momentum. The same applies to the set $\mathcal{S}^{(2)}_{\ell,j}$ in the second equation, with the difference that in this case there are $2j+1$ shift for the momentum operator.
\par
To better understand the above solution, let us consider the case in which the initial CQ state is $\delta(q) \delta(p)
\ket{0}\bra{0}$, so that at $t=0$ only the level $\ket{0}$ is populated. Additionally, we fix $\omega_0 = - \omega_1
= \omega > 0$, so that opposite shifts in momentum cancel each others, see Fig.~\ref{fig:SG_non_diag_evo}.
At time $t$ we find that both quantum levels are populated, and in partiular we have
\begin{subequations}
\begin{align}
u_0(q,p,t)
=&
\sum_{\ell, j = 0}^{\infty}
\frac{1}{\binom{\ell + 2j}{\ell}} P_0(\ell) \, P_1(2j) \times\nonumber\\
&\times\sum_{\pi \in \mathcal{S}^{(1)}_{\ell,j}} \pi
\left(
\left(
e^{- \frac{p}{m} \tau_0 \partial_q}
\right)^{\ell}
\left(
e^{B \omega \tau \partial_p}
\,
e^{- B \omega \tau \partial_p}
\right)^j
\right)
\delta(q) \delta(p), \\
u_1(q,p,t)
=&
\sum_{\ell , j = 0}^{\infty}
\frac{1}{\binom{\ell + 2j + 1}{\ell}} P_0(\ell) \, P_1(2j+1) \times\nonumber\\
&\times\sum_{\pi \in \mathcal{S}^{(2)}_{\ell,j}} \pi
\left(
\left(
e^{- \frac{p}{m} \tau_0 \partial_q}
\right)^{\ell}
\left(
e^{B \omega \tau \partial_p}
\,
e^{- B \omega \tau \partial_p}
\right)^j
e^{B \omega \tau \partial_p}
\right)
\delta(q) \delta(p).
\end{align}
\end{subequations}
By counting the number of shift operators in momentum, we can see that the population $u_0$ spreads
in position $q$ while keeping the momentum $p$ fixed at $0$. On the other hand, the population $u_1$
has an odd number of momentum shifts, and therefore its momentum is fixed at $-B \omega \tau$.
\par
We can also use the unravelling technique to obtain a numerical solution for the CQ dynamics of
Eq.~\eqref{eq:evo_pop_SG_non_diag}. In this equation, we see that the CQ state can evolve either
continuously, or through a jump in both the quantum and classical degrees of freedom. When a jump
occur, the quantum state is projected in either $\ket{0}$ or $\ket{1}$, and this change is accompanied
by a positive or negative shift in momentum, respectively. Furthermore, due to the form of the Lindblad
operators, we have that a CQ state whose quantum degree of freedom is described by $\ket{0}$ can
only jump when it is hit by the operator $L_0$, and vice versa for $\ket{1}$. In our simulation, we require
these jumps to be associated with an opposite change in momentum. In this case, an initial CQ state with
well-defined momentum (for instance the one used in Fig.~\ref{fig:SG_non_diag_evo}) can spread to at
most 3 different values of momentum.
\subsubsection{Diffusion in phase-space}
\begin{figure}[!ht]
\center
\includegraphics[width=0.4\textwidth]{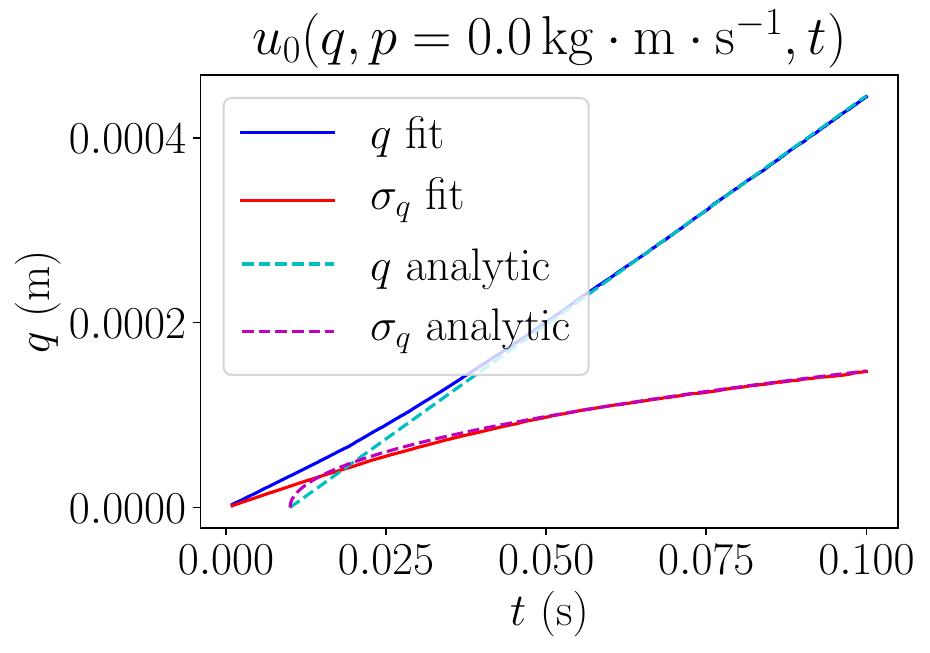}
\includegraphics[width=0.4\textwidth]{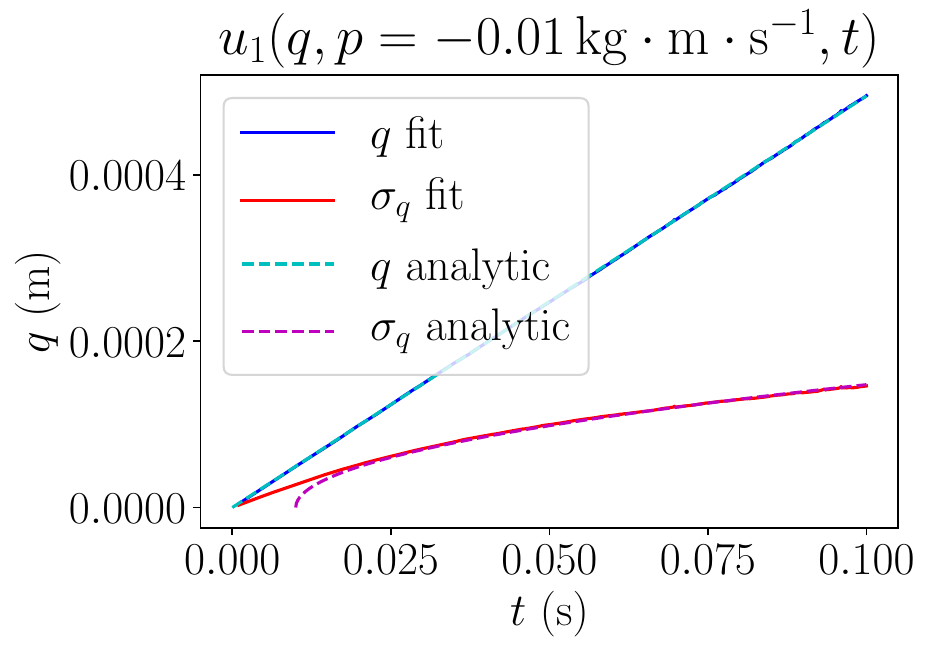}
\caption{The mean value $\langle q \rangle$ (or better, its absolute value) and standard deviation $\sigma_q$ of populations $u_0$ and $u_1$. For
the numerical simulation we choose the frequencies $\omega_0 = -\omega_1 = \omega = 1 \, \mathrm{s}^{-1}$, the
interaction constant $B = 1 \, \mathrm{J} \cdot \mathrm{s} \cdot \mathrm{m}^{-1}$, the mass $m =  1 \, \mathrm{kg}$,
and a jump rate $\tau = 10^{-2} \, \mathrm{s}$. The time step used in the simulation is $\delta t = 10^{-4} \, \mathrm{s}$.
The solid lines are obtained by fitting the numerical data with a normal distribution, while the dashed lines are obtained
using analytical considerations discussed in the main text. It is interesting to notice that, while $\sigma_q$ is roughly the
same for the two populations, the mean position is shifted. This is because the distribution of $u_1$ is the first to be
populated, since only one jump in momentum is needed when starting from the initial quantum state $\ket{0} \bra{0}$.
On the other side, the distribution of $u_0$ is populated (for values of $q \neq 0$) only after two jumps in momentum
are preformed. Since the rate of jump is given by $\tau$, we find that the mean position of the two populations is
actually shifted (in time) by this parameter.}
\label{fig:SG_non_diag_var}
\end{figure}

We can additionally study the spreading, due to the interaction between classical and quantum degrees of freedom,
of the position of the particle as time passes by. The numerical simulation shows that at later times, the populations
$u_0(q,p,t)$ and $u_1(q,p,t)$ divide into two Gaussian distributions, one with zero momentum, the other with non-zero
momentum, see Fig.~\ref{fig:SG_non_diag_evo}. Here, we consider the same scenario as in the previous section, where
the initial CQ state is $\cqstate(q,p) = \delta(q) \delta(p) \ket{0}\bra{0}$. As we noticed before, this state evolves into two
different ensembles, one associated with the quantum level $\ket{0}$, and the other with the quantum level $\ket{1}$.
In particular, the population $u_0$ spreads in position while keeping the momentum fixed at $p = 0$, while the
population $u_1$ spreads in position with a fixed momentum $p = - \omega B \tau$. In Fig.~\ref{fig:SG_non_diag_var},
we show the average position and standard deviation of the two populations as a function of time. We obtain these
values by fitting the numerical data, and by the following analytical considerations.
\par
Let us consider the evolution in phase space of the initial state under consideration. At each time step, this state
has a probability $P_{\text{jump}} = \frac{\delta t}{\tau}$ of jumping in momentum, and a probability $1 - P_{\text{jump}}$
of jumping in position. As we noticed before, given the initial state under consideration, the momentum of the CQ state at
time $t$ can be either $0$ or $p_{\min} = - B \omega \tau$. When the momentum is non-zero, a position jump modifies the
position of the state by $\Delta q = p_{\min} \delta t = - B \omega \tau \delta t$. We can then compute the average change
in position between two jumps, when the momentum is $p_{\min}$. The (normalised) probability that $n$ time steps separate
two jumps is given by
\begin{equation}
\text{prob}\left(\# \, \text{steps} = n \right) = P_{\text{jump}} (1 - P_{\text{jump}} )^n.
\end{equation}
Using the above distribution, we can compute the average distance travelled by the state in between two momentum jumps,
which is given by
\begin{equation}
\Delta Q
=
\sum_{n=0}^{\infty} n \, \Delta q \, \text{prob}\left(\# \, \text{steps} = n \right)
=
\frac{1 - P_{\text{jump}}}{P_{\text{jump}}} \Delta q
=
- B \omega \tau \left( \tau - \delta t \right).
\end{equation}
After $n$ time steps, the average number of jumps is $N_{\text{jump}} = n P_{\text{jump}}$. Furthermore, the
position of the state can only change after an odd number of jumps, since only in that case the momentum is non-zero.
As a result, we find that the average position of the state at time $t$ is
\begin{equation}
\label{eq:average_pos_SG_nondiag}
\langle q \rangle (t) = \frac{1}{2} N_{\text{jump}} \Delta Q + \langle q \rangle (0) = \frac{1}{2} \left( \tau - \delta t \right) t + \langle q \rangle (0),
\end{equation}
where we define $t = n \, \delta t$. It is worth noting that the above result only applies to the average position of
the population $u_1$. In this case, since the initial quantum state is $\ket{0}$, a single jump in momentum is required
for the CQ state to populate $\ket{1}$. However, the CQ state starts to populate $\ket{0}$ (with a position $q \neq 0$),
only after two jumps in momentum are performed. Since the rate of jumping is given by the parameter
$\tau$, we have that the average position of the population $u_0$ is roughly delayed by this amount of time, and it is
therefore equal to $\langle q \rangle (t-\tau)$, see Fig.~\ref{fig:SG_non_diag_var}.
\par
\begin{figure}[!ht]
\center
\includegraphics[width=0.7\textwidth]{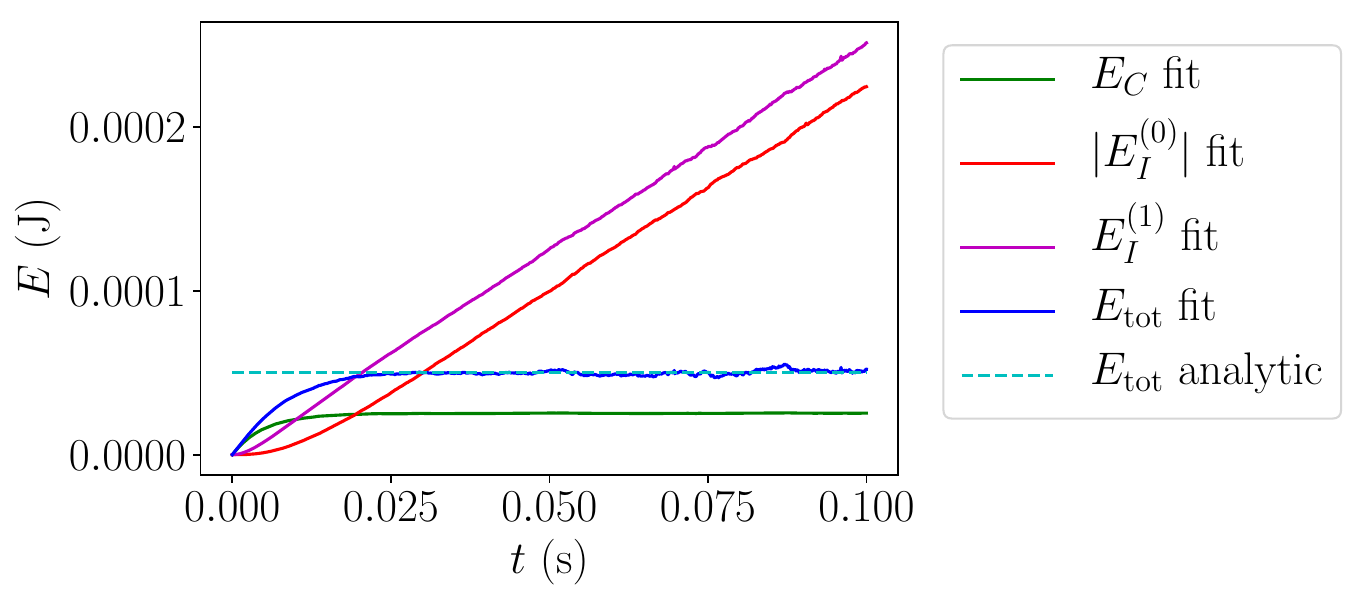}
\caption{The energy of the CQ state as a function of time, when the initial state is $\delta(q) \delta(p) \ket{0}\bra{0}$.
The different contributions to the energy are shown, such as the kinetic energy $E_C = \langle H_C \rangle$ and the
potential energies $E_I^{(i)} = \langle H_I^{(i)} \rangle$. As in the first model in the previous section, the total energy of the system
is not conserved. However, in this case the total energy $E_{\text{tot}} = \langle H_C + H_I \rangle$ increases during
the initial part of the evolution, until it reaches the maximum value $\frac{1}{2} \tau^2$.}
\label{fig:SG_nondiag_energy}
\end{figure}

To compute the variance of the position for the populations, we notice that the number of jumps $N_{\text{jump}}$ is
distributed according to a binomial distribution, and its variance is given by $\sigma_{\text{jump}}^2 = n P_{\text{jump}}
\left( 1 - P_{\text{jump}} \right)$. Then, keeping the other terms constant in Eq.~\eqref{eq:average_pos_SG_nondiag}, we
have
\begin{equation}
\sigma_q (t) = \frac{1}{2} \sigma_{\text{jump}} \left| \Delta Q \right| = \frac{B \omega}{2} (\tau - \delta t)^{\frac{3}{2}} \sqrt{t}.
\end{equation}
As we noticed before, the CQ state starts in position $q=0$ with momentum $p=0$, and roughly starts moving only
after $t = \tau$, that is, the time interval before a momentum jumps occurs (on average). As a result, the standard
deviation of the populations needs to be modified so as to account for this delay, and we find that for both $u_0$ and
$u_1$ the standard deviation is given by $\sigma_q (t - \tau)$, at least for $t \gg \tau$. As a final remark we notice that also in this model, like in the previous one, the average energy, as computed by averaging the operator $H(z) = H_C(z) + H_I(z)$ over the classical and quantum degrees of freedom, is not conserved. Perhaps surprisingly, though, the energy increases from the initial time until it saturates at a positive value, see Fig.~\ref{fig:SG_nondiag_energy}. This is due to the form of the Lindblad operators considered, which flip a $\ket{0}$ state into a $\ket{1}$ and vice versa, and thus create, after an initial transient, an equilibrium between four possible states in the classical phase-space, highlighted in the last panel of Fig.~\ref{fig:SG_non_diag_evo}. In this equilibrium, we find that the kinetic energy receives a contribution by those states associated with non-zero momentum, while the potential energy equilibrates since the contribution from the term $u_0$ is proportional to that of $u_1$, but with opposite sign (see the first panel of Fig.~\ref{fig:SG_non_diag_var}, and the above discussion on the expected value of the position).
\section{Quantum harmonic oscillator in a classical potential}
\label{sec:HO_evo}

In this section, we will consider the dynamics of coupling two degrees of freedom, one quantum and one classical. It is a particle whose internal degree of freedom is a quantum harmonic oscillator, coupled to its classical position and momentum. The model we focus on here demonstrates some basic features of the classical-quantum dynamics, though it describes a somewhat artificial situation where raising the energy level of the harmonic oscillator kicks the momentum towards the left, and lowering the energy kicks the momentum towards right. One quickly sees that energy will not be conserved.

The simpler model we explore in this section has one important and universal feature: starting from a state that is a delta function in momentum and position, and a superposition of two states in the harmonic oscillator, as expected, the numerical simulations show the initially pure state going through a process of decoherence, with the parts of the state associated to different energy levels becoming more and more distinguishable in the phase space as time goes on. We numerically simulated the dynamics and analytically found the decoherence rate, for the large $n$ approximation (where $n$ marks the energy levels of the harmonic oscillator).

We will begin with a simple model with interaction Hamiltonian, depending on both lowering and raising operators, which are quantum operators acting on the internal quantum degree of freedom, and the classical position $q$ and momentum $p$ of the particle carrying it:

\begin{equation}
H_I(q,p) = B(q) \left( \omega_0 a a^{\dagger}
- \omega_1 a^{\dagger} a \right).    
\end{equation}

This simple model is analogous to the one used for the spin-$\frac{1}{2}$ particle. We consider a model with
two Lindblad operators, one of which is proportional to the creation operator  and the other is proportional to the annihilation operator: 
\begin{subequations}
\label{eq:lindblad_diag_OS}
\begin{align}
L_{\alpha=1} &= \sqrt{|\omega_0|} \, a, \\
L_{\alpha=2} &= \sqrt{|\omega_1|} \, a^{\dagger}.
\end{align}
\end{subequations}
As a result, the functions $h^{\alpha}(q,p)$ are defined as
\begin{subequations}
\begin{align}
h^{\alpha=1}(q,p) &= \text{sign}(\omega_0) \, B(q), \\
h^{\alpha=2}(q,p) &= \text{sign}(\omega_1) \, B(q).
\end{align}
\end{subequations}
In the following, we focus on the case in which $B(q) = q B$, where $B$ is a constant, and for simplicity we set the coupling
constant $\omega_0 = - \omega_1 = 1 \, \mathrm{s}^{-1}$. Notice that, for this choice of constants, the
interaction Hamiltonian reduces to  $H_I(q,p) = q B \, \Id$, and the commutator in Eq.~\eqref{eq:general_SG_dynamics}
drops out. We will discuss more general models, including two coupled oscillators in Sec.~\ref{sec:con}.

\par
The resulting CQ master equation, where we additionally account for the free-particle classical Hamiltonian $H_C
= \frac{p^2}{2 m}$, is given by
\begin{align}
\label{eq:HO_master_equation}
\frac{\partial \cqstate(q,p,t)}{\partial t} = &- \frac{p}{m} \frac{\partial \, \cqstate(q,p,t)}{\partial q} \nonumber\\
&+\frac{1}{\tau} \left(\gamma_{\uparrow} a^{\dagger} \, \cqstate(q,p-B\tau,t) \, a +\gamma_{\downarrow} a \, \cqstate(q,p+B\tau,t) \, a^{\dagger}\right)\nonumber\\
&\hphantom{+\frac{1}{\tau} (+}\left( - \frac{1}{2}\acom{\gamma_{\downarrow} a^{\dagger} a + \gamma_{\uparrow} a a^{\dagger}}{\cqstate(q,p,t)} \right),
\end{align}
where $\tau$ is the rate of jump in both the classical and quantum degrees of freedom. The constants $\gamma_\uparrow$, $\gamma_\downarrow$ determine the rate of damping ($\gamma_\downarrow$) of the quantum oscillator compared with the rate of pumping ($\gamma_\uparrow$). For simplicity we henceforth take $\gamma_\uparrow=\gamma_\downarrow=1$ although from a physical point of view, one might want to have the damping term be larger, to drive the quantum oscillator to its ground state. In a field theory, taking the damping term to be larger than the anti-damping term, can result in the theory violating causality\cite{poulinKITP}, a state of affairs which the authors of \cite{poulinKITP} tune down by allowing some degree of energy non-conservation.

\begin{figure}[ht!]
\center
\includegraphics[width=0.3\textwidth]{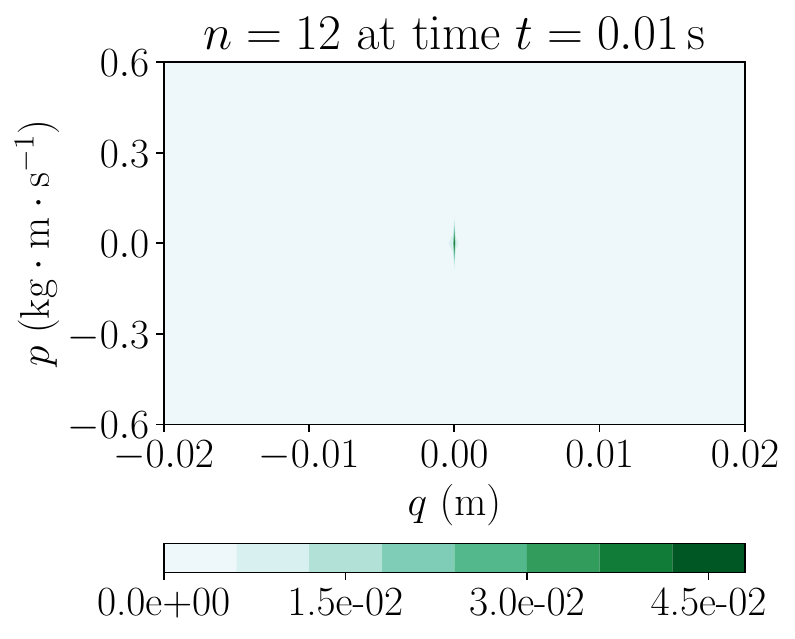}
\includegraphics[width=0.3\textwidth]{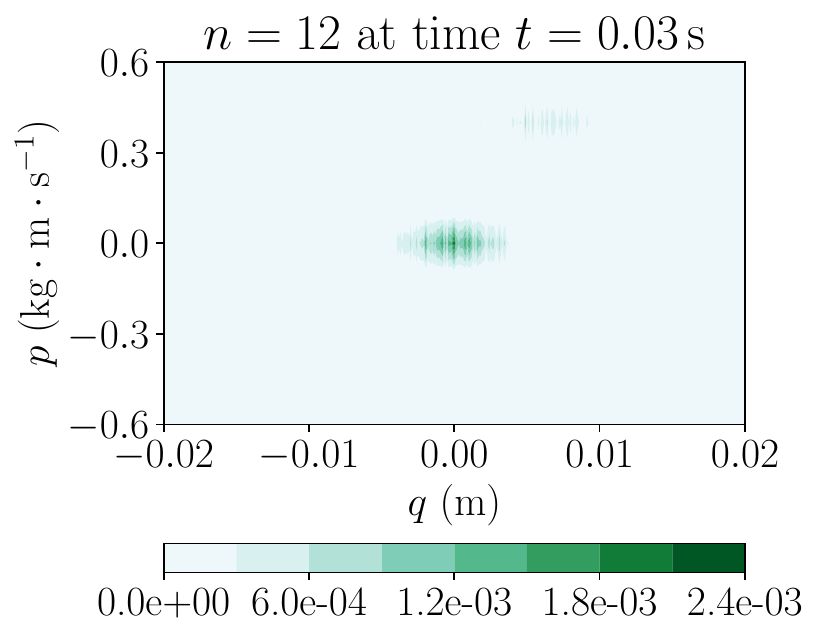}
\includegraphics[width=0.3\textwidth]{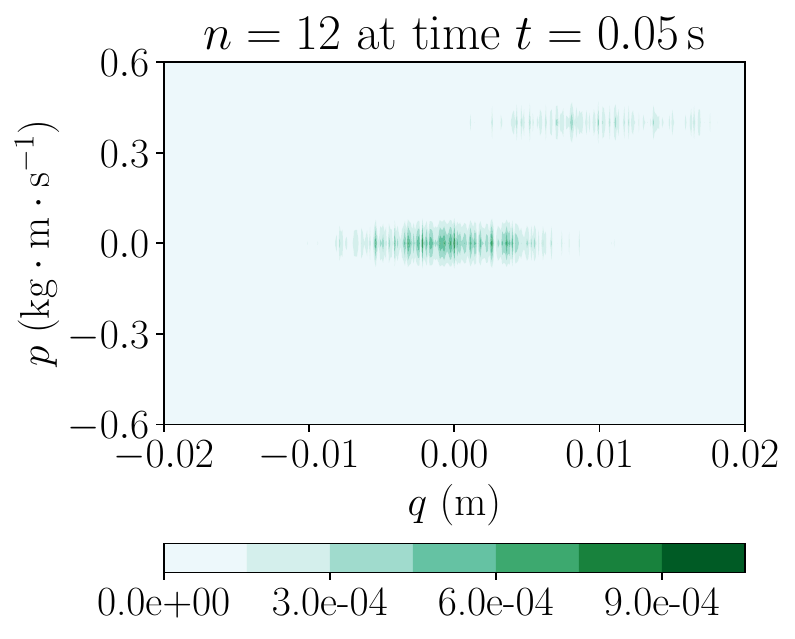}
\includegraphics[width=0.3\textwidth]{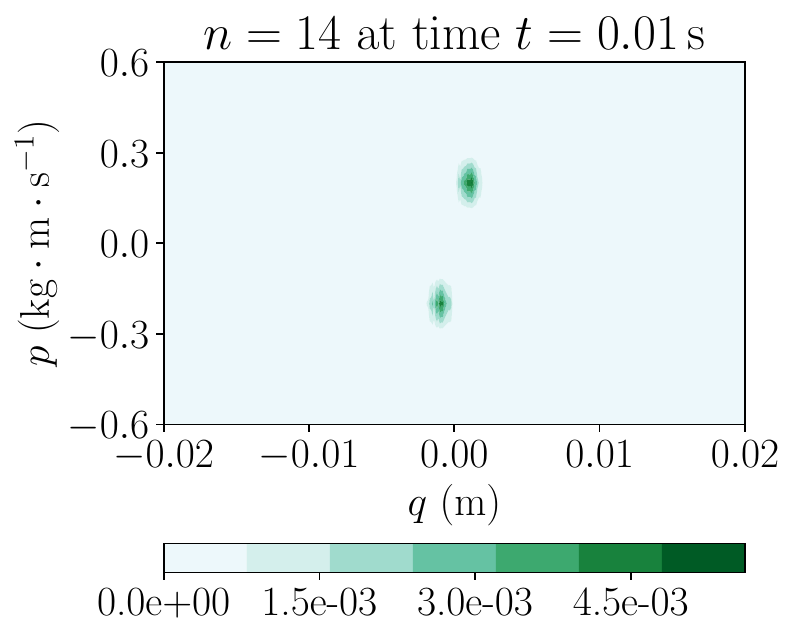}
\includegraphics[width=0.3\textwidth]{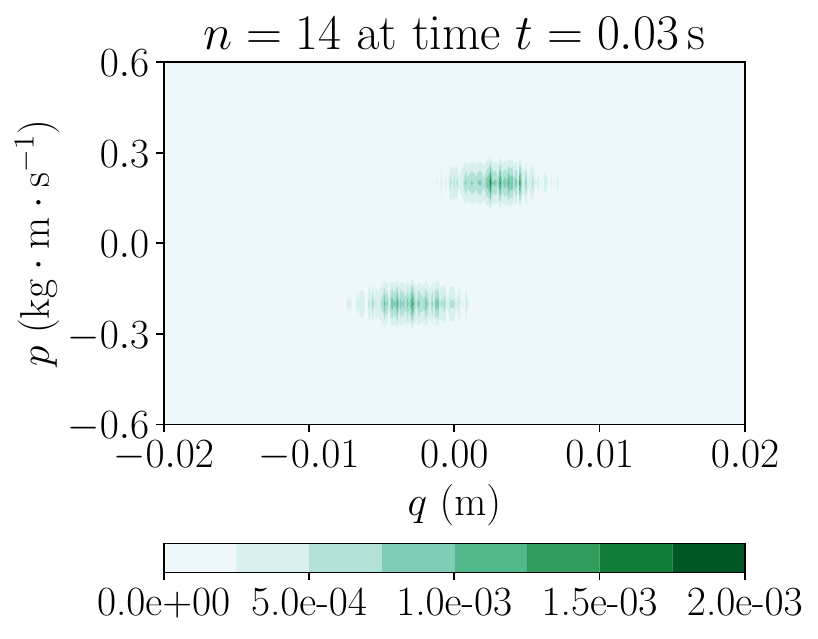}
\includegraphics[width=0.3\textwidth]{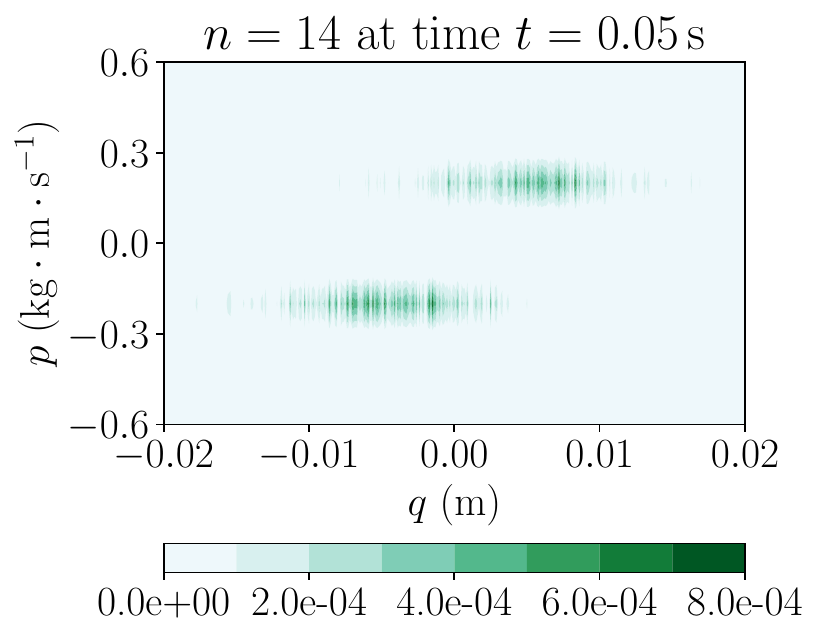}
\includegraphics[width=0.3\textwidth]{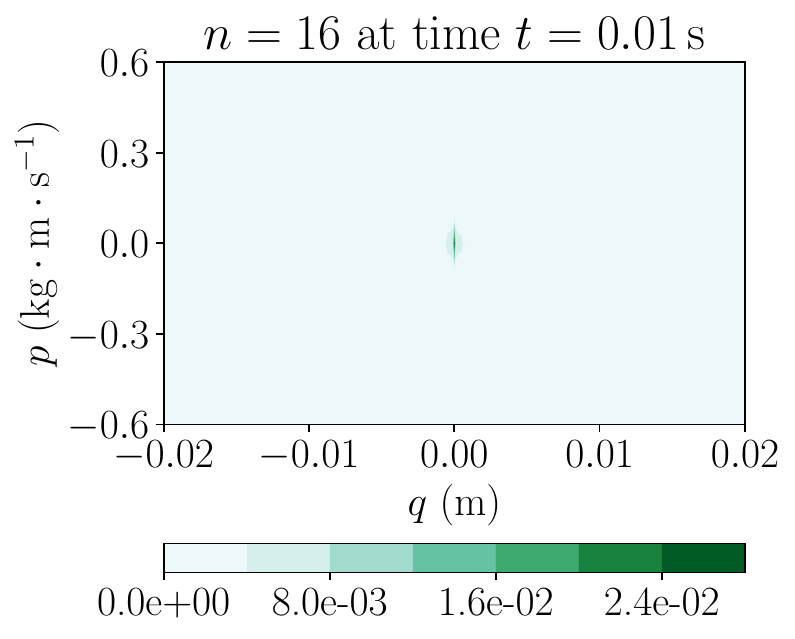}
\includegraphics[width=0.3\textwidth]{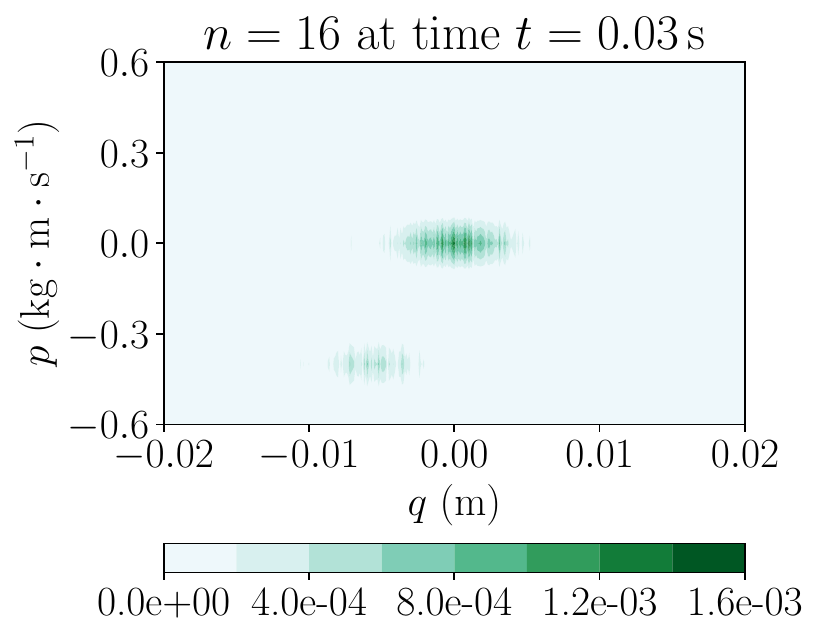}
\includegraphics[width=0.3\textwidth]{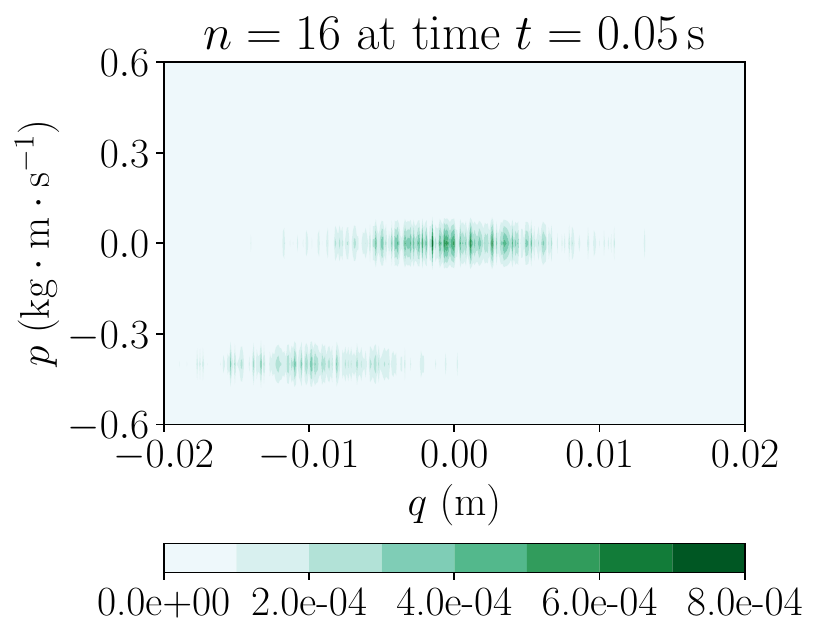}
\caption{The evolution of the populations $u_{n}(q,p,t)$ of the harmonic oscillator in phase space,. The system is initially
in a superposition over the state $\ket{n_1 = 12}$ and $\ket{n_2 = 16}$, centered in the origin of the phase space.
The evolution of the populations is shown for $t = 0.01 \, \mathrm{s}$, $0.03  \, \mathrm{s}$ and $0.5 \, \mathrm{s}$.
The state is updated every $\delta t = 2.5 \cdot 10^{-5}  \, \mathrm{s}$, and the jump rate of the model is $\tau = 0.1
\, \mathrm{s}$. The constant $B = 1  \, \mathrm{J} \cdot \mathrm{s} \cdot \mathrm{m}^{-1}$, and the mass of the
harmonic oscillator is $m =  1  \, \mathrm{kg}$. {\bf Top panels.} The evolution of the population for $n = n_1$ is shown.
Initially, this population is peaked around the origin of phase space, and it starts to diffuse in position due to the action of the
rising and lowering operators. At later time, a second packet can be seen, with momentum $p = 0.4 \, \mathrm{kg} \cdot
\mathrm{m} \cdot \mathrm{s}^{-1}$. This packet originated from the population at $n = n_2$, and reaches $n = n_1$
thanks to the repeated action of lowering operators. {\bf Central panels.} The evolution of the population for $n = \frac{n_1 +
n_2}{2}$ is shown. The symmetric packets with negative and positive momentum ($p = \pm 0.2 \, \mathrm{kg} \cdot
\mathrm{m} \cdot \mathrm{s}^{-1}$) originated from $n = n_1$ and $n = n_2$, respectively. {\bf Bottom panels.} The
evolution of the population for $n = n_2$ is shown. The general behaviour in phase space is analogous to that of the population
for $n = n_1$, but with a second packet with negative momentum.}
\label{fig:ps_evo_HO}
\end{figure}

In this model, when the energy
of the particle is increased with the raising operator $a^{\dagger}$, its momentum is lowered, and vice versa, when the
energy is lowered with the annihilation operator $a$, its momentum is increased. It is worth noting that the momentum
is here one dimensional; thus, increasing the momentum does not necessarily mean increasing also its absolute value,
and therefore, for the free-particle Hamiltonian $H_C$, we cannot conserve the energy in this model. An example of the
dynamics generated by this master equation is shown in Fig.~\ref{fig:ps_evo_HO}.

\par
Because the CQ master equation introduces a new contribution to decoherence, coming from the interaction between
the classical and quantum degrees of freedom, we now aim to study the decoherence rate of the harmonic oscillator.
In order to analytically find this rate, we have to compute the time evolution of the CQ state; to simplify the analysis,
however, we will neglect the contribution of the Poisson bracket with the classical Hamiltonian $H_C$, and only focus
on the jumping dynamics, which is here responsible for the decoherence of the system. First, we write out the CQ state
$\cqstate(q,p,t)$ in terms of the Fock basis,
\begin{equation}
\cqstate(q,p,t)=\sum_{n,m} u_{n,m} (q,p,t) \ket{n}\bra{m}.
\end{equation}
We can additionally perform the Fourier transform of the above state with respect to the momentum $p$. Thus, we get the
following equation
\begin{align}
\frac{\partial \tilde{\cqstate}(q,x,t)}{\partial t} =
\frac{1}{\tau} \left( a^{\dagger} \, \tilde{\cqstate}(q,x,t) \, e^{i B \tau x} \, a + a \, \tilde{\cqstate}(q,x,t) \, e^{-i B \tau x} \, a^{\dagger}
- \frac{1}{2}\acom{a^{\dagger} a + a a^{\dagger}}{\tilde{\cqstate}(q,x,t)} \right).
\end{align}
This step simplifies our analysis, since in this way we remove from Eq.~\eqref{eq:HO_master_equation} the non-locality in
momentum. However, at the end of our analysis we will have to perform an inverse-Fourier transform to recover the state
in the momentum domain.
\par
If we sandwich the above equation with two energy eigenstates $\bra{n} \, \cdot \, \ket{m}$, we obtain the following equation
for the elements $u_{n,m}(q,x,t)$ of the CQ state,
\begin{align*}
\frac{\partial u_{n,m}(q,x,t)}{\partial t} = &
\frac{\sqrt{n m}}{\tau} \, u_{n-1,m-1}(q,x,t ) \, e^{i B \tau x}\\
&+
\frac{\sqrt{(n+1)(m+1)}}{\tau} \, u_{n+1,m+1}(q,x,t ) \, e^{-i B \tau x}\\
&- \frac{n+m+1}{\tau} \, u_{n,m}(q,x,t).
\end{align*}
The fact that the matrix element $(n,m)$ interacts only with itself, and its nearest neighbours (on a line parallel to the matrix
diagonal) $(n-1,m-1)$ and $(n+1,m+1)$ will prove to be convenient. Indeed, this implies that if we are interested in the problem
where the initial state of the CQ density matrix is a superposition of two energy eigenstates,
\begin{equation}
\label{eq:init_cond_HO}
\cqstate(q,p,t=0) = \delta(p) \, \delta(q) \, \frac{1}{2} (\ket{n_1}+\ket{n_2})(\bra{n_1}+\bra{n_2}),
\end{equation}
then we only need to solve three differential vector equations, where the vector takes values along the three diagonal lines involved;
the main diagonal, among whose elements we find $(n_1,n_1)$ and $(n_2,n_2)$, the second line parallel to the diagonal marked by
the matrix element $(n_1,n_2)$, and the third one marked by $(n_2,n_1)$. 
\subsection{Solving the differential equations for the CQ harmonic oscillator}
We now attempt to solve the differential equation with the previously mentioned initial conditions, where we know that
the three diagonals along which we have to solve are the ones marked with the matrix elements $(n_2,n_1)$, $(n_1,n_1)$
and $(n_1,n_2)$. For example, we can focus on the differential equation of the third diagonal,
\begin{align}
\frac{\partial u_{n_1+k,n_2+k}(q,x,t)}{\partial t} =
&\frac{\sqrt{(n_1+k)(n_2+k)}}{\tau} \, u_{n_1+k-1,n_2+k-1}(q,x,t ) \, e^{i B \tau x} \nonumber \\
- &\frac{n_1+n_2+1}{\tau} \, u_{n_1+k,n_2+k}(q,x,t) \nonumber \\
+ &\frac{\sqrt{(n_1+k+1)(n_2+k+1)}}{\tau} \, u_{n_1+k+1,n_2+k+1}(q,x,t ) \, e^{-i B \tau x}. \label{matrixM}
\end{align}
Assuming that $n_1 < n_2$, we can now rearrange the functions $u_{n_1+k,n_2+k}(q,x,t)$, for $k \in \{-n_1, -n_1+1,
\ldots \}$, into a vector
\begin{equation}
\label{eq:coherence_vec}
\vec{v}=(u_{0,n_2-n_1}, u_{1,n_2-n_1+1},...)^{T}= \sum_{k} u_{n_1+k,n_2+k} \, \vec{e}_k,
\end{equation}
where the $ \{\vec{e}_k\}_k$ are unit vectors, and we suppressed the time and phase space dependence for simplicity.
The goal is then to find the solution of the equation
\begin{equation}
\label{dv_dt}
\frac{\partial \vec{v}}{\partial t} = \mathcal{M} \, \vec{v},
\end{equation}
where $\mathcal{M}$ is a tridiagonal matrix, whose entries are determined by Eq.~\eqref{matrixM}. The size of this matrix
is infinite, but we can obtain an approximate solution to the above equation by considering only part of the elements in the
vector $\vec{v}$. In particular, we require the index $k$ to take values in an interval $\left( -\frac{N}{2}, \frac{N}{2} \right]$,
for a given integer $N$. This approximation is  acceptable for short enough times, when the coherence $u_{n_1,n_2}$ has not
yet spread too much. By imposing these lower and upper bounds to the $k$ index, we make the system effectively finite dimensional,
and therefore the finite-dimensional matrix $\mathcal{M}$ can be diagonalised. In the following, we refer to the $i$-th eigenvector
of this matrix as $\vec{w}_i$, with eigenvalue $\lambda_i$. 
\par
The equation regulating the dynamics of the eigenvectors $\vec{w}_i$ is then given by
\begin{equation}
\frac{\partial \vec{w}_i}{\partial t}= \lambda_i \vec{w}_i,
\end{equation}
and the solution is given by $\vec{w}_i(t) = e^{\lambda_i \, t} \vec{w}_i(0)$. The same procedure can be followed for the
main diagonal, marked with the matrix elements $(n_1,n_1)$ and $(n_2,n_2)$, so as to obtain the evolution for the populations.
The diagonalisation of the matrix $\mathcal{M}$ can be performed numerically, so as to obtain the decay rates for the coherences
of the (Fourier transformed) CQ state. However, obtaining the evolution of the actual CQ state $\rho(q,p,t)$ would require us to
Fourier transform back to the momentum domain, which can be computational expensive, given that such transformation should
be performed on every element $u_{n,m}(q,x,t)$ of the CQ state. For this reason, we preferred to make use of the unravelling
technique to obtain the evolution of the CQ harmonic oscillator, and to compute the decoherence rate.
\subsection{Decoherence rate in the large \texorpdfstring{$n$}{n} approximation}
We now compute an analytical expression for the decoherence rate in the case in which the Fock numbers $n_1$ and $n_2$ are
large compared to the range of the $k$ index. In this situation, we can approximate the matrix elements along the diagonals of
$\mathcal{M}$ to be equal, so as to obtain a Toeplitz matrix. Concretely, if the $k$ index takes value between $\left( -\frac{N}{2},
\frac{N}{2} \right]$, where $N \ll n_1, n_2$ is an even number, then the matrix $\mathcal{M}$ can be approximated as
\begin{equation}
\mathcal{M}\approx
\begin{pmatrix}
\dots & \dots & 0& \dots & \dots \\
\dots & -\frac{n_1+n_2}{\tau} & \frac{e^{i \phi}}{\tau} \sqrt{n_1 n_2} & 0 &  \dots \\
   0    & \frac{e^{-i \phi}}{\tau} \sqrt{n_1 n_2}&-\frac{n_1+n_2}{\tau} &\frac{e^{i \phi}}{\tau} \sqrt{n_1 n_2} & \dots \\
   0    & 0 & \frac{e^{-i \phi}}{\tau} \sqrt{n_1 n_2} & -\frac{n_1+n_2}{\tau}& \dots\\
\dots & 0 & 0 & \dots & \dots
\end{pmatrix},
\end{equation}
where we define $e^{i \phi} = e^{i B \tau x}$ for convenience. We can now find the eigenvalues $\lambda_m$ of the above
Toeplitz matrix, which are given by~\cite{toep}
\begin{equation}
\lambda_m = -\frac{n_1+n_2}{\tau}+\frac{2}{\tau} \sqrt{n_1 n_2} \cos{\frac{m \, \pi}{N+1}} , \quad m = 1, \ldots, N.
\end{equation}
The eigenvectors of the matrix $\left\{ \vec{\omega}_m \right\}^N_{m=1}$ are instead given by
\begin{equation}\label{eigvectors_omega}
\vec{\omega}_m = \sum_{r=1}^N B_{m,r} \, \vec{e}_{r - \frac{N}{2}} ,
\end{equation}
where $B_{m,r} = \sqrt{\frac{2}{N+1}} \, e^{-i r \phi} \, \sin{\frac{m \, r \, \pi}{N+1}}$ is the unitary matrix
diagonalizing $\mathcal{M}$.
\par
Under the dynamics described by Eq.~\eqref{dv_dt}, the eigenvectors evolve as
$\vec{\omega}_m(t) = e^{\lambda_m \, t} \, \vec{\omega}_m(0)$. Thus, by re-writing the vector $\vec{v}$,
defined in Eq.~\eqref{eq:coherence_vec}, in terms of the eigenvectors of $\mathcal{M}$, we can compute the
approximate time evolution of the coherence of the state, which is given by
\begin{equation}
\label{eq:coherence_evo_HO_a}
u_{n_1-\frac{N}{2}+\ell, n_2-\frac{N}{2}+\ell}(q,x,t)
=
\sum_{r,m=1}^N u_{n_1-\frac{N}{2}+r, n_2-\frac{N}{2}+r}(q,x,0)
\, B^{-1}_{r,m} \, e^{\lambda_m \, t} \, B_{m,\ell}, \quad \ell = 1, \ldots , N.
\end{equation}

\begin{figure}[ht!]
\center
\includegraphics[width=0.32\textwidth]{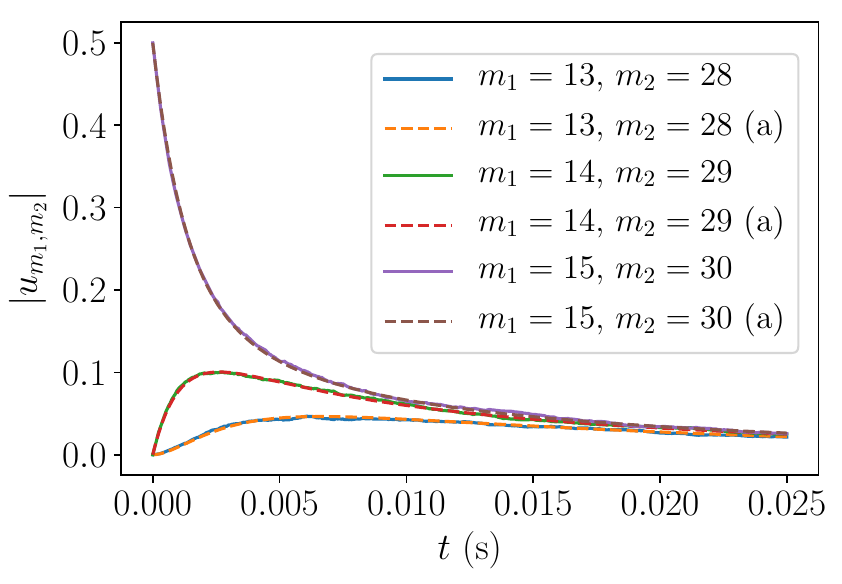}
\includegraphics[width=0.32\textwidth]{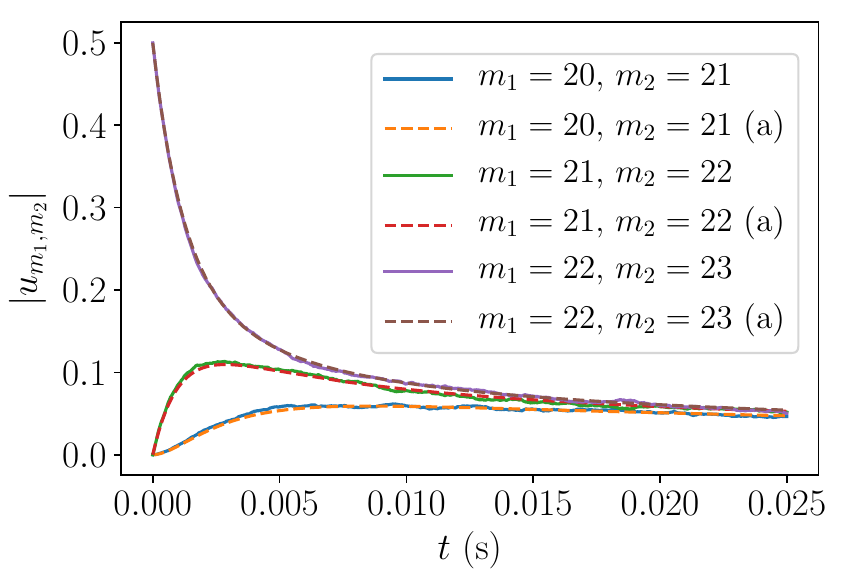}
\includegraphics[width=0.32\textwidth]{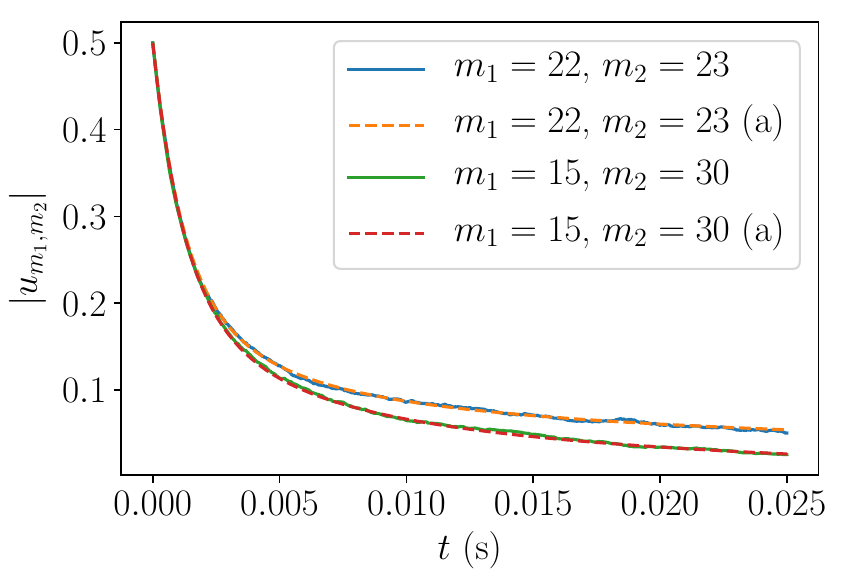}
\caption{The evolution of the coherence for two different choices of initial state for the CQ system. In both cases,
the system is evolved up to the time $t = 2.5 \cdot 10^{-2} \, \mathrm{s}$, with time steps of $\delta t = 2.5 \cdot 10^{-6}
\, \mathrm{s}$, and a jumping rate $\tau = 2.5 \cdot 10^{-1} \, \mathrm{s}$. The interaction constant is $B = 1 \,
\mathrm{J} \cdot \mathrm{s} \cdot \mathrm{m}^{-1}$, and the mass is $m =  1 \, \mathrm{kg}$. {\bf Left.} The
initial quantum state is an even superposition between $\ket{n_1 = 15}$ and $\ket{n_1 = 30}$. The evolution of the
quantum coherence $\left| u_{m_1,m_2} \right|$ is shown for different values
of $m_1$ and $m_2$, and the classical
degrees of freedom have been traced out. The numerical evolution is represented with the solid line, and the dashed
line is the analytical solution given in Eq.~\eqref{eq:coherence_evo_HO_a}, for $N=10$. {\bf Centre.} The initial quantum
state is a even superposition between $\ket{n_1 = 22}$ and $\ket{n_1 = 23}$, and the evolution of the coherence is
shown for those $m_1$ and $m_2$ close to $n_1$ and $n_2$. Again, the solid lines are obtained through the numerical
simulation, using the unravelling code, and the dashed ones are the (approximate) analytical solutions.
{\bf Right} The evolution of the coherence $\left| u_{n_1,n_2} \right|$ for the two choices of initial state. When the
difference between $n_1$ and $n_2$ is big, the coherence reaches lower values than in the case in which $n_1$ and
$n_2$ are close together; this behaviour of the coherence is analogous to the case in which the dynamics is fully
quantum, and the system decoheres due to the coupling with an external bath, rather than to the interaction with the
classical degrees of freedom.}
\label{fig:decoh_ho}
\end{figure}

We can now specialise the above approximate solution to the initial condition we introduced in Eq.~\eqref{eq:init_cond_HO}.
In this case, the sole contribution to the coherence at time $t=0$ is $u_{n_1,n_2}(q,x,0) = \frac{1}{2} \, \delta(q)\delta(p)=\frac{1}{2} \, \delta(q) \int dx e^{i p x}$, and its
time evolution is given by
\begin{equation}
\label{eq:analytic_decoh_HO}
u_{n_1,n_2}(q,x,t) = \frac{e^{-\frac{n_1+n_2}{\tau}t}}{N+1}
\sum_{m=1}^N
\sin^2\left( \frac{N}{2 (N+1)} \, m \, \pi\right)
e^{\frac{2}{\tau} \sqrt{n_1 n_2} \cos\left({\frac{m \pi}{N+1}}\right)t}
\, \delta(q)\delta(p),
\end{equation}
where the dominant rate of decoherence is given by $\Gamma = \frac{n_1+n_2}{\tau}$, with corrections of the order
$\sqrt{n_1 n_2}$. The above analytical result can be compared to the numerical one obtained with the unravelling technique,
see Fig.~\ref{fig:decoh_ho}.
\par
Notice that,
in order to obtain the decoherence as a function of phase space variables, we have to perform the inverse Fourier transform
on the coherence given in Eq.~\eqref{eq:analytic_decoh_HO}. However, since the time evolution of $u_{n_1,n_2}$ is constant
in the $x$-space, we find that in $p$-space the solution picks up a delta function $\delta(p)$, and the decoherence rate is not
modified. The presence of a delta function for the momentum is consistent with the dynamics we are considering. Indeed, from
Eq.~\eqref{eq:HO_master_equation} it is easy to see that the coherence $u_{n_1,n_2}$ is non-zero for $p=0$ only, while
for any $k \neq 0$ the coherence $u_{n_1+k,n_2+k}$ is non-zero for $p = -k B \tau$. Likewise, the position dependence is
described by $\delta(q)$, since the master equation we are considering does not include the classical Hamiltonian $H_C =
\frac{p^2}{2m}$.
\par
It is worth noting that the evolution of the coherence induced by the CQ dynamics described in Eq.~\eqref{eq:HO_master_equation}
is analogous to the evolution induced by a fully quantum master equation where the phase space degrees of freedom are neglected.
The decoherence can however be interpreted in two different ways; for the CQ dynamics, this is due to the interaction with the classical
degrees of freedom, and the fact that (quantum) information is leaking in the classical phase space. For the fully quantum dynamics,
instead, the decoherence is due to the interaction with an external environment. The coherence evolution for these two settings can
still be distinguished if we introduce the classical Hamiltonian $H_C$ in the CQ dynamics, since this term provides a non-trivial dependence
on the position $q$ in phase space. As a final remark, we notice that coherence in this model is ``moving'' from the initial superposition between $\ket{n_1}$ and $\ket{n_2}$ to superposition of nearby Fock states. This is shown in the left and centre plots of Fig.~\ref{fig:decoh_ho}; the coherence of the initial state $\ket{n_1}+\ket{n_2}$ monotonically decreases, while the coherence of the superposition $\ket{n_1 \pm 1}+\ket{n_2 \pm 1}$ increases at early times and then start decreasing as well.

\section{Discussion}
\label{sec:con}

\subsection{Uniqueness of unravelling}

We have here considered general dynamics which consistently couples a quantum system with classical degrees of freedom. The combined system is described by a hybrid-density which gives the probability of the system to be at any point in phase space and if so, what the density matrix of the quantum system is.
We introduced an unravelling approach to solve the equations of motion. In this approach, both the classical and quantum systems evolve along a particular trajectory with some probability, and when we average over these trajectories, we recover the dynamics for the hybrid-density. Unlike the purely quantum case, the trajectories considered here can lead to objectively certain state transformations, in which the quantum system transitions to a particular pure state, conditioned on the classical degrees of freedom. This happens when the Lindblad operators $L_\alpha$ are uniquely determined by the classical transition $W^{\alpha}(z|z-\Delta)$ rate, so that the observation of a particular classical transition unambiguously informs us about the quantum transition. Since the classical system has a definite trajectory, the trajectory of the quantum system is also definite. This is contrast to the purely quantum case, where the decomposition of a density matrix into an ensemble of pure states is not generally unique. A particularly simple and illustrative example is the master equation
\begin{align}
    \frac{\partial \cqstate}{\partial t}=L_\alpha \rho(q,p-\Delta_\alpha) L^\dagger_\alpha-
    \frac{1}{2}\acom{L^\dagger_\alpha L_\alpha}{\cqstate(q,p)}
\end{align}
where, if the $\Delta_\alpha$ are different for each $\alpha$, observing a sudden jump in momentum by an amount $\Delta_\alpha$ implies that the quantum state has changed from $\ket{\psi}$ to a suitably normalized $L_\alpha \ket{\psi}$. The classical system encodes the trajectory of the quantum system.

Furthermore, for the class of hybrid master equations we considered, there is a very real sense in which the quantum system jumps from being in a superposition of several states, to being in one of those states. Ironically, this is in contrast to {\it spontaneous collapse models}~\cite{ghirardi_unified_1986,pearle_combining_1989,ghirardi_markov_1990} where collapse is meant to be caused by an unobservable field.  In these theories, as in ordinary quantum theory, the density matrix of the quantum system fully describes the system, and so, one cannot distinguish between decoherence (which we think of as the gradual decay of off-diagonal elements of the density matrix in a chosen basis), and the case where the wavefunction has a probability per unit time of being projected to a particular states, such that on average, off-diagonal matrix elements of the density matrix decay with time. In both cases, the density matrix of the quantum system is the same. In contrast, in the theory studied here,  by monitoring the classical degree of freedom one can determine that a sudden change has occurred.

The system which is treated classically could be merely a large quantum system that is effectively a classical system such as an environment or a measuring apparatus. But it could also be fundamentally classical, for example, space-time, or it could be a system whose classical nature is somehow assumed, such as the experimenter herself.
This provides a potential solution to the measurement or ''reality'' problem of quantum theory. Namely, quantum systems in superposition collapse to a particular state in that superposition because they interact with a system which is classical. Which set of possible states they can collapse to (sometimes called the {\it pointer basis}) is determined by the interaction.

Turning towards more pedestrian phenomena, we have found that the toy models we studied exhibited a trade-off between the amount of classical diffusion that occurred in phase space and the amount of quantum decoherence or noise that occurred in the quantum system. To minimise the disturbance on the quantum state, there needs to be a lot of diffusion in the classical system. In Ref.~\cite{dec_Vs_diff} we will formally prove that this is indeed the case. 

\subsection{Discussion on energy non-conservation}

We also found as in \cite{bps}, that because the Hamiltonian was not the generator of time-translations, it typically is not conserved. We saw that this effect does not have to be large, and there are a number of ways to keep it under control so that we can have stable solutions. Violations of energy conservation present a possible experimental signature of fundamental hybrid dynamics, but also feature in spontaneous collapse models, and gravitationally induced decoherence~\cite{diosi1987universal,penrose1996gravity}.
A full study on the issue of energy conservation is beyond the scope of the current work, but in the following we present a brief discussion of the problem, at leading order.

It is instructive to recall the analogous issue in classical mechanics.  If one has diffusion as in the Fokker-Planck equation, then the magnitude of the momentum will gradually drift and increase, resulting in energy increase. The introduction of a friction term prevents this, by causing the system to settle down to an equilibrium state. The Fokker-Planck equation for the phase space density $\rho(q,p)$ is of the form
\begin{align}
\frac{\partial \, \rho(q,p)}{\partial t}
=&
\pb{H_C(q,p)}{\rho}
+
\frac{\partial^2D(q,p)\rho(q,p)}{\partial p^2}
+\frac{\partial}{\partial p}\big[\eta(q,p)\rho(q,p)\big] ,
\label{eq:FP_with_friction}
\end{align}
with the friction coefficient $\eta:=\gamma\frac{\partial H_C}{\partial p}$. This is just the theory of Brownian motion as developed by Einstein and Smoluchowski, with a thermal stationary distribution given by
\begin{align}
\rho_\beta=\frac{e^{-\beta H_C}}{Z} ,
\end{align}
where $Z$ is the partition function and the inverse temperature is 
\begin{align}
\beta=\gamma/D .
\label{eq:temp}
\end{align}
By adding a friction term, we can stop the diffusion in the classical degrees of freedom.

For the type of hybrid model presented in Sec.~\ref{ssec:SG_diag_lin}, with Lindblad operators that are a basis of orthogonal projectors $L_\alpha=\proj{\alpha}$ onto energy eigenstates of $\hq$, and $h^\alpha=\omega Bq$ (to take a simple example) we may add a friction term such as
\begin{align}
\cal{F}(\rho)&=\frac{\partial}{\partial p}\big[\gamma(q)\frac{\partial H_C}{\partial p} \rho \big]
=\frac{\gamma(q)}{m}\Big[p \frac{\partial \rho}{\partial p}+\rho\Big]
\end{align}
to the right hand side of the evolution law of Eq.~\eqref{eq:general_SG_dynamics}. The friction coefficient $\gamma$ can be chosen to depend on $q$. To first order in $\tau$, the diffusion term is given by Eq.~\eqref{eq:SG-diffusion}, and so from Eq.~\eqref{eq:temp}, we expect that to first order, the classical system will equilibriate to a thermal state at inverse temperature $\beta=\gamma(q)/(B\omega \langle \alpha \rangle)^2\tau$. Because the Lindblad operators project onto eigenstates of $\hq$, the quantum system also equilibrates by decohering into one of it's energy eigenstates.

The case where the Lindblad operations cause noise is more interesting, since then one has diffusion not only in phase space but also in the Hilbert space. What's more, the amount of diffusion in phase space can depend on the quantum state; for example, for a quantum harmonic oscillator the diffusion can grow with $n$, the energy level of the oscillator. This means that the diffusion term can be arbitrarily large, and overcome any classical friction term. We would thus like the friction term to also increase with $n$ so that the classical system can reach a steady state. To explore this, let us consider a more sophisticated model than Eq.~\eqref{eq:FP_with_friction}, where a classical harmonic oscillator with Hamiltonian $H_C=\omega_c (p^2+q^2)/2$ is coupled to a quantum harmonic oscillators with free Hamiltonian $\hq=\omega_Q a^\dagger a$; the Lindblad operators are now $L_\alpha=a$, the annihilation operator. A natural coupling between the two oscillators would be to take the interaction Hamiltonian to be $H_I=-2\omega_c\omega_Q qQ$ with $Q=\frac{1}{\sqrt{2}}(a+a^\dagger)$, but it is hard to make the hybrid dynamics of such a coupling completely positive. An alternative, is to take the quantum part to be positive definite (while leaving the classical part arbitrary), e.g. $H_I=B(q) Q^2$, which is still a local coupling. Let us be slightly  more general, and consider coupling to $Q^2+\kappa P^2$ with $P=\frac{i}{\sqrt{2}}(a^\dagger-a)$ and $\kappa\geq 0$.  It will prove convenient to use $[a,a^\dagger]=1$ to rewrite this as
\begin{align}
H_I=B(q)\frac{1}{2} \Big((1-\kappa) (aa+a^\dagger a^\dagger)+(1+\kappa)(a^\dagger a + 1)\Big)
    \end{align}
where we have dropped the purely classical terms from the interaction. A master equation for such a theory is then
\begin{align}
\label{eq:HO_master_equation-with_CHO}
\frac{\partial \cqstate(q,p)}{\partial t} 
&= 
\{H_c(q,p),\cqstate(q,p)\}
+\mathcal{F}(\cqstate(q,p))
-i \omega_Q \big[ a^\dagger a,\cqstate(q,p)\big]
-i B(q) \big[Q^2+k P^2,\cqstate(q,p,t)\big]\nonumber\\
&+
\int d\Delta W_\uparrow (q,p|q,p-\Delta)a^{\dagger} \, \cqstate(q,p-\Delta) \, a + \int d\Delta W_\downarrow (q,p|q,p-\Delta) a \, \cqstate(q,p-\Delta) \, a^{\dagger} \nonumber \\
&- \frac{1}{2} \acom{W_\downarrow(q,p) \, a^{\dagger} a + W_\uparrow(q,p) \, a a^{\dagger}}{\cqstate(q,p)}
\end{align}
where we are still to specify the transition rates $W_{\uparrow}$ and $W_\downarrow$, associated to the pumping and dumping of the quantum oscillator, respectively, and the classical friction term $\mathcal{F}$. The last term in the first line is the influence of the classical system on the harmonic oscillator, while the second and third lines is the influence of the quantum system on the classical one. In the following, we will require the expansion of $W_{\uparrow}$ and $W_\downarrow$ to contain a term proportional to $\tau B'(q) \frac{\partial \cqstate(q,p)}{\partial p}$, so as to balance the energy increase of the classical system due to the interaction with the quantum one.

On the other hand, at leading order in $\Delta$ the second and third lines give rise to a Lindblad equation with a damping and pumping term,
\begin{align}
    \mathcal{D}(\cqstate):= W_\downarrow(z) \,
    a^{\dagger} \, \cqstate \, a + W_\downarrow (z) \, a \, \cqstate\, a^{\dagger}
- \frac{1}{2}\acom{W_\downarrow(z) a^{\dagger} a + W_\uparrow(z) a a^{\dagger}}{\cqstate} 
\label{eq:HOLindblad}
\end{align}
and so when $W_{\uparrow}(z) \geq W_\downarrow(z)$, the state of the harmonic oscillator will continue to increase in energy. The quantum analogue of adding a friction term is to take the damping term to be strictly larger than the pumping term, and if the classical system were to remain at rest, the quantum oscillator will thermalise to an inverse temperature proportional to $\log{W_\downarrow(z)/W_\uparrow(z)}$, at which point one can verify that $\mathcal{D}(\cqstate)=0$. We can thus have the system reach equilibrium by taking $\log{W_\downarrow(z)/W_\uparrow(z)}$ to be strictly positive and adding a purely classical friction term $\mathcal{F}(\cqstate)$.

If on the other hand, we take $B(q)$ and the rates $W_\downarrow(z)$ and $W_\uparrow(z)$ such that the classical system equilibriates in a region of phase space with sufficient probability of having  $W_\downarrow(z)\leq W_\uparrow(z)$, then energy can be pumped into the harmonic oscillator indefinitely. In such a case, we would like to control this, so that it happens slowly.  To explore this, let us consider the specific realisations,
\begin{subequations}
\begin{align}
 \int d\Delta W_\uparrow (p|p-\Delta)\, \cqstate(q,p-\Delta)
 &= \frac{1}{\tau} h_\uparrow(q) e^{\tau\partial_p ( X_\uparrow \cdot )}\cqstate(q,p),
\\
 \int d\Delta W_\downarrow (p|p-\Delta)\, \cqstate(q,p-\Delta)
 &= \frac{1}{\tau} h_\downarrow(q) e^{\tau\partial_p ( X_\downarrow \cdot )}\cqstate(q,p),
\end{align}
\end{subequations}
where the differential operator $\partial_p ( X \cdot )$ acts over an hybrid operator $\rho(q,p)$
as $\partial_p ( X \, \rho(q,p) )$. The two realisations are positive as long as $h_\uparrow$ and $h_\downarrow$ is everywhere positive, and the master equation will preserve normalisation with $W_\uparrow(q,p)=h_\uparrow(q)/\tau$ and $W_\downarrow(q,p)=h_\downarrow(q)/\tau$.  
Let us now choose $X_\downarrow=\Delta_\downarrow(q)+\eta_\downarrow$, $X_\uparrow = \Delta_\uparrow(q)+\eta_\uparrow$, where $\eta_\uparrow$ and $\eta_\downarrow$ are friction terms as in Eq.~\eqref{eq:FP_with_friction}. Then, at zeroth order in $\tau$ we have 
the Lindblad equation~\eqref{eq:HOLindblad} with temperature proportional to $\log{h_\downarrow(q)/h_\uparrow(q)}$. However, even if we don't have $h_\downarrow(q)\geq h_\uparrow(q)$, the rate at which energy is pumped into the oscillator is
\begin{equation}
\label{eq:energy_rate}
\delta \hq = \omega \frac{d{\langle a^\dagger a\rangle}}{d t} = \frac{2}{\tau}\omega^2(h_\uparrow(q)-h_\downarrow(q))n- \frac{\omega}{\tau} h_\uparrow(q),
\end{equation}
which is easily calculated from the master equation. This term can potentially be made small in the phase space region where the classical system equilibriates. The next order terms give,
\begin{align}
    \ldots &+ h_\downarrow(q)\Delta_\downarrow(q) \, a \frac{\partial \cqstate}{\partial p} a^\dagger
    + h_\uparrow(q)\Delta_\uparrow(q) \, a^\dagger \frac{\partial \cqstate}{\partial p} a\nonumber\\
    &+\frac{\tau}{2} h_\downarrow(q)\Delta_\downarrow(q)^2 \, a \frac{\partial^2 \cqstate}{\partial p^2} a^\dagger
    + \frac{\tau}{2} h_\uparrow(q)\Delta_\uparrow(q)^2 \, a^\dagger \frac{\partial^2 \cqstate}{\partial p^2}
    a\nonumber\\
    &+ h_\downarrow(q) \, a \frac{\partial\gamma_\downarrow\cqstate }{\partial p}a^\dagger
     + h_\uparrow(q) \, a^\dagger\frac{\partial\gamma_\uparrow \cqstate}{\partial p}a
     + \ldots \label{eq:expansion_friction}
\end{align}
where we take the last two terms to be of roughly the same order as the previous diffusion terms.

The $\kappa=1$ case is a natural one to consider, since it is reminiscent of the master equation of general relativity coupled to a scalar field~\cite{oppenheim_post-quantum_2018}. The classical oscillator couples to the energy of the quantum one, as the classical gravitational field couples to the scalar field expanded in terms of momentum modes. Likewise, analogy with the gravitational case suggests taking the pure commutator term of Eq.~\eqref{eq:HO_master_equation-with_CHO} since gravity is always coupling to the energy. In this case, we can take
\begin{align}
    h_\downarrow(q)\Delta_\downarrow(q)= h_\uparrow(q)\Delta_\uparrow(q)=\frac{\omega}{2}B'(q)
    \label{eq:DDtrade-off}
\end{align}
so that the first two terms in Eq.~\eqref{eq:expansion_friction} will give the Poisson bracket $\{B(q)\langle\frac{1}{2}Q^2+\frac{1}{2}P^2\rangle,\cqstate\}$ in the classical limit. When we put this all together, the master equation~\eqref{eq:HO_master_equation-with_CHO} expands as
\begin{align}
\label{eq:HO_master_equation-with_CHO_expanded}
\frac{\partial \cqstate(q,p)}{\partial t} 
&=  
\{H_c(q,p),\cqstate\}
+\mathcal{F}(\cqstate)
-i B(q) \, \omega \, \big[ a^\dagger a,\cqstate(q,p,t)\big] \nonumber \\
&+ \frac{\omega}{\tau} \Big(
   h_\uparrow(q) a^{\dagger} \, \cqstate \, a +h_\downarrow(q) a \, \cqstate\, a^{\dagger}
-\frac{1}{2} \acom{\left( h_\uparrow(q)+h_\downarrow(q) \right) a^{\dagger} a + h_\downarrow(q)}{\cqstate} \Big)\nonumber\\ &+\frac{B'(q)\,\omega}{2} \Big( a^{\dagger} \, \frac{\partial\cqstate}{\partial p} \, a + a \, \frac{\partial\cqstate}{\partial p}\, a^{\dagger}\Big)+
\frac{\tau}{4}B'(q)\omega \Big(\Delta_\uparrow(q) a^{\dagger} \, \frac{\partial^2\cqstate}{\partial p^2} \, a + \Delta_\downarrow(q) a \, \frac{\partial^2\cqstate}{\partial p^2}\, a^{\dagger}\Big)\nonumber\\
&+\omega \Big(h_\uparrow(q) a^{\dagger} \, \frac{\partial\eta_\uparrow\cqstate}{\partial p} \, a +
h_\downarrow(q) a \, \frac{\partial\eta_\downarrow\cqstate}{\partial p}\, a^{\dagger}\Big)
\end{align}

In analogy with the field theory case, we would like to take $h_\downarrow(q)=h_\uparrow(q)=h(q)$, since in that context, the theory is local with the consequence that the oscillator will not equilibrate and will instead increase in energy~\cite{poulinKITP}. This can be slowed down by any desired amount by taking $h(q)/\tau$ to be small. However, anticipating the discussion in Ref.~\cite{dec_Vs_diff}, slowing the energy increase down results in a greater amount of classical diffusion. In particular, the requirement that Eq.~\eqref{eq:DDtrade-off} be satisfied, which in this case is
\begin{align}
    h(q)\Delta(q)=\frac{\omega}{2}B'(q)
    \label{eq:equal_trade-off}
\end{align}
encodes this trade-off.
The right hand side governs the rate of evolution of $p$ and is set by the dynamics, and the left hand side then requires that a realisation of this form have a trade-off between $h(q)$ and $\Delta(q)$. 

Let us consider the simplest example of $B(q)=Gq$, where $G$ is a positive constant, so that the classical system acts as if it is subject to a constant force $F\approx\omega G n $ when the quantum system is in the Fock state $\ket{n}$, with $n$ large. The rate at which energy is pumped into the quantum harmonic oscillator is $\delta \hq =\omega h(q)/\tau$, as shown in Eq.~\eqref{eq:energy_rate}.
Recalling the Fokker-Planck equation, the effective diffusion coefficient $D$ is the term in front of $\frac{1}{2}\frac{\partial^2\cqstate}{\partial p^2}$ in Eq.~\eqref{eq:HO_master_equation-with_CHO_expanded}, going as $D\approx F\tau\Delta(q)/2$. Then, Eq.~\eqref{eq:equal_trade-off} gives $\delta \hq D \approx \frac{1}{4n}\omega F^2$ so that
a small amount of energy being pumped into the oscillator requires large diffusion in the classical system in relation to the total back-reaction exerted by the quantum system on the classical one. As $n$ increases the relative trade-off, when the force $F$ is held fixed, becomes less pronounced.

There appears to be significant freedom in the choice of $h(q)$ and $\Delta(q)$, the main requirement other than Eq.~\eqref{eq:equal_trade-off} being that $h(q)$ be positive. One can even take $h(q)=|B'(q)|$ and $\Delta(q)$ proportional to $\mathrm{sign}(B'(q))$. If on the other hand, $\Delta(q)$ is large, we can then get energy being pumped into the classical system, since large amounts of diffusion in momentum will increase the energy of the classical system. This can be controlled by increasing the friction term, to keep the momentum low. However, since the diffusion term goes like $D \propto n \omega G \tau \Delta(q)$ and thus increases with $n$, the friction term would also need to scale with $n$ or risk getting overwhelmed by the diffusion. For this reason, the purely classical friction term $\mathcal{F}(\cqstate)$ will not be sufficient, while the hybrid friction terms in Eq.~\eqref{eq:HO_master_equation-with_CHO_expanded},
\begin{equation}
\mathcal{F}_H(\cqstate)
=
\omega \Big(h(q) a^{\dagger} \, \frac{\partial\eta_\uparrow\cqstate}{\partial p} \, a +
h(q) a \, \frac{\partial\eta_\downarrow\cqstate}{\partial p}\, a^{\dagger}\Big),
\end{equation}
should allow us to compensate the increase. This is seen explicitly by tracing out the oscillator, resulting in dynamics for the classical phase space density $\rho_n$ conditioned on the the oscillator being in state $n$
\begin{align}
\label{eq:HO_master_equation-for rhon}
\frac{\partial \rho_n(q,p)}{\partial t} 
&=  
\{H_c,\rho_n\}
+\mathcal{F}(\rho_n)\nonumber\\
&+B'(q)\omega (n+\frac{1}{2})\frac{\partial\rho_n}{\partial p}
+\frac{\tau\Delta}{2}B'(q)\omega(n+\frac{1}{2}) \frac{\partial^2\rho_n}{\partial p^2} \nonumber\\
&+2h(q)\omega (n+\frac{1}{2}) \frac{\partial\eta\rho_n}{\partial p}+...
\end{align}
where we have taken $\eta_\uparrow=\eta_\downarrow=\eta$ for simplicity. To this order, this is just the Fokker-Planck equation, and if $\eta=\tilde{\gamma}\frac{\partial H_c}{\partial p}$, then we expect the effective diffusion coefficient to be $D\approx\tau\Delta(q) B'(q) \omega n$, and the effective friction coefficient to be $\gamma\approx n h(q)\tilde{\gamma}(q)$. Eq.~\eqref{eq:temp} would then give the effective inverse ``temperature'' that the classical system is held at, although here, since it could depend on $q$, it is not a temperature but still defines the equilibrium state $\beta =h(g)\tilde{\gamma}(q)/\tau\Delta(q) B'(q) \omega $, which is now independent of $n$. Note that this system could act as if the classical system is being held at one temperature while the oscillator is held at a different temperature. Were we to have access to the environment that each system appears immersed in, we would presumably see heat flow from one  environment to the other. Understanding how this would work in detail, either analytically or numerically would be an interesting research direction to pursue. One would like to better understand  the various trade-offs involved in systems like this, and what realisations lead to equilibrium or steady states, or states which only change slowly.
\section*{Acknowledgements}
We would like to thank Robert Alicki and Joan Camps for valuable discussions, and Juan Rocamonde Quintela for comments on a draft version of this paper. JO is supported by an EPSRC Established Career Fellowship, a Royal Society Wolfson Merit Award, C.S. and Z.W.D.~acknowledges financial support from EPSRC. This research was supported by the National Science Foundation under Grant No. NSF PHY11-25915 and by the Simons Foundation {\it It from Qubit} Network.  Research at Perimeter Institute is supported in part by the Government of Canada through the Department of Innovation, Science and Economic Development Canada and by the Province of Ontario through the Ministry of Economic Development, Job Creation and Trade.

\bibliography{./biblio_unrav,./refgrav2,./refcq,./refjono2,./rmp12}
\bibliographystyle{plainnat}
\appendix
\section{The CQ master equation as a stochastic equation}
\label{app:master_stoch}
In this appendix, we provide an analytical solution for the CQ master equations considered in Sec.~\ref{sec:SG_evo}.
These equations are special cases of the more general one shown in Eq.~\eqref{eq:general_SG_dynamics}, which
we reproduce here for simplifying the exposition,
\begin{align*}
\frac{\partial \, \cqstate(z,t)}{\partial t}
&=
-i \left[ H_I(z) , \cqstate(z,t) \right]
+
\frac{1}{\tau_0} \left( e^{\tau_0 \, \pb{ H_C (z) }{ \, \cdot \, }} - 1 \right) \cqstate(z,t) \\
&+
\frac{1}{\tau} \sum_{\alpha \neq 0} 
\left(
e^{\tau \, \pb{ h^{\alpha} (z) }{ \, \cdot \, }} \, L_{\alpha} \, \cqstate(z,t) \, L^{\dagger}_{\alpha}
-
\frac{1}{2} \acom{ L^{\dagger}_{\alpha} L_{\alpha} }{ \cqstate(z,t) }
\right),
\end{align*}
where we recall that the interaction Hamiltonian $H_I(z) = \sum_{\alpha \neq 0} h^{\alpha} (z) L^{\dagger}_{\alpha}
L_{\alpha}$ drives a stochastic evolution of the classical and quantum degrees of freedom with a rate $\tau > 0$, while
the classical Hamiltonian $H_C (z)$ governs the stochastic evolution of the sole classical degrees of freedom with a
different rate $\tau_0$.
\subsection{Solution of a stochastic equation}
\label{apps:stoc_eq}
We first present the solution of Eq.~\eqref{eq:general_SG_dynamics} in the special case in which no
classical-quantum coupling is present, and the whole master equation only describes the evolution of
the classical degrees of freedom. The classical system is here described by the ensemble $\rho(q,p,t)$,
and its evolution is given by
\begin{equation}
\label{eq:stoc_eq}
\frac{\partial \rho(q,p,t)}{\partial t}
=
\frac{1}{\tau_0}
\left(
e^{\tau_0 \pb{ H_C(q,p) }{ \cdot }} - 1
 \right)
 \rho(q,p,t),
\end{equation}
where the finite parameter $\tau_0 > 0$ can be understood as the rate of jumping in phase space, due to the effect
of the shift operator $e^{\tau \pb{ H_C(q,p) }{ \cdot }}$.
\par
It is easy to show that the solution of Eq.~\eqref{eq:stoc_eq} is given by
\begin{equation} \label{eq:solution_easy_stoc}
\rho(q,p,t) = \sum_{k=0}^{\infty} P(k) \, \rho(k \tau_0),
\end{equation}
where $P(k)$ is the Poisson distribution which carries the time dependence of the solution,
\begin{equation}
P(k) = \left( \frac{t}{\tau_0} \right)^k \frac{e^{-\frac{t}{\tau_0}}}{k!}
\end{equation}
and the ensemble $\rho(k \tau_0)$ is obtained from the initial ensemble $\rho(q,p,t=0)$ by applying
the operator $k$ times,
\begin{equation} \label{eq:liouv_eq_sol}
\rho(k \tau_0) = e^{k \tau_0 \pb{ H_0(q,p) }{ \cdot }} \rho(q,p,0).
\end{equation}
It is worth noting that each ensemble $\rho(k \tau_0)$ is the solution of the Liouville equation
\begin{equation}
\label{eq:liouv_eq}
\frac{\partial \rho(q,p,t)}{\partial t}
=
\pb{ H_C(q,p) }{ \rho(q,p,t) },
\end{equation}
at time $t = k \tau_0$, so that we can interpret it as the state of the system after $k$ jumps in
phase space. Thus, the solution of Eq.~\eqref{eq:stoc_eq} can be understood as a mixture
of different ensembles, where each of these ensemble are the solution of Eq.~\eqref{eq:liouv_eq}
at discrete times $k \tau_0$, for $k \in \N$, and the probability of making $k$ jumps is given by
the Poisson distribution whose mean value is $\frac{t}{\tau_0}$.
\begin{figure}[ht!]
\includegraphics[width=0.5\textwidth]{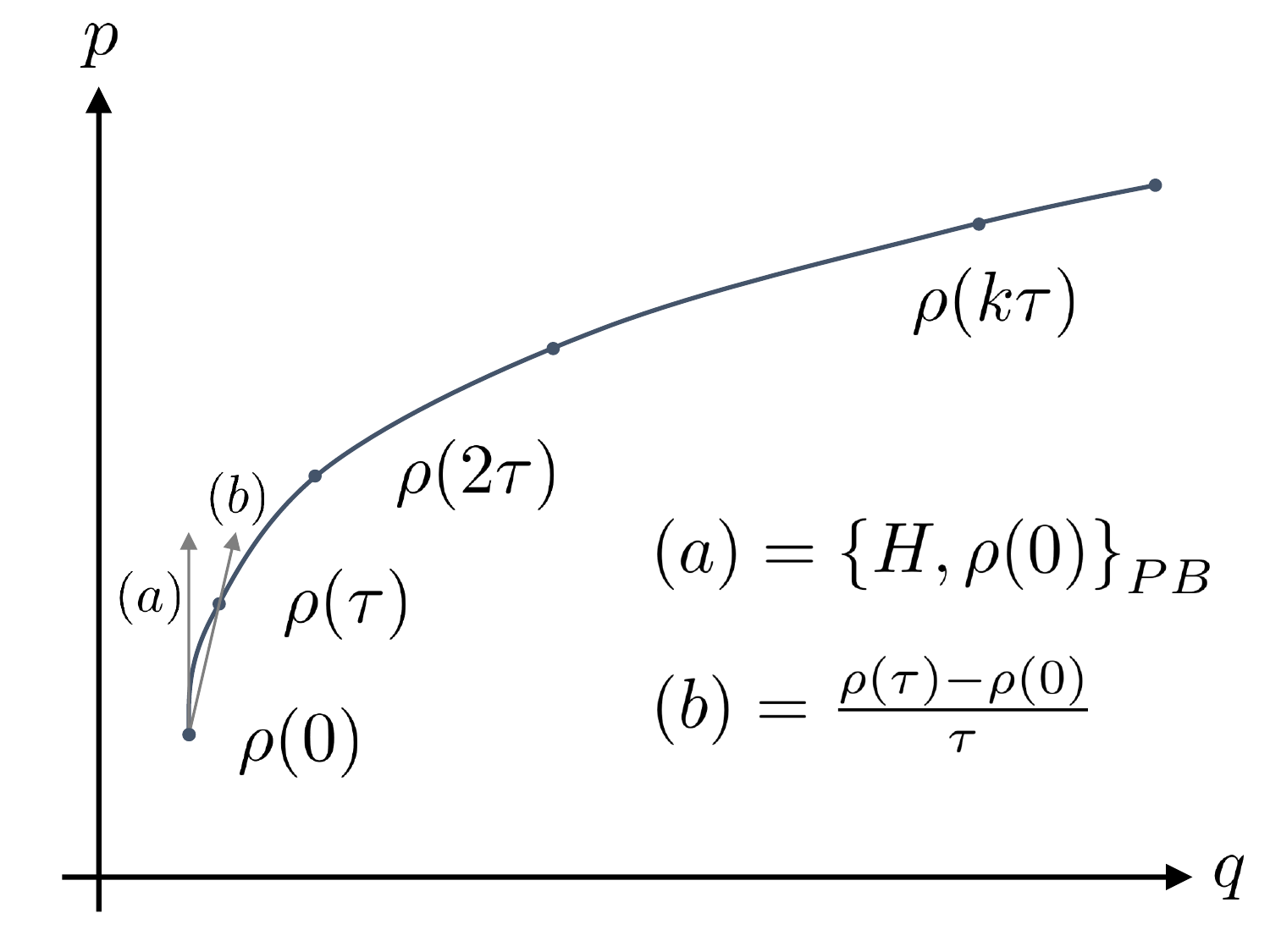}
\caption{The phase-space of a classical system with Hamiltonian $H_C(q,p)$. The blue line
represent the continuous trajectory of the ensemble $\rho(q,p,t)$, which initially is described
by $\rho(q,p,0) = \delta(q-\bar{q}) \delta(p-\bar{p})$. This trajectory is obtained through the
Liouville equation, see Eq.~\eqref{eq:liouv_eq}. Each point on the trajectory corresponds to the position
of the ensemble at discrete times $t = k \tau_0$, $k \in \N$. The solution of Eq.~\eqref{eq:stoc_eq} is a
mixture of the ensembles at these discrete times, weighted by a Poisson distribution with mean value
$\frac{t}{\tau_0}$. This equation can be understood as a constraint on the instantaneous derivative of the
ensemble, as shown in Eq.~\eqref{eq:stoch_as_deriv_const}. In particular, the derivative at time $t$
needs to be equal to $\frac{\rho(q,p,t+\tau_0) - \rho(q,p,t)}{\tau_0}$, for a finite value of $\tau > 0$.}
\label{fig:liuv_vs_jumping}
\end{figure}
\par
We can additionally use the simple master equation here introduced to better understand the role of the diffusion term discussed in Sec.~\ref{sec:unravelling}, and more generally the role of the infinite number of corrections we obtain when the hybrid master equation is expanded with respect to $\tau_0$. Let us first notice that Eq.~\eqref{eq:stoc_eq} can be re-express as
\begin{equation}
\label{eq:stoch_as_deriv_const}
\frac{\partial \rho(q,p,t)}{\partial t}
=
\frac{\rho(q,p,t+\tau_0) - \rho(q,p,t)}{\tau_0},
\end{equation}
where $\rho(q,p,t+\tau_0)$ is the solution of the Liouville equation at time $t+\tau_0$, as we show in
Eq.~\eqref{eq:liouv_eq_sol}. Thus, the solution of the master equation is given by an ensemble whose
instantaneous derivative at time $t$ depends on the solution of the Liouville equation at time $t+\tau_0$,
and more in general on the phase-space trajectory imposed by the Hamiltonian $H_C(q,p)$ through the
Liouville equation, see Fig.~\ref{fig:liuv_vs_jumping}. As a result, the instantaneous derivative of the
solution of the master equation cannot be given by the sole Poisson bracket of the Hamiltonian, but needs
to be suitably corrected. These corrections are reflected in the infinite expansion of the hybrid dynamics, see Eq.~\eqref{eq:cq-expanded} in the main text.
\subsection{Solution to the stochastic CQ dynamics with diagonal Lindblad operators}
We can now provide a solution for Eq.~\eqref{eq:general_SG_dynamics} in a simple case, which
nevertheless turns out to be useful for obtaining an analytical solution in Sec.~\ref{ssec:SG_diag_lin}.
Specifically, we study the case in which a single population $u(q,p,t)$ of the CQ state is considered,
and only two shifts are considered, one associated with the classical Hamiltonian $H_C(q,p)$, and
one associated with the quantum coupling $H_I(q,p)$. The equation we consider is then
\begin{equation}
\label{eq:stoch_two_ham}
\frac{\partial u(q,p,t)}{\partial t}
=
\frac{1}{\tau_0} \left( e^{\tau_0 \pb{ H_C(q,p) }{ \cdot }} \ u(q, p, t) - u(q,p,t) \right)
+
\frac{1}{\tau_1} \left( e^{\tau_1 \pb{ H_I(q,p) }{ \cdot }} \ u(q, p, t) - u(q,p,t) \right),
\end{equation}
where the action of the two Hamiltonians can either commute or not.
\par
If the action of the two Hamiltonians commutes, $\pb{ H_I(q,p) }{ H_C(q,p) } = 0$, then
the solution of the above equation can be easily obtained, and it is given by
\begin{equation}
\label{eq:sol_stoch_two_ham}
u(q,p,t) = \sum_{k,n = 0}^{\infty} P_0(k) \, P_1(n) \, u(n,k),
\end{equation}
where $k$ is the number of jumps associated with $H_C$, and $n$ is the number of jumps associated
with $H_I$. Both $P_0(k)$ and $P_1(n)$ are Poission distributions, defined as
\begin{equation}
P_i(m) = \left( \frac{t}{\tau_i} \right)^{m} \frac{e^{-\frac{t}{\tau_i}}}{m!} , \quad i = 0,1,
\end{equation}
and the population $u(n,k)$ is obtained from the initial one $u(q,p,t)$ by applying $k$ classical
jumps and $n$ quantum jumps, that is,
\begin{equation}
u(k,n) = e^{\tau_0 k \pb{ H_C(q,p) }{ \cdot } + \tau_1 n \pb{ H_I(q,p) }{ \cdot }} \ u(q, p, 0).
\end{equation}
It is easy to check that this is the solution of Eq.~\eqref{eq:stoch_two_ham}.
\par
When the action of the two Hamiltonians does not commute, i.e., $\pb{ H_I(q,p) }{ H_C(q,p) } \neq 0$,
the solution of the stochastic CQ equation is still in the form of Eq.~\eqref{eq:sol_stoch_two_ham}, with $P_0(k)$ and
$P_1(n)$ Poisson distributions. However, the population $u(n,k)$ is now different, since the order of the shifts is now important,
and we write it as
\begin{equation}
\label{eq:shifted_evolution}
u(k,n) = \frac{1}{\binom{n+k}{k}} \sum_i \pi_i
\left( 
\underbrace{e^{\tau_0 \pb{ H_C(q,p) }{ \cdot }} \ldots e^{\tau_0 \pb{ H_C(q,p) }{ \cdot }}}_k
 \,
\underbrace{e^{\tau_1 \pb{ H_I(q,p) }{ \cdot }} \ldots e^{\tau_1 \pb{ H_I(q,p) }{ \cdot }}}_n
\right)
u(q, p, 0),
\end{equation}
where $\pi_i$ is a permutation on the $n+k$ shifts which create a different combination, and $i = 1, \ldots, \binom{n+k}{k}$.
\subsubsection{Solution to the stochastic CQ dynamics for a qubit in linear potential}
\label{app:sol_SG_diag_lin}
Let us now use the results obtained in the previous section to study the analytical solution of the qubit
in linear potential, where the Lindblad operators are diagonal, see Sec.~\ref{ssec:SG_diag_lin}. It is easy to
see that the differential equations for the populations, Eq.~\eqref{eq:evo_pop_SG_diag}, are almost in the same
form of the one we have studied in the previous section. Indeed, one can obtain these equations from the stochastic
one in Eq.~\eqref{eq:stoch_two_ham} by setting the classical Hamiltonian $H_C = \frac{p^2}{2\,m}$, the quantum
coupling $H_I = q \, B \omega$, and sending $\tau_0 \rightarrow 0$.
\par
Notice that the action of the two Hamiltonians does not commute, since their Poisson bracket is not zero, and the
two shifts act as follow on the population,
\begin{subequations}
\begin{align}
e^{\tau_0 \pb{ H_C(q,p) }{ \cdot }} \ u(q,p) &= u(q - \frac{p}{m} \tau_0, p), \\
\label{eq:mom_jump}
e^{\tau_1 \pb{ H_I(q,p) }{ \cdot }} \ u(q,p) &= u(q, p + B \omega \tau_1). 
\end{align}
\end{subequations}
It is clear that shifts in momentum affect the subsequent shifts in position, since the latter depend on the
value of the momentum. We can define the jump units for position and momentum as $\Delta p = B \omega \tau_1$,
and $\Delta q = \frac{B \omega \tau_1}{m} \tau_0$.
\par
We are now interested in expressing the population $u(k,n)$ of Eq.~\eqref{eq:shifted_evolution} in terms
of its position and momentum, rather than in terms of its classical and quantum jumps. We can write it as
\begin{equation}
u(k,n) = \int \rmd q \, \rmd p \ P(q,p|k,n) u(q,p)
\end{equation}
where the conditional probability distribution $P(q,p|k,n)$ can be divided into two distributions, one for the
position and one for the momentum,
\begin{equation}
P(q,p|k,n) = P(q|k,n) \, P(p|n).
\end{equation}
By looking at the effect of the quantum jump on the state, see Eq.~\eqref{eq:mom_jump}, it is easy to show
that the distribution of the momentum, conditioned on the number of quantum shifts $n$, is given by
\begin{equation}
P(p|n) = \delta(p - n \Delta p).
\end{equation}
The conditional probability distribution for the position, instead, is less straightforward to obtain.
\par
In order to get the mean value and variance of the distribution $P(q|k,n)$, we express the problem in a
different way. For a fixed value of classical and quantum jumps $k$ and $n$, we consider all possible
combinations of jumps, or \emph{histories}. We represent one such history with a $n+k$-bit string $x$,
where $0$ is a position jump, and $1$ is a momentum jump. The final position of the particle for a given
history can then be computed using the following formula, 
\begin{equation}
\label{eq:pos_nkx}
q_{k,n}(x) = n k + \frac{n \left( n - 1 \right)}{2} - \sum_{\ell = 0}^{n+k-1} \ell \, x_{\ell},
\end{equation}
where $x_{\ell}$ is the $\ell$-th bit of the string $x$.
Eq.~\eqref{eq:pos_nkx} provides a link between a history $x$ and the final position $q$, and can
therefore be used to get the conditional distribution we are interested in. In particular, we can make
the simplifying assumption that each $x_{\ell}$ is an \emph{independent} random variable,
\begin{equation}
x_{\ell} =
\begin{cases}
1 \quad \text{with} \ p = \frac{n}{n+k},\\
0 \quad \text{with} \ 1-p = \frac{k}{n+k}.
\end{cases}
\end{equation}
The mean value and variance of $x_{\ell}$ is the same for all $\ell$, and they are given by
$\langle x_{\ell} \rangle = \frac{n}{n+k}$ and $\sigma_{\ell}^2 = \frac{n k}{(n+k)^2}$,
respectively. If we now use Eq.~\eqref{eq:pos_nkx}, together with the fact that each $x_{\ell}$
is independent, we find that
\begin{subequations}	
\begin{align}
\langle q_{k,n} \rangle &= n k + \frac{n \left( n - 1 \right)}{2} - \sum_{\ell = 0}^{n+k-1} \ell \, \langle x_{\ell} \rangle = \frac{n k}{2} , \\
\sigma_{q_{k,n}}^2 &= \sum_{\ell = 0}^{n+k-1} \ell^2 \, \sigma_{\ell}^2 = \frac{n k (n+k-1) (2 (n+k)-1)}{6 (n+k)},
\end{align}
\end{subequations}	
where in the second equation we have used the fact that, given two random variables $X$, $Y$ and
two real numbers $a$, $b$, the variance of a random variable $Z = a X + b Y$ is given by $\sigma^2_Z
= a^2 \sigma^2_X + b^2 \sigma^2_Y$.
\par
We can now use the above results for estimating the mean value and variance of the marginal distributions
for position and momentum. These distributions are given by, respectively, 
\begin{subequations}	
\begin{align}
P(q) &= \sum_{n,k=0}^{\infty} P(q|k,n) \, P_0(k) \, P_1(n) , \\
P(p) &= \sum_{n=0}^{\infty} P(p|n) \, P_1(n) ,
\end{align}
\end{subequations}	
The mean value and variance for momentum are given by the Poisson distribution $P_1(n)$, since
$P(p|n) = \delta(p - n \Delta p)$, and we get
\begin{subequations}	
\label{eqs:mean_var_mom}
\begin{align}
\langle p \rangle &= \frac{\Delta p \, t}{\tau_1} = B \omega t , \\
\sigma_{p}^2 &= \frac{\Delta p^2 \, t}{\tau_1} = (B \omega)^2 \tau_1 t .
\end{align}
\end{subequations}	
To compute the mean value and variance of position, instead, we can make use of the law of total expectation
and the law of total variance. In the limit of $n,k \gg 1$, we have that
$\sigma_{q_{k,n}}^2 \approx \frac{1}{3} n k (n+k)$, and after straightforward calculations we find
\begin{subequations}	
\label{eqs:mean_var_pos}
\begin{align}
\langle q \rangle &= \frac{1}{2} \frac{t}{\tau_0} \frac{t}{\tau_1} \Delta q  = \frac{1}{2} \frac{B \omega}{m} t^2 , \\
\sigma_{q}^2 &= \frac{1}{3} \frac{t^2}{\tau_0 \, \tau_1} \left( \frac{t}{\tau_0} + \frac{t}{\tau_1} + 5 \right) \Delta q^2.
\end{align}
\end{subequations}	
The results of Eqs.~\eqref{eqs:mean_var_mom} and \eqref{eqs:mean_var_pos} are then compared,
in Fig.~\ref{fig:SG_diag_variance}, to the values obtained by numerical simulation. 
\subsection{Solution to the stochastic CQ dynamics with non-diagonal Lindblad operators}
\label{app:stoc_sol_SG_nondiag}
We now consider the case in which the Lindblad operators of Eq.~\eqref{eq:general_SG_dynamics} are non-diagonal,
and therefore the evolution of each population depends on the others. We consider the case of a two-level
quantum system, since this is the situation studied in the main text, see Sec.~\ref{ssec:SG_nondiag}. In
particular, we study the evolution of the populations of a CQ system with classical Hamiltonian $H_C(q,p)$ and
interaction Hamiltonian $H_I(q,p)$ (whose action we assume does not commute), and with Lindbald operators
connecting the two levels (see Eqs.~\eqref{eq:lindblad_non_diag} for an example). The stochastic equation for
the population $u_i$, where $i = 0,1$, is given by
\begin{align}
\label{eq:stoc_SG_nondiag}
\frac{\partial u_i(q,p,t)}{\partial t}
=&
\frac{1}{\tau_0} \left( e^{\tau_0 \pb{ H_C(q,p) }{ \cdot }} \ u_i(q, p, t) - u_i(q,p,t) \right)\nonumber\\
&+
\frac{1}{\tau_1} \left( e^{\tau_1 \pb{ H^{(i)}_Q(q,p) }{ \cdot }} \ u_{i \oplus 1}(q, p, t) - u_i(q,p,t) \right),
\end{align}
where the interaction Hamiltonian additionally depends on the level is acting on. 
\par
In order to solve this equation, we re-write the time derivative as $\left( u_i(t+\delta t) - u_i(t) \right)
\frac{1}{\delta t}$, and we express Eq.~\eqref{eq:stoc_SG_nondiag} as
\begin{equation}
u_i(t+\delta t) =
\frac{\delta t}{\tau_0} e^{\tau_0 \pb{ H_C }{ \cdot }} \, u_i(t)
+
\frac{\delta t}{\tau_1} e^{\tau_1 \pb{ H^{(i)}_Q }{ \cdot }} \, u_{i \oplus 1}(t)
+
\left( 1 - \left( \frac{\delta t}{\tau_0} + \frac{\delta t}{\tau_1} \right) \right) u_i(t)
\end{equation}
where we suppressed the dependence on the phase-space variables for compactness. If we
now group the terms in the above equation into the following operators, acting on $u_i$
and $u_{i \oplus 1}$ respectively,
\begin{align*}
A &=
\frac{\delta t}{\tau_0} e^{\tau_0 \pb{ H_C }{ \cdot }}
+
\left( 1 - \left( \frac{\delta t}{\tau_0} + \frac{\delta t}{\tau_1} \right) \right), \\
B_i &=
\frac{\delta t}{\tau_1} e^{\tau_1 \pb{ H^{(i)}_Q }{ \cdot }} \quad i = 0,1 
\end{align*}
we can express the evolution of the populations as
\begin{align}
u_0(t+\delta t) = A \, u_0(t) + B_0 \, u_1(t), \\
u_1(t+\delta t) = A \, u_1(t) + B_1 \, u_0(t).
\end{align}
By considering a sequence of times $t = \left\{k \, \delta t \right\}_{k \in \N}$, we can recursively solve the
above equations, and we obtain
\begin{equation}
u_i(k \, \delta t) = c^{(i)}_0(k) \, u_i(0) + c^{(i)}_1(k) \, u_{i \oplus 1}(0),
\end{equation}
where the coefficients (when $k$ is even) are given by
\begin{subequations}
\begin{align}
c^{(i)}_0(k) &=
\sum_{j=0}^{\frac{k}{2}} \sum_{m=1}^{\binom{k}{2j}}
\pi_m \left( A^{k - 2j} \left(B_i B_{i \oplus 1}\right)^{j} \right), \\
c^{(i)}_1(k) &=
\sum_{j=0}^{\frac{k}{2} - 1} \sum_{m=1}^{\binom{k}{2j+1}}
\pi_m \left( A^{k - (2j+1)} \left(B_i B_{i \oplus 1}\right)^{j} B_i \right).
\end{align}
\end{subequations}
In the above equations, $\pi_m \in S_k$ are permutations of the operators $A$’s and $B_i$’s such that
each combination obtained is different from the others, and the relative order between the $B_i$’s is
not modified.
\par
If we now replace the operators $A$ and $B_i$, and we send $k \rightarrow \infty$ and $\delta t \rightarrow 0$
while asking for $k \, \delta t = t$, we find that the solution of Eq.~\eqref{eq:stoc_SG_nondiag} is
\begin{equation}
u_i(q,p,t)
=
\sum_{\ell = 0}^{\infty} \sum_{j = 0}^{\infty} P_0(\ell) \, P_1(2j) \, u_i(\ell,j)
+
\sum_{\ell^{\prime} = 0}^{\infty} \sum_{j^{\prime} = 0}^{\infty} P_0(\ell^{\prime}) \, P_1(2j^{\prime}+1) \, u_{i \oplus 1}(\ell^{\prime},j^{\prime}),
\end{equation}
where both $P_0$ and $P_1$ are Poisson distributions with mean value $\frac{t}{\tau_0}$ and $\frac{t}{\tau_1}$
respectively, and
\begin{subequations}
\begin{align}
u_i(\ell,j) &=
\frac{1}{\binom{\ell + 2j}{\ell}}
\sum_m \pi_m
\left(
\left(
e^{\tau_0 \pb{ H_C }{ \cdot }}
\right)^{\ell}
\left(
e^{\tau_1 \pb{ H^{(i)}_Q }{ \cdot }}
\,
e^{\tau_1 \pb{ H^{(i \oplus 1)}_Q }{ \cdot }}
\right)^j
\right)
u_i(q,p,t=0), \\
u_{i \oplus 1}(\ell,j) &=
\frac{1}{\binom{\ell + 2j + 1}{\ell}}
\sum_m \pi_m
\left(
\left(
e^{\tau_0 \pb{ H_C }{ \cdot }}
\right)^{\ell}
\left(
e^{\tau_1 \pb{ H^{(i)}_Q }{ \cdot }}
\,
e^{\tau_1 \pb{ H^{(i \oplus 1)}_Q }{ \cdot }}
\right)^j
e^{\tau_1 \pb{ H^{(i)}_Q }{ \cdot }}
\right) \times \nonumber\\
&\hphantom{\frac{1}{\binom{\ell + 2j + 1}{\ell}}
\sum_m \pi_m
\left(
\left(
e^{\tau_0 \pb{ H_C }{ \cdot }}
\right)^{\ell}
\left(
e^{\tau_1 \pb{ H^{(i)}_Q }{ \cdot }}
\,
e^{\tau_1 \pb{ H^{(i \oplus 1)}_Q }{ \cdot }}
\right)^j\right)}\times u_{i \oplus 1}(q,p,t=0).
\end{align}
\end{subequations}
In the above solution, the probability distributions $P_0$ and $P_1$ carry the time dependence,
while the functions $u_i(\ell,j)$ and $u_{i \oplus 1}(\ell,j)$ contain the phase-space information.
In order to find a more explicit solution, one needs to consider specific Hamiltonian operators
$H_C(q,p)$ and $H_I(q,p)$, as we do in the main text in Sec.~\ref{ssec:SG_nondiag}.
\section{Unravelling code for CQ dynamics}
\label{app:unravellign_code}
The unravelling code is tailored to solve a specific model of CQ dynamics, whose master
equation is given in Eq.~\eqref{eq:general_SG_dynamics} for a rate of classical jump
$\tau_0 \rightarrow 0$, that is,
\begin{align}
\label{eq:master_eq_code}
\frac{\partial \, \cqstate(z,t)}{\partial t}
=&
-i \left[ H_I(z) , \cqstate(z,t) \right]
+
\pb{ H_C (z) }{ \cqstate(z,t) }\nonumber\\
&+\frac{1}{\tau} \sum_{\alpha} 
\left(
e^{\tau \, \pb{ h^{\alpha} (z) }{ \, \cdot \, }} \, L_{\alpha} \, \cqstate(z,t) \, L^{\dagger}_{\alpha}
-
\frac{1}{2} \acom{ L^{\dagger}_{\alpha} L_{\alpha} }{ \cqstate(z,t) }
\right),
\end{align}
where $H_I(z) = \sum_{\alpha} h^{\alpha} (z) \, L^{\dagger}_{\alpha} L_{\alpha}$ is the interaction Hamiltonian
between the classical and quantum degrees of freedom. For simplicity, we consider the case in which $z = \left( q, p \right)$,
that is, we consider a single system in phase space. The unravelling of this master equation can be obtained following
the steps shown in Sec.~\ref{sec:unravelling_CQ} of the main text. However, here we will modify the updating rule of
Eq.~\eqref{eq:unrav_evo} to include a continuous evolution of the classical degrees of freedom together with that of
the quantum degrees of freedom.
\par
As stressed in the main text, the unravelling code evolves the CQ pure state, see Eq.~\eqref{eq:CQ_pure_state}, which
we represent with the tuple $\left( \ket{\phi} , \ q , \ p , \ t \right)$. The first element of the tuple, $\ket{\phi} \in \C^d$
is the pure state of a qudit, while $q \in \R$ is the position and $p \in \R$ is the momentum of the system, and $t \in \R_+$
is time. To update the state, we first sample from a uniform distribution over the interval $\left[0, 1 \right]$, and check if the
outcome $p^{\star}$ is lower or equal that
\begin{equation}
p_0 = 1 - \frac{\delta t}{\tau} \sum_{\alpha} \bra{\phi} L_{\alpha}^{\dagger} L_{\alpha} \ket{\phi}.
\end{equation}
If this is the case, that is, $p^{\star} \leq p_0$, we apply the continuous update to the CQ state, and we obtain the following
tuple,
\begin{equation}
\label{eq:cont_evo_code}
\left(
\frac{1}{\sqrt{N_0}} \, \left( 1 - i \, \delta t \, H_{\text{eff}} \right) \ket{\phi} , \
q + \frac{\partial H_C}{\partial p} \, \delta t , \
p - \frac{\partial H_C}{\partial q} \, \delta t , \
t + \delta t
\right),
\end{equation}
where $N_0$ is the normalisation of the updated quantum state, and $H_{\text{eff}} = H_I(z)
- \frac{i}{2 \tau} \sum_{\alpha} L_{\alpha}^{\dagger} L_{\alpha}$. If $p^{\star} > p_0$, instead,
we randomly choose an $\alpha$-jump according to the distribution,
\begin{equation}
p_{\alpha} = \frac{\delta t}{\tau} \bra{\phi} L_{\alpha}^{\dagger} L_{\alpha} \ket{\phi}, \quad \forall \, \alpha.
\end{equation}
In this case, we evolve the state using the following update rule,
\begin{equation}
\left(
\frac{1}{\sqrt{N_{\alpha}}} \, L_{\alpha} \ket{\phi} , \
q + \frac{\partial h^{\alpha}}{\partial p} \, \tau , \
p - \frac{\partial h^{\alpha}}{\partial q} \, \tau , \
t + \delta t
\right),
\end{equation}
where $N_{\alpha}$ is the normalisation of the new quantum state.
\par
It is worth noting that the updating rule of Eq.~\eqref{eq:cont_evo_code} includes the continuous
evolution of the classical degrees of freedom, as opposed to the case shown in the main text,
Eq.~\eqref{eq:unrav_evo}. This addition in the updating rule is due to presence, in the above
master equation, of the Poisson bracket between the classical Hamiltonian $H_C$ and the CQ state.
In fact, it is easy to see that the first order in $\delta t$ of the phase space shift operator coincides
with the action of the Poisson bracket, since for any (analytic) function $f$ over the phase space we
have
\begin{align}
\label{eq:approx_shift}
&f \left( q - \frac{\partial H_C}{\partial p} \, \delta t , p + \frac{\partial H_C}{\partial q} \, \delta t \right)
=\nonumber\\
&=e^{\delta t \left( \frac{\partial H_C}{\partial q} \, \frac{\partial}{\partial p}
- \frac{\partial H_C}{\partial p} \, \frac{\partial}{\partial q} \right)} f(q, p)
=
\big( 1 + \delta t \pb{ H_C }{ \, \cdot \, } + O(\delta t^2) \big) f(q,p).
\end{align}
Using the above equation together with the fact that the unravelling provides a first order approximation
of the evolution given by the master equation, one can show that the updating rule in this section correctly
reproduces the dynamics of Eq.~\eqref{eq:master_eq_code}.
\par
The code for the CQ unravelling can be found on \href{https://github.com/carlosparaciari/unravelling}{GitHub}.

\end{document}